\documentclass[aps,pre,twocolumn,showpacs,superscriptaddress,%
               showkeys,reprint]{revtex4-1}
\usepackage{graphicx}
\usepackage{bm}
\usepackage{amsfonts,amssymb,amsmath}
\usepackage{color}
\usepackage[colorlinks=true,%
            citecolor=black,%
            linkcolor=black,%
            urlcolor=blue]{hyperref}

\usepackage{graphicx}
\usepackage{amsmath,amsfonts}
\usepackage[english]{babel}
\usepackage{bm}    
\usepackage{eepic} 

\newcommand{\be}{\begin{equation}}
\newcommand{\ee}{\end{equation}}
\newcommand{\<}{\langle}
\renewcommand{\>}{\rangle}

\DeclareMathOperator{\cov}{\mathrm{cov}}

\def\R{{\mathbb R}}
\def\S{{\mathbb S}}

\def\N{{\mathbb N}}
\def\C{{\mathbb C}}

\newcommand{\arxiv}[1]{\href{http://arxiv.org/abs/#1}{\texttt{arXiv:#1}}}
\begin{document}

\title{Phase diagram for the bisected-hexagonal-lattice five-state 
       Potts antiferromagnet}

\author{Jes\'us Salas}
\email{jsalas@math.uc3m.es}
\affiliation{
  Departamento de Matem\'aticas,
  Escuela Polit\'ecnica Superior,
  Universidad Carlos III de Madrid,
  28911 Legan\'es, Spain}
\affiliation{
  Grupo de Teor\'{\i}as de Campos y F\'{\i}sica Estad\'{\i}stica,
  Associate Unit Instituto Gregorio Mill\'an (UC3M)--Instituto de Estructura de
  la Materia (CSIC), Madrid, Spain}

\date{June 5, 2020. Revised August 28, 2020}

\begin{abstract}
In this paper we study the phase diagram of the five-state Potts antiferromagnet
on the bisected-hexagonal lattice. This question is important since 
Delfino and Tartaglia recently showed that a second-order transition in 
a five-state Potts antiferromagnet is allowed, and the bisected-hexagonal 
lattice had emerged as a candidate for such a transition on numerical grounds. 
By using high-precision Monte Carlo simulations and two complementary 
analysis methods, we conclude 
that there is a finite-temperature first-order transition point. This one 
separates a paramagnetic high-temperature phase, and a low-temperature 
phase where five phases coexist. This phase transition is very weak 
in the sense that its latent heat (per edge) is two orders of magnitude 
smaller than that of other well-known weak first-order phase transitions.  
\end{abstract}



\maketitle

%
%
\section{Introduction} \label{sec.intro}

The $q$-state Potts model \cite{Potts_52, Wu_82, Baxter_book, Wu_84} 
is one of most studied models in statistical mechanics, and plays an 
important role in the theory of phase transitions, especially for 
two-dimensional (2D) models. Despite its apparent simplicity, no exact 
solution is known in the whole $(q,T)$ plane, 
where $q$ is an integer $\ge 2$, and $T\in\R$ is the temperature. 
Instead of the temperature $T$, we will use in this paper the variable 
\be
v \;=\; v(T) \;=\; e^{J/(k_B\, T)} -1\,, 
\label{def_v}  
\ee
where $J$ is the coupling constant of the Potts model 
(see Sec.~\ref{sec.potts}), and $k_B$ the Boltzmann constant.

Baxter \cite{Baxter_73,Baxter_82,Baxter_book} found the exact free energy on 
the curve $v = \sqrt{q}$ for the square lattice. This curve has been identified
with the critical curve for the ferromagnetic (FM) regime of the model.  
By universality (see, e.g., Ref.~\cite{Fisher_98} and references therein), 
Baxter's solution implies that the transition is second order for $q\le 4$,
and first order for $q>4$ for any $q$-state FM Potts model defined on any
translation-invariant lattice. Universality has allowed researchers to
understand the phase diagram of the FM regime of this model, their critical
exponents when $q\le 4$, and their connection with conformal field
theories (CFTs) \cite{Nienhaus_84,DiFrancesco_97}.

From a more practical point of view, the Potts model has applications in 
condensed-matter systems \cite{Wu_82,Wu_84,Chalker}. 
From a more abstract point of 
view, the partition function for the $q$-state Potts model on a graph $G$ 
(see Sec.~\ref{sec.potts}) is essentially the same as the so-called 
Tutte polynomial for the graph $G$ \cite{Welsh_00,Sokal_00,handbook_Tutte}. 
This is an object of great interest in combinatorics, as it contains many 
combinatorial information on the graph $G$. This close connection has 
allowed the interchange of methods and ideas from one field to the
other (see, e.g., Refs.~\cite{JSS_forests,JS_flow}). 

Unfortunately, the antiferromagnetic regime $v\in [-1,0)$ of the $q$-state
Potts model is less well understood, as universality does not hold in 
general: the phase diagram depends not only on the dimensionality $D=2$ and
the number of states $q$, but also on the microscopic structure of the 
lattice. This implies that the study of this regime has to be done on a 
case-by-case basis. 

There is some kind of ``\emph{poor man}'' universality in AF Potts models  
\cite{SS_97a} due to the fact that when $q$ is large enough, the system is 
disordered \emph{even} at $T=0$. More precisely, for each 
translation-invariant lattice ${\mathcal G}$, 
there is a value $q_c(\mathcal{G})$ such that:
\begin{itemize}
  \item If $q > q_c(\mathcal{G})$, the system is disordered at all 
             temperatures $T\ge 0$.
  \item If $q = q_c(\mathcal{G})$, the system is critical at $T=0$, and 
             disordered at all positive temperatures $T> 0$. 
  \item If $q < q_c(\mathcal{G})$, any behavior is possible: 
       (a) It can be disordered at all $T\ge 0$ (kagome lattice with $q=2$
           \cite{Suto1,Suto2}).
       (b) It can display critical point at $T=0$, and be disordered at 
           any $T>0$ (triangular lattice with $q=2$ \cite{Stephenson}). 
       (c) It can undergo a finite-$T$ first-order transition between an
           ordered phase and a disordered one (triangular lattice with 
           $q=3$ \cite{Adler_95}).
       (d) It can undergo a finite-$T$ second-order transition between an
           ordered phase and a disordered one (any bipartite lattice 
           with $q=2$, or the diced lattice with $q=3$ \cite{KSS}).
\end{itemize}
This value $q_c(\mathcal{G})$ can be an integer value (like for the square,
and kagome lattices with $q_c=3$, and for the triangular lattice with 
$q_c=4$); but it can be also a noninteger value (like for the hexagonal 
lattice: $q_c=(3+\sqrt{5})/2$). In the latter case, 
we should use the Fortuin--Kasteleyn representation \cite{FK1,FK2}
of the $q$-state 
Potts model to give a rigorous meaning to a $q$-state Potts model with 
a noninteger value of states. For many lattices, this value is 
known only approximately: e.g., the diced lattice 
$q_c(\text{diced})\approx 3.45$ \cite{JS_unpub}, the Union Jack lattice 
$q_c(\text{UJ})=4.326(5)$ \cite{union-jack}, or the lattices shown in 
Ref.~\cite{planar_AF_largeq}. Although, it was expected that the maximum
value were $q_c=4$, this conjecture turned out to be false: there are 
infinite classes of lattices on which $q_c$ is arbitrary large
\cite{KSS,planar_AF_largeq}. This observation is important in the motivation 
of this paper.   

%
%
\begin{figure}[htb]
  \centering
  \begin{tabular}{c} 
  \includegraphics[width=0.8\columnwidth]{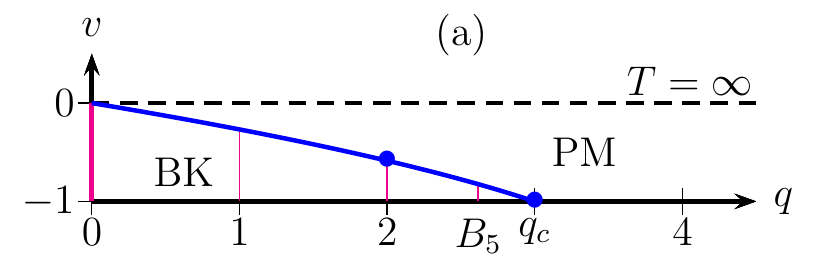} \\
  \includegraphics[width=0.8\columnwidth]{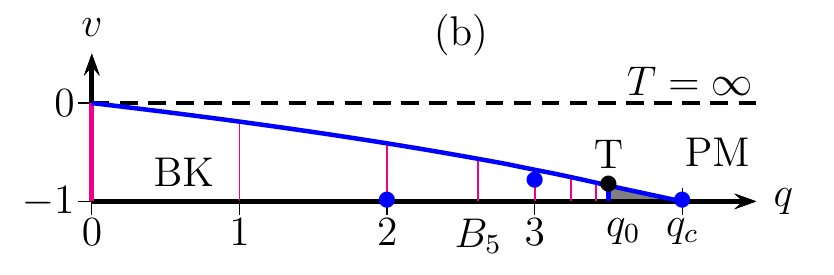} \\
  \end{tabular}
  \vspace*{-2mm}
  \caption{%
  Phase diagram for the AF $q$-state Potts model on the square 
  (a) and on the triangular (b) lattices. The phase labeled PM (resp.\/ 
  BK) represents the paramagnetic (resp.\/ Berker--Kadanoff) phase.
  Notice that in (b), the region for $q\gtrsim 3.5$ has been distorted 
  to show the diagram more clearly. Regime~IV (see text) is colored 
  gray.  
}
\label{fig_phase_diagram}
\end{figure}

The simplest phase diagram for the 2D $q$-state $\mathcal{G}$-lattice 
AF Potts model is qualitatively similar to that of the square lattice 
[see Fig.~\ref{fig_phase_diagram}(a)].
There is a simple AF critical curve $v_\text{AF}(q)$ starting at $(q,v)=(0,0)$
and ending at $(q_c(\mathcal{G}),-1)$. The region above this curve and below
to the line $v=0$ corresponds to a paramagnetic phase, which is disordered.
The region below the curve $v_\text{AF}(q)$ and above the line $v=-1$ 
corresponds to the Berker--Kadanoff phase \cite{Saleur_90,Saleur_91}.   
This is a massless phase with algebraic decay of correlations. In fact,
this phase exists except when $q$ is a Beraha number $B_k$:
\be
B_k \;=\; 4 \cos^2 (\pi/k) \,, \qquad k \;=\; \N \setminus \{1\} \,.
\label{def_beraha}
\ee
At these values, there are massive cancellations of eigenvalues and amplitudes
(in the transfer-matrix formalism; see Refs.~\cite{transfer1,transfer2,%
transfer3,transfer4,transfer_torus,transfer5,transfer6} 
and references therein), so that the dominant eigenvalue is
buried deep inside the spectrum of the corresponding transfer matrix. 
These values are
represented in Fig.~\ref{fig_phase_diagram}(a) by pink vertical lines at
$B_2=0, B_3=1, B_4=2$, and $B_5=(3+\sqrt{5})/2\approx 2.618\, 033\ldots$. 
At these values the thermodynamic limit of the free energy and its derivatives
do not commute with the limit $q\to B_k$. 
This phase diagram is also valid for the diced, hexagonal, Union Jack, 
and BH lattices. 

The phase diagram for the 2D $q$-state triangular-lattice 
AF Potts model is more involved, as it contains an additional element
[see Fig.~\ref{fig_phase_diagram}(b)].
The AF critical curve $v_\text{AF}(q)$ starts at $(q,v)=(0,0)$ with
a slope $dv_\text{AF}/dq |_{q=0} = -0.1753(2)$ \cite{JSS_forests}, and 
moves towards $(q_c(\mathcal{G}),-1)$. However, there is a T-point 
located at $(v_T,q_T)$ with $v_T=-0.95(2)$ and $q_T=3.77(3)$ 
\cite{JSS_tri}. At this T-point, the curve $v_\text{AF}(q)$ splits into
two branches: one goes to the point $(q_0,-1)$ where 
$q_0=3.819\, 671\ldots$ \cite{Baxter_86,Baxter_87}, and the other branch goes
to the point $(q_c(\text{tri}),-1)$. This critical curve was first 
numerically obtained in Ref.~\cite{JSS_tri} using transfer-matrix and 
critical-polynomial methods 
\cite{Jacobsen_12,Scullard_12,Jacobsen_13,Jacobsen_14,Scullard_16}. 
The Berker--Kadanoff phase has the same properties as for the 
square-lattice case, and 
it does not exist for the Beraha numbers $B_k$ \eqref{def_beraha} [in
Fig.~\ref{fig_phase_diagram}(b) we show these numbers up to $B_8$]. 
At $q=B_4=2$ and $q=B_6=3$, it is clear that the thermodynamic limit 
does not commute with the limit $q\to B_k$: the AF critical curve does not
go through the known critical points for $q=2,3$. 
In Fig.~\ref{fig_phase_diagram}(b), the region for $q\gtrsim 3.5$ 
has been distorted
so that the T-point, as well as the two branches, were visible. The region 
enclosed by these points correspond to the so-called Regime~IV in 
Ref.~\cite{Baxter_87}. The conformal properties of this phase were 
first obtained using a Bethe-\emph{Ansatz} approach in Ref.~\cite{Vernier}.  

In recent years, several ``universality classes'' in Potts AF have been found:
\begin{itemize}
  \item[(a)] If $\mathcal{G}$ is a plane Eulerian triangulation (see 
        Sec.~\ref{sec.BH} for details) with one sublattice 
        consisting of vertices of degree $4$, and the other two sublattices 
        are self-dual, then the four-state Potts AF Potts model 
        has a finite-temperature critical point governed by a CFT with 
        central charge $c=3/2$ corresponding to a four-state Potts model 
        and one Ising model (decoupled) \cite{union-jack}. 
  \item[(b)] For the same type of triangulations with the exception that the 
        other two sublattices are not self-dual, then the four-state Potts 
        AF Potts model has a finite-temperature critical point governed 
        by a CFT with central charge
        $c=1$ corresponding to a four-state Potts model \cite{union-jack}. 
  \item[(c)] If $\mathcal{G}$ is plane quadrangulation of \emph{self-dual type} 
        (see Refs.~\cite{selfdual1,selfdual2} for technical details), the 
        three-state Potts AF Potts model has a critical point at $T=0$, and 
        it is disordered for $T>0$.
  \item[(d)] If $\mathcal{G}$ is plane quadrangulation of 
        \emph{non-self-dual type},
        the three-state Potts AF Potts model has a finite-temperature phase
        transition. If this is of second order, it belongs to the 
        universality class of the three-state FM Potts model 
        \cite{selfdual1,selfdual2}, which is governed by a CFT with $c=4/5$. 
\end{itemize} 
The first two examples show that $q_c(\mathcal{G}) > 4$ for any plane Eulerian 
triangulation $\mathcal{G}$; while the last two examples, show that 
$q_c(\mathcal{G}) \ge 3$ for any plane quadrangulation $\mathcal{G}$.

In 2017, Delfino and Tartaglia \cite{Delfino_17} used exact methods of 2D 
field theory to classify the second-order transition points allowed in models 
with $\S_q$ symmetry (i.e., $q$-state Potts models). Here $q$ is assumed to 
be a real parameter. They found several solutions labeled I, II${}_\pm$, 
III${}_\pm$, IV${}_\pm$, and V${}_\pm$ (Ref.~\cite{Delfino_17}, Table~I).  
Solution~III${}_-$ exists for $q\in[0,4]$ and was identified with both the FM
critical and tricritical curves of the Potts model.
Solution~I exists for $q=3$, and is described by a CFT with $c=1$.
Actually, a lattice realization of this solution corresponds to the infinitely
many models already described in point (c) above \cite{selfdual1,selfdual2}.

Their solution~V is a particularly interesting result: it shows that a 
second-oder phase transition is allowed for $q\in [4, (7+\sqrt{17})/2]$,
where $(7+\sqrt{17})/2 \approx 5.561\, 552$; and then, in a five-state AF Potts
model. The latter can occur on one of the infinitely many possible 2D lattices
with $q_c > 5$ and, since they work in field theory (i.e., directly in the
continuum), there is no prediction about which lattice could be a ``good'' one.
While \emph{a priori} the identification of a ``good'' lattice seems quite 
difficult, the results of Ref.~\cite{union-jack} suggested that the 
bisected-hexagonal (BH) lattice was a good candidate, as 
$q_c(\text{BH})= 5.397(5)$, and the results obtained by Monte Carlo (MC) 
\cite{Binder_Landau,Madras} and critical-polynomial methods  
supported the second-order nature of the transition point. There is a 
critical point at $v_c=-0.915\, 32(2)$ with $\gamma/\nu=1.777(3)$, and 
$\alpha/\nu=1.01(1)$. On the other hand, Ref.~\cite{planar_AF_largeq}
contained a detailed study of families of lattices for which $q_c$ is 
arbitrary large. 
Moreover, for $q \gtrsim 8$, the specific heat diverges, close to the
corresponding transition points, like $L^{\approx 2}$.  
This is the signature of a first-order phase transition for a 2D system 
\cite{Nauenberg_74,Klein_76,Fisher_82}. However, the behavior for 
$5 \lesssim q \lesssim 8$ was unclear. Therefore, the question of 
whether the five-state BH-lattice AF Potts model has a second- or first-order
phase transition at $v=v_c$ is still open.  
 
The goal of this paper is to clarify the order of the transition of the 
BH-lattice five-state AF Potts model. If second-order, the result would 
provide a lattice realization of the solution~V of Delfino and Tartaglia. 
If first-order, that would be in accordance with the general behavior 
found in Ref.~\cite{planar_AF_largeq}. Indeed, if this is the case, it would 
not mean that the result of Delfino--Tartaglia was false. On the contrary, it 
would only show that the BH lattice is not a ``good'' lattice in the 
above sense, and one has to look for another lattice $\mathcal{G}$ with
$q_c(\mathcal{G})>5$ that realizes the second-order transition at $q=5$. 

We have studied this model by high-precision
MC simulations using the well-known and efficient Wang--Swendsen--Koteck\'y
(WSK) algorithm \cite{WSK1,WSK2}. Unfortunately, the large value of the
autocorrelation times close to the transition point (namely, 
$\tau_\text{int} \sim 10^4$ for systems of linear size $L=510$), severely 
limited the maximum size we could simulate with at least 
$10^5\, \tau_\text{int}$ Monte Carlo steps (MCS).  

First, we have obtained by preliminary MC simulations a ``rough''
description of the phase diagram of this model. There is a disordered 
paramagnetic phase when the
temperature is large enough (i.e., $v\gtrsim -0.9$). At low temperature,
actually at $T=0$, there are exponentially many ground states which lead to a 
nonzero entropy density. (This provides an exception to the third
law of thermodynamics \cite{Aizenman,Chow}.) However, at low temperatures,
the system is not disordered (as when $q > q_c$); but it effectively behaves  
as having five coexisting phases. The analysis is explained in 
detail in Appendix~\ref{appen.ground}, and it is based on previous 
models \cite{selfdual1,selfdual2} for which there is a ``height'' 
representation of the corresponding $T=0$ spin models 
\cite{Henley_unpub,Kondev_96,Burton_Henley_97,SS_98,Jacobsen_09}. Actually,
the situation is very close to the low-$T$ phase of the three-state 
diced-lattice AF Potts model \cite{KSS}. Therefore, we have two distinct 
regimes separated 
by a transition point. For $v<v_c$, five phases coexist, and for 
$v>v_c$ there is a unique paramagnetic phase. The question is now to 
determine the type of the transition at $v=v_c$.  

The analysis of the MC data has been performed in two complementary ways. 
On one side, we have analyzed the data using the ``standard'' approach: i.e., 
using a general finite-size-scaling (FSS) \emph{Ansatz} 
\cite{Privmann,SS_98} (see also Ref.~\cite{KSS} for a more modern application) 
to fit universal amplitudes like the Binder cumulants or $\xi/L$, where  
$\xi$ is the (second-moment) correlation length. Indeed, the error bars
of all physical quantities were evaluated using the method introduced by Madras
and Sokal \cite{Madras_Sokal,Sokal_97,Madras} that takes into account the 
correlation among successive measurements. From this analysis we obtained
a more precise determination of the transition point $v_c$, and an estimate
for $y_t=2.0(1)$, which agrees well with the predicted value for a 
first-order phase transition on a 2D system  
\cite{Nauenberg_74,Klein_76,Fisher_82}. We also estimated the 
critical-exponent rations $\gamma/\nu$ and $\alpha/\nu$; as well as the 
dynamic critical exponent $z_\text{int}$. However, the results for these
exponents were not conclusive. 
 
The other method of analysis is based on the histogram method 
(see Ref.~\cite{Lee_91} and references therein), as well as on the use of 
reweighting techniques (like the Ferrenberg--Swendsen 
algorithm \cite{FS_89}) and the jackknife method to compute error bars for
correlated data \cite{Weigel,Young}. For a certain models undergoing a 
first-order phase transition (including the $q$-state FM Potts model with
$q$ large enough), there is a rigorous theory \cite{Borgs_90,Borgs_91},
which provides support to a previous phenomenological approach 
\cite{Binder_84,Challa_86}. Using this approach, we located the 
(pseudocritical) temperature $v_\circ(L)$ for which the energy histogram
showed two peaks of \emph{equal} length. It is interesting to note that
for $L\lesssim 48$, this two-peak structure does not exists; and it
appears only for $L\gtrsim 96$. By studying the properties of these
two-peak histograms, we concluded that the system undergoes a first-order
phase transition at $v_c$ with a very small latent heat (per edge) 
$\Delta E = 0.000\, 48(1)$. This number could be compared to the exact latent
heat (per edge) for the five-state FM Potts model on the square lattice
$\Delta E = 0.026 459\ldots$ \cite{Baxter_73}, which is the ``canonical''
example of a weak first-order transition. Moreover, the latent heat (per
edge) for the three-state triangular-lattice AF Potts model is
$\Delta E = 0.0219(5)$ \cite{Adler_95}, which is another example of weak 
first-order transition.  

The plan of the paper is as follows. Section~\ref{sec.setup} contains the
necessary background to make this paper as self-contained as possible. 
We describe briefly the Potts model, and the geometric properties of the
BH lattice. Section~\ref{sec.MC} contains information about the physical 
observables we are going to measure in the MC simulations, as well as details
about the efficiency of the WSK algorithm.  
In Sec.~\ref{sec.res}, we analyze the MC data 
using the standard FSS approach without assuming the order of the transition.
As the results were not completely conclusive, we include in 
Sec.~\ref{sec.histo} another data analysis based on the histogram method,
which has been used quite often to distinguish the order of a transition. 
Finally in Sec.~\ref{sec.conclusions} we discuss our findings. 
In Appendix~\ref{appen.stag}, we discuss the question about the 
right staggering to use in our case. In Appendix~\ref{appen.ground}, we 
study carefully the ground state of our model. This analysis will
provide useful insights to understand the physics of this model.  

%
%
\section{Basic setup} \label{sec.setup}

In this section we will discuss the main topics we will need in the next
sections. In Sec.~\ref{sec.potts}, we summarize the definitions 
about the $q$-state Potts model, and in Sec.~\ref{sec.BH}, we 
describe the BH lattice. 

%
%
\subsection{\texorpdfstring{The $\bm{q}$-state Potts model}
                           {The q-state Potts model}} \label{sec.potts}

The $q$-state Potts model \cite{Potts_52, Wu_82, Baxter_book, Wu_84} is defined 
on any undirected graph $G=(V,E)$ with vertex set $V$ and edge set $E$ (in 
statistical mechanics such graph is usually a finite subset of a regular 
lattice with certain boundary conditions). On each vertex $x\in V$ of 
the graph, we place a spin $\sigma_x$ that can take any value (or `color')
in the set $\Omega = \{1,2,\ldots, q\}$, where $q$ is an integer $q\ge 2$. 
Each spin $\sigma_x$ interacts with the spins $\sigma_y$ located on the 
nearest-neighbor vertices of $x$ with a coupling constant $K\in\R$. 
(Two vertices $x,y\in V$ are nearest neighbors if there is an edge 
$\{x,y\}\in E$.) The partition function of this model is 
\be
Z_G(q,J) \;=\; \sum\limits_{\sigma \colon V \mapsto \Omega} 
\exp \left( \beta \sum\limits_{ \{x,y\} \in E } \delta_{\sigma_x,\sigma_y} 
     \right) \,,
\label{def.Z.Potts}
\ee
where the outer sum is over all spin configurations, the sum inside the
exponential is over all edges of the graph, $\delta_{a,b}$ is the usual 
Kronecker $\delta$, and the coupling constant is given by $\beta = J/(k_B T)$ 
(where $k_B$ is the Boltzmann constant and $T\ge 0$ is the
temperature). If $\beta >0$ (resp.\/ $\beta<0$), the system is in the FM 
(resp.\/ AF) regime. 
As we are interested in the AF regime, we will use the more convenient 
temperature-like parameter $v$ \eqref{def_v}, which appears naturally in 
the Fortuin--Kasteleyn representation of the Potts model \cite{FK1,FK2}. 
In the AF regime, $v\in [-1,0)$. 

%
%
\subsection{The bisected-hexagonal lattice} \label{sec.BH} 

The BH lattice is the Laves lattice $[4,6,12]$, or the
dual of the Archimedean lattice $(4,6,12)$ \cite{tilings}. A BH graph 
$G_{\text{BH},L}$ of size $L\times L$ and embedded on the torus can be 
obtained by following the procedure outlined in Ref.~\cite{union-jack} 
(see also Ref.~\cite{selfdual2}, Sec.~2.2) with a finite triangular graph 
$G_{\text{tri},L}$ of size $L\times L$ and embedded on the torus as the 
starting graph. 

%
%
\begin{figure}[htb]
  \centering
  \includegraphics[width=0.9\columnwidth]{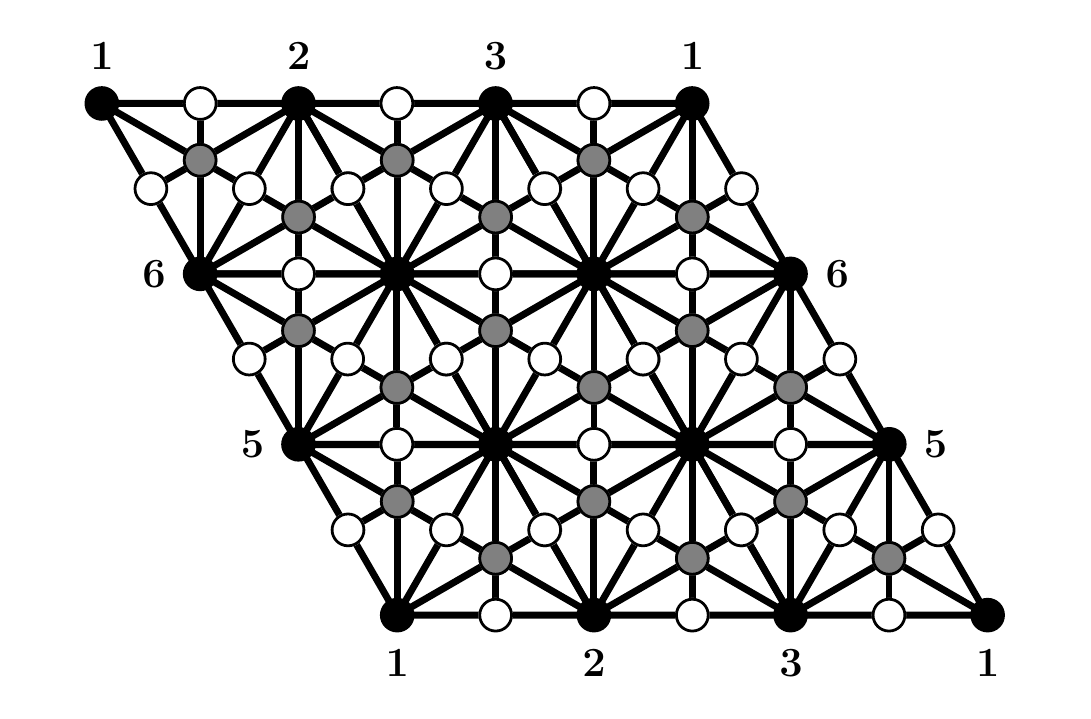}
  \caption{%
   The BH lattice of size $3\times 3$ and periodic boundary conditions. 
   Vertices with the same label should be identified. The
   three sublattices are shown with different colors (black, gray, and white).
   }
  \label{fig.bh}
\end{figure}

The BH graph $G_{\text{BH},L} = (V_\text{BH},E_\text{BH})$ is an Eulerian
triangulation: i.e., all faces are triangles and all vertices have even 
degree. The vertex set $V_\text{BH}$ can be partitioned into three disjoint 
subsets $V_i$ such that if $\{x,y\}\in E_\text{BH}$, then $x,y$ cannot belong 
to the same subset. Each subset forms a sublattice of $G_{\text{BH},L}$.   
In Fig.~\ref{fig.bh}, these sublattices are depicted as black, gray, and 
white dots, respectively. The properties of these sublattices are the 
following:  
\begin{itemize}
\item {\em Sublattice} $A$ contains the $L^2$ vertices of degree $12$, 
      which correspond to the original triangular graph $G_{\text{tri},L}$ 
      of size $L\times L$ (black dots in Fig.~\ref{fig.bh}).
\item {\em Sublattice} $B$ contains the $2L^2$ vertices of degree $6$, and 
      they form the hexagonal graph dual to $G_{\text{tri},L}$ 
      (gray dots in Fig.~\ref{fig.bh}). 
\item {\em Sublattice} $C$ contains the $3L^2$ vertices of degree $4$ 
      (white dots in Fig.~\ref{fig.bh}). 
\end{itemize}
To summarize, the linear size of the BH graph $G_{\text{BH},L}$ is defined
to be the size of the triangular sublattice~$A$. For future convenience, we 
will assume that $L \equiv 0 \pmod{3}$, so that sublattice~$A$
is 3-colorable. The smallest of such graphs is 
depicted in Fig.~\ref{fig.bh}. Then, the number of vertices of 
$G_{\text{BH},L}$ is $|V_\text{BH}|=6L^2$, the number of edges is
$|E_\text{BH}|=18L^2$, and the number of triangular faces is $12L^2$. 
In the following, we will denote as $V_k$ the vertex set of sublattice 
$k\in \{A,B,C\}$. 

\begin{figure}[htb]
  \centering
  \includegraphics[width=0.9\columnwidth]{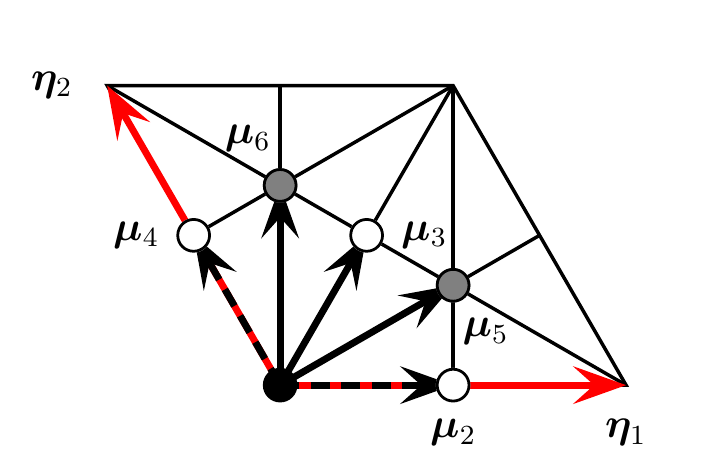}
  \caption{%
   Unit cell of the BH graph. We show the unit vectors $\bm{\eta}_1$ and 
   $\bm{\eta}_2$ that span the underlying triangular sublattice~$A$. 
   We also show the six vectors $\bm{\mu}_i$ that define the basis of the 
   this unit cell. The color code of the vertices is as in Fig.~\ref{fig.bh}. 
   }
  \label{fig.bh.geom}
\end{figure}

The BH graph can be regarded as a triangular Bravais graph (corresponding to 
sublattice~$A$) with a six-site basis (see Fig.~\ref{fig.bh.geom}). 
In order to define some physical observables, it is useful to embed the 
torus in $\R^2$ with the usual Euclidean distance, and draw the BH graph 
$G_{\text{BH},L}$ in such a way that it is not distorted and keeps its
original symmetries.

%
%

A generic vertex $\bm{x}_i$ of the BH graph $G_{\text{BH},L}$ of size 
$L\times L$ is described geometrically by three numbers $(x'_1,x'_2,k)$ as
\begin{equation}
\bm{x}_k \;=\; x'_1 \bm{\eta}_1 + x'_2 \bm{\eta}_2 + 
       \sum\limits_{j=1}^6 \delta_{k,j}\, \bm{\mu}_j , 
\;\; x'_1,x'_2=1,\ldots L . 
\label{def_x_BH}
\end{equation}
The vector $\bm{x}'=(x'_1,x'_2)^t$ lives on a the triangular sublattice~$A$  
spanned by the (unit) vectors:
\begin{equation}
\bm{\eta}_1 \;=\; \left(1,0\right)^t\,, \qquad  
\bm{\eta}_2 \;=\; \left(-\frac{1}{2},\frac{\sqrt{3}}{2} \right)^t 
\label{def_etas} 
\end{equation}
and marks the position of the unit cells. The last term in \eqref{def_x_BH} 
shows the relative position w.r.t. $\bm{x}'$ of the vertices in the 
corresponding unit cell. Thus, the subindex $k$ in $\bm{x}_k$ indicates 
the sublattice the vertex belongs to.
Indeed, $\bm{\mu}_1 =(0,0)^t$ for all vertices in sublattice $A$. 
The two vectors (of norm $1/\sqrt{3}$) associated to the vertices in 
sublattice~$B$ are
\be
\bm{\mu}_5 \;=\; \frac{1}{\sqrt{3}}\,
                  \left(\frac{\sqrt{3}}{2}, \frac{1}{2} \right)^t\,, \qquad
\bm{\mu}_6 \;=\; \frac{1}{\sqrt{3}}\, \left(0,1\right)^t\,.
\label{def_mu6}
\ee
The three vectors (of norm $1/2$) associated to the vertices in  
sublattice~$C$ are:
\be
\bm{\mu}_2 \;=\; \frac{1}{2}\, \bm{\eta}_1 \,, \quad
\bm{\mu}_3 \;=\; \frac{1}{2}\, (\bm{\eta}_1 + \bm{\eta}_2) \,, \quad  
\bm{\mu}_4 \;=\; \frac{1}{2}\, \bm{\eta}_2 \,.
\label{def_mu4}
\ee
Therefore, in this geometric representation not all edges have the same length.
We will denote a vertex $\bm{x}$ or $x$ depending on whether we are using this
geometric representation or not. 

%
%
\section{Monte Carlo simulations} \label{sec.MC}

This section is devoted to describe the MC simulations we have performed. 
First, in Sec.~\ref{sec.observables}, we describe the observables we have
measured. In Sec.~\ref{sec.WSK}, we discuss the MC algorithm we
have used. 

%
%
\subsection{Physical observables} \label{sec.observables}

The simplest observable is the internal energy
\begin{equation}
\mathcal{E} \;=\; \sum\limits_{\{x,y\}\in E_\text{BH}} 
                  \delta_{\sigma_x\sigma_y} \,,
\label{def_E}
\end{equation}
where the sum is over all edges in the BH graph.  

For the magnetic observables, it is convenient to use the tetrahedral 
(vector) representation of the spins $\bm{\sigma}_{\bm{x}}\in \S^{q-1}$:
\begin{equation}
\bm{\sigma}_{\bm{x}} \;=\; \sum\limits_{\alpha=1}^q 
\delta_{\sigma_{\bm{x}},\alpha} \bm{e}^{(\alpha)} \,,
\label{def_bsigma}
\end{equation}
where the vectors $\bm{e}^{(\alpha)}$ satisfy
\begin{equation}
\bm{e}^{(\alpha)} \cdot \bm{e}^{(\beta)} \;=\; 
\frac{q \delta_{\alpha \beta}-1}{q-1} \,.
\label{eq_e}
\end{equation}

The second observable is the staggered magnetization. In general,
the staggering assigns a phase $e^{i \phi_k}\in \C$ to every vertex
belonging to the $k$-th sublattice. Unfortunately, it is not clear 
\emph{a priori} what is the right staggering $\phi_k$ to choose for the BH
lattice when $q=5$. In Appendix~\ref{appen.stag}, we have found that a good
choice is to consider the magnetization of the spins in sublattice~$A$ 
[cf., Eq.~\eqref{def_mag_gen_OK}]. Then, we will consider 
\be
\bm{\mathcal{M}} \;=\; \bm{\mathcal{M}}_A \;=\; 
           \frac{1}{V_A} \, \sum\limits_{x\in V_A} \bm{\sigma}_{x} \,.
\label{def.M.OK}
\ee
We are interested in the squared magnetization given by 
[cf., \eqref{def_bsigma} and~\eqref{eq_e}]   
\begin{equation}
\bm{\mathcal{M}}^2 \;=\; 
 \frac{q}{q-1} \, \frac{1}{V_A^2}\, \sum\limits_{\alpha=1}^q \left( 
         \sum\limits_{x\in V_A} \delta_{\sigma_x,\alpha} \right)^2 
       - \frac{1}{q-1} \,. 
\label{def.M2.OK}
\end{equation}

The above observable is a ``zero-momentum'' one. In order to 
compute the second-moment correlation length $\xi$, we need to consider the 
Fourier transform of the spin variables and define the observable
$\widetilde{\bm{\mathcal{M}}}(\bm{k})$:   
\be
\widetilde{\bm{\mathcal{M}}}(\bm{k}) \;=\; \frac{1}{V_A} \, 
\sum_{\bm{x}\in V_A} \bm{\sigma}_{\bm{x}} \, e^{i \bm{k}.\bm{x} }  
\label{def.Fourier.spin}
\ee
evaluated at the smallest allowed nonzero momenta $\bm{k}$. 
As seen in Sec.~\ref{sec.BH}, the translation invariance of the BH graph 
$G_{\text{BH},L}$ is that of the underlying triangular Bravais sublattice~$A$. 
Therefore, the allowed momenta for $G_{\text{BH},L}$ are given by   
\begin{equation}
\bm{k} \;=\; \frac{2\pi}{L} \bigl( m_1 \bm{\rho}_1 + m_2 \bm{\rho}_2\bigr)\,, 
\quad m_1,m_2 = 1,\ldots,L , 
\label{def_momenta}
\end{equation}
where the momenta basis is given by
\be
\bm{\rho}_1 \;=\; \frac{2}{\sqrt{3}} \,
            \left(\frac{\sqrt{3}}{2},\frac{1}{2} \right)^t\,, \quad
\bm{\rho}_2 \;=\; \frac{2}{\sqrt{3}} \, \left(0,1\right)^t \,,
\label{def_rhos}
\ee
and satisfy [cf., \eqref{def_etas}]: 
\begin{equation}
\bm{\eta}_i \cdot \bm{\rho}_j \;=\; \delta_{ij} \,.
\end{equation}
The set of the smallest nonzero momenta is  
\begin{equation}
\mathcal{K} \;=\; \left\{ \pm \frac{2\pi}{L}\bm{\rho}_1, 
                     \pm \frac{2\pi}{L}\bm{\rho}_2,
            \pm \frac{2\pi}{L}(\bm{\rho}_1-\bm{\rho}_2) \right\} \,. 
\label{def_K}
\end{equation}
The six momenta $\bm{k}\in \mathcal{K}$ have a norm $4\pi/(\sqrt{3}L)$.

Thus, the square of the Fourier transform \eqref{def.Fourier.spin} evaluated 
at the smallest nonzero momenta is given by 
\begin{eqnarray}
\mathcal{F} &=& \frac{1}{6} \, \sum\limits_{\bm{k}\in \mathcal{K}} 
                \widetilde{\bm{\mathcal{M}}}(\bm{k})^* \cdot  
                \widetilde{\bm{\mathcal{M}}}(\bm{k}) \nonumber \\ 
          &=& \frac{q}{q-1}\, \frac{1}{3}\, \frac{1}{V_A^2}
           \sum\limits_{\bm{k}\in \mathcal{K_+}} \, 
           \sum\limits_{\alpha=1}^q \left|
           \sum\limits_{\bm{x}\in V_A} e^{i \bm{k}\cdot \bm{x}}
           \, \delta_{\sigma_{\bm{x}},\alpha} \right|^2 , \quad 
\label{def.F.OK}
\end{eqnarray}
where in the last equation, $\mathcal{K}_+ \subset \mathcal{K}$ contains the 
three momenta with positive sign. We have included \emph{all} the allowed 
nonzero momenta in $\mathcal{K}_+$ to increase the statistics of this 
observable. [The three momenta with negative sign in $\mathcal{K}$ 
\eqref{def_K} do not add additional information due to the exact symmetry
$\bm{k}\to-\bm{k}$, and can be eliminated to save CPU time.] 
To obtain the final result \eqref{def.F.OK}, we have made use of 
Eqs.~\eqref{def_bsigma} and~\eqref{eq_e}, and the fact that 
$\bm{k}\neq \bm{0}$ for all $\bm{k}\in \mathcal{K}$. 

Starting from the energy \eqref{def_E}, we can compute several mean  
values of interest: the energy density (per edge) $E$, the specific heat
$C_H$, and the thermal Binder-like ratio $U_4$ 
(Ref.~\cite{Billoire_Lacaze_Morel_92}, footnote on p.~776): 
\begin{subequations}
\label{def_thermal.obs}
\begin{align}
\label{def_E_mean}
E(v;L)   &\;=\; \frac{1}{|E_\text{BH}|} \, \left\< \mathcal{E} \right\>\,, \\
\label{def_CH}
C_H(v;L) &\;=\; \frac{1}{|E_\text{BH}|} \,
        \left\< (\mathcal{E} - \< \mathcal{E}\>)^2 \right\>\,, \\
\label{def_U4}
U_4(v;L) &\;=\; \frac{\left\< (\mathcal{E} - \< \mathcal{E}\>)^4 \right\> }
               { \left\< (\mathcal{E} - \< \mathcal{E}\>)^2 \right\>^{2}}\,,
\end{align}
\end{subequations}
where $|E_\text{BH}|=18L^2$.  

The values of $E(v;L)$ and $C_H(v;L)$ are easy to obtain in the thermodynamic
limit at the extreme cases $v=0$ and $v=-1$. At $v=0$, when the spins are 
completely uncorrelated, we have that 
\begin{subequations}
\begin{align}
\label{def_E_v=0}
E(0;\infty)   &\;=\; \frac{1}{q} \;=\; \frac{1}{5} \,, \\ 
\label{def_CH_v=0}
C_H(0;\infty) &\;=\; \frac{q-1}{q^2} \;=\; \frac{4}{25} \,. 
\end{align}
\end{subequations}
At $v=-1$, the spin configurations are just proper 5-colorings of the graphs 
$G_{\text{BH},L}$, and because they are 3-colorable,   
then $E(-1,\infty)=C_H(-1,\infty)=0$.
The values of $U_4$ \eqref{def_U4} are easy to compute when the energy
density can be approximated by a single Gaussian. This is true when the
system is in a disordered phase or in an ordered one, but not at the 
transition point. In the former cases, it attains the same value  
\cite{Tsai_98}: 
\be
U_4(v;\infty) \;=\; 3 \,, \qquad v \;\neq\; v_c \,. 
\label{def_U4_v_neq_vc}
\ee
At finite $L$, $U_4(v;L)$ displays a minimum close to $v=v_c$
\cite{Billoire_Lacaze_Morel_92}. If the transition is first order, then 
this minimum converges to $1$, and to a nontrivial value, if the transition 
is second order.  

In the magnetic sector, we define from \eqref{def.M2.OK} and \eqref{def.F.OK} 
the susceptibility $\chi$, the corresponding ``nonzero-momenta'' quantity $F$, 
and the magnetic Binder ratio $R$:
\begin{subequations}
\label{def_magnetic.obs}
\begin{align}
\label{def_chi}
\chi(v;L) &\;=\; |V_A| \, \left\< \bm{\mathcal{M}}^2 \right\>\,,\\
\label{def_F}
F(v;L)    &\;=\; |V_A| \, \left\< \mathcal{F} \right\>\,,\\
\label{def_R}
R(v;L)    &\;=\; \frac{\left\< (\bm{\mathcal{M}}^2)^2 \right\>  }
                 {\left\<  \bm{\mathcal{M}}^2    \right\>^2}\,, 
\end{align}
\end{subequations}
where $|V_A|=L^2$.  
 
Notice that our definition of the susceptibility does \emph{not} contain
the connected part as $\< \bm{\mathcal{M}} \> = \bm{0}$ for an infinitely long  
MC simulation. Therefore, at $v=0$, the susceptibility \eqref{def_chi} takes
the value
\be
\chi(0;\infty) \;=\; 1 \,,
\label{def_chi_v=0}
\ee
and at $v=-1$, it should grow like $L^2$. 
On the other hand, the Binder ratio $R$ has the following limiting values 
\cite{KSS}:
\begin{subequations}
\begin{align}
\label{def_R_v=0}
R(0;\infty)  &\;=\; \frac{q+1}{q-1} \;=\; \frac{3}{2} \,, \\
R(-1;\infty) &\;=\; 1 \,, 
\label{def_R_v=-1}
\end{align}
\end{subequations}
where we have assumed that at $T=0$ the system is ordered. 

Finally, the second-moment correlation length $\xi$ is defined as
\begin{equation}
\xi(v;L) \;=\; \frac{1}{2\sin(\pi/L)}\, \sqrt{\frac{\chi(v;L)}{F(v;L)} -1} \,.
\label{def_xi}
\end{equation}
Indeed, due to our definition of the susceptibility \eqref{def_chi}, this
formula gives the right correlation length only in the disordered phase. At
$v=0$, the correlation length $\xi(v;\infty)$ vanishes. If there is an 
ordered low-$T$ phase, then we expect that our definition 
\eqref{def_xi} implies that $\xi(-1;L)$ should grow like $L^2$.

In order to compute the error bars of $U_4$, $R$ and $\xi$, we first 
compute the variance of the observables \cite{SS_00}:
\begin{subequations}
\begin{align}
\mathcal{O}_4 &\;=\; \frac{ (\mathcal{E} - \< \mathcal{E}\>)^4 }
                       { \< (\mathcal{E} - \< \mathcal{E})^4 \> }
  -          2 \, \frac{    (\mathcal{E} - \< \mathcal{E}\>)^2 }
                       { \< (\mathcal{E} - \< \mathcal{E})^2 \> }
  + 1 \,, \\[2mm]
\mathcal{O}_R &\;=\; \frac{\bm{\mathcal{M}}^4 }
                       { \< \bm{\mathcal{M}}^4 \> }
  -          2 \, \frac{    \bm{\mathcal{M}}^2 }
                       { \< \bm{\mathcal{M}}^2 \> }
  + 1 \,, \\[2mm]
\mathcal{O}_\xi &\;=\; \frac{\bm{\mathcal{M}}^2 }
                         { \< \bm{\mathcal{M}}^2 \> }
  -                  \, \frac{   \mathcal{F}}
                         { \<    \mathcal{F} \> } \,.
\end{align}
\end{subequations}
Please note that they all have zero mean values. Then the desired standard
deviations are given by the following expressions:  
\begin{subequations}
\begin{align}
\sigma(U_4) &\;=\; U_4\, \sigma(\mathcal{O}_4)\,, \\
\sigma(R)   &\;=\; R  \, \sigma(\mathcal{O}_R)\,, \\[2mm]
\sigma(\xi) &\;=\; \frac{1}{4 \sin(\pi/L)} \, \frac{\chi}{F} \,
      \left(\frac{\chi}{F} -1\right)^{-1/2}\, \sigma(\mathcal{O}_\xi) \,.
\end{align}
\end{subequations}

%
%
\subsection{The Wang--Swendsen--Kotek\'y algorithm} \label{sec.WSK}

In order to simulate the five-state AF Potts model on the BH lattice, we have 
used the algorithm of choice: the Wang--Swendsen--Kotek\'y (WSK) algorithm  
\cite{WSK1,WSK2}. This cluster algorithm is a legitimate one for any 
graph and any \emph{positive} temperature. The main trouble with the WSK 
algorithm is that it is \emph{not} generically ergodic (or irreducible) at 
$T=0$, where many AF models show interesting physical phenomena. 
Moreover, it is well known that the WSK algorithm is ergodic at any temperature
$T\ge 0$ for any \emph{bipartite} graph $G$ and for any integer $q\ge 2$ 
\cite{Mohar,Burton_Henley_97,Ferreira_Sokal}. 

For the majority of the critical points that have being studied using the WSK 
algorithm, its dynamic behavior overcomes that of single-site algorithms 
(e.g., Metropolis), which is given by a dynamic critical exponent 
$z_{\text{int}},z_{\text{exp}} \gtrsim 2$ (see, e.g., Ref.~\cite{Sokal_97}).  
 
In particular, for $q=3$ and $G$ being a certain bipartite class of 
quadrangulations $\mathcal{Q}$ on the torus \cite{selfdual1,selfdual2}, 
it has been conjectured (based on strong numerical support) that there is a 
critical point at $T=0$, but WSK shows no critical slowing down: 
i.e., $\tau_{\text{int}},\tau_{\text{exp}} \le A$ uniformly 
in $v$ and $L$. This was first discovered on the square lattice 
\cite{Ferreira_Sokal,SS_98}. 

Moreover, for $q=3$ and $G$ being any bipartite quadrangulation on the torus
not belonging to $\mathcal{Q}$ \cite{selfdual1,selfdual2}, a similar conjecture
has been claimed: there is a finite-temperature critical point, and WSK 
belongs to the same dynamic universality class as the Swendsen--Wang 
\cite{SW} cluster algorithm for the three-state FM Potts model 
(with $z_{\text{int},\mathcal{M}^2} = 0.475(6)$ 
\cite{SS_97,Garoni_11}). This phenomenon was first found on the diced lattice
\cite{KSS}. 

In addition to these cases, for $q=3$ and the hexagonal lattice it was also
found that WSK has no critical slowing down \cite{JS_98}. 
(But this should be expected, as $q_c(\text{hex}) < 3$).  

Finally, the three-state AF Potts model on the triangular lattice displays a 
finite-temperature \emph{weak} first-order phase transition 
\cite{Adler_95}, so that WSK is expected to perform poorly: i.e., the
autocorrelation times are expected to grow exponentially fast 
$\tau_{\text{int}},\tau_{\text{exp}} \sim e^{2\sigma_{o,d} L}$ for 2D systems,
where $\sigma_{o,d}$ is the interface tension \cite{Sokal_97}.
 
For nonbipartite graphs, the question whether WSK is irreducible or not
should be investigated on a case-by-case basis. It is worth to note that the 
lack of ergodicity is usually not a ``big'' problem on planar graphs: if
$G$ is a planar graph with chromatic number $\chi(G)$, then WSK is ergodic 
for any $q > \chi(G)$ \cite{Mohar}. In particular, for any finite 3-colorable 
subset of the BH lattice with e.g., free boundary conditions, WSK is  
ergodic for any $q\ge 4$; in particular, for $q=5$.   

In statistical mechanics, one is interested in graphs embedded on a torus 
(i.e., periodic boundary conditions) to get rid of surface effects. 
If we consider finite subsets of nonbipartite translation-invariant 
lattices $\mathcal{L}$ on the torus, then WSK is not ergodic for important 
physical cases at $q=q_c(\mathcal{L})$: triangular lattices 
$G_{\text{tri},3L}$ (with $L\ge 2$) at $q=4$ \cite{MS1}, kagome lattices 
$G_{\text{kag},3L}$ (with $L\ge 1$) at $q=3$ \cite{MS2}, and even 
nonbipartite square lattices at $q=3$ \cite{Lubin_Sokal}.  

\begin{figure}[htb]
  \centering
  \includegraphics[width=0.8\columnwidth]{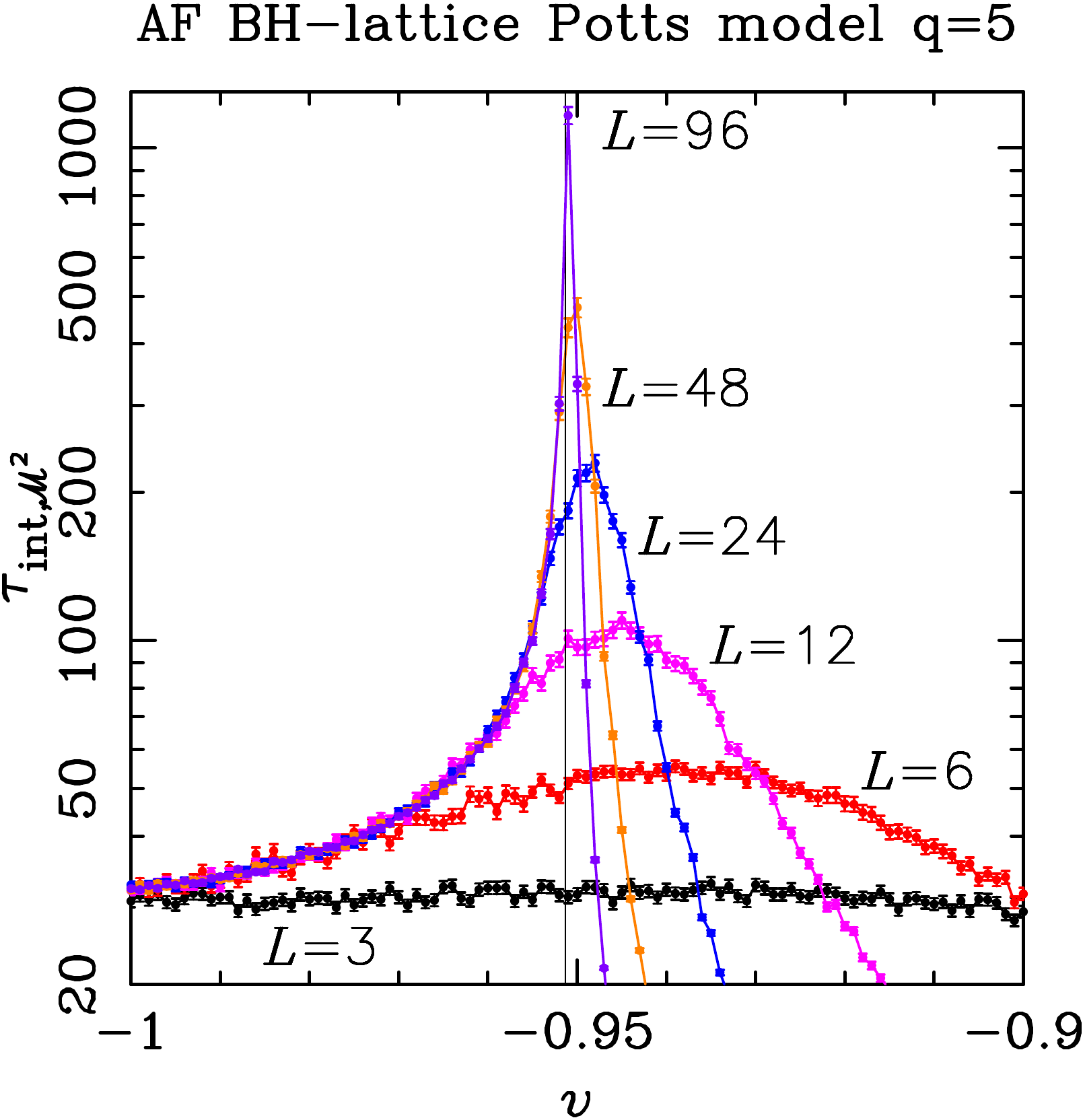}
  \vspace*{-2mm}
  \caption{%
   Integrated autocorrelation time $\tau_{\text{int},\bm{\mathcal{M}}^2}$ 
   for the five-state BH-lattice AF Potts model in the interval $v\in[-1,-0.9]$.
   We show data for $L=3$ (black), $L=6$ (red), $L=12$ (pink), $L=24$ (blue),
   $L=48$ (brown), and $L=96$ (violet). The vertical thin line marks the 
   estimate $v_c=-0.95132(2)$ \cite{union-jack}. Points have been joined 
   with lines to guide the eyes. 
   }
  \label{fig.tauM2.fine}
\end{figure}

Moreover, for larger values of $q$, ergodicity may be ``restored'': 
If $\Delta$ is the maximum degree of a graph $G$, then WSK is ergodic for 
any $q \ge \Delta +1$. In addition, if $G$ is connected and contains a vertex 
of degree $< \Delta$, then WSK is ergodic for any $q\ge \Delta$ 
\cite{Mohar,Jerrum}. This result implies that WSK is ergodic at $T=0$ on 
the BH graphs $G_{\text{BH},L}$ for any $q\ge 12 \gg 5$. Furthermore,
it has been proven that WSK is also ergodic on any triangular (resp.\/ kagome)
graph for $q\ge 6$ (res.\/ $q\ge 4$) \cite{McDonald,Bonamy_19}, and there is 
strong numerical evidence that it is also ergodic for $G_{\text{tri},3L}$ and
$q=5$ \cite{SS_tri}.   
For the BH lattice, previous MC simulations have located its critical 
point at $v_c=-0.95132(2) > -1$ \cite{union-jack}, so in principle, we 
do not have to worry about WSK not being ergodic at $T=0$. However, our 
past experience with MC simulations for the four-state triangular-lattice 
AF Potts model \cite{SS_tri}, has shown that nonergodicity at $T=0$ 
may induce systematic errors at $T>0$. As $|v_c + 1| \approx 0.05 \ll 1$, this 
phenomenon cannot be discarded. The slowest mode of the WSK dynamics for
the BH graphs $G_{\text{BH},L}$ and $q=5$ is $\bm{\mathcal{M}}^2$, 
as in other similar models \cite{Ferreira_Sokal,SS_98,KSS,selfdual1,selfdual2}. 
Therefore, for each value of $L=3,6,12,24,48,96$, we have made $101$ MC 
simulations at equidistant values in the range $v\in [-1,-0.9]$ 
to investigate how the autocorrelation time 
$\tau_{\text{int},\bm{\mathcal{M}}^2}$ behaves as a function
of $v$ and $L$ (see Fig.~\ref{fig.tauM2.fine}). 
The length of these simulations are in the range 
$10^6$ to $8\times 10^6$ MCS; we discarded the first $10\%$ MCS 
($\gtrsim 1.2\times 10^3 \tau_{\text{int},\bm{\mathcal{M}}^2}$) to eliminate
the initialization bias, and the number of measurements was 
$\gtrsim 1.1\times 10^4 \tau_{\text{int},\bm{\mathcal{M}}^2}$.
 
Figure~\ref{fig.tauM2.fine} shows that $\tau_{\text{int},\bm{\mathcal{M}}^2}$ 
displays a peak that, as $L$ increases, moves towards the estimate for
$v_c=-0.95132(2)$, while its width becomes narrower. In the ordered phase, the
behavior is rather smooth and the curves for different values of $L$ 
converge to a value $\tau_{\text{int},\bm{\mathcal{M}}^2} \approx 31$ at $T=0$. 
In the disordered phase, the curves converge as $v\to 0$ (not shown 
in the figure) to a value $\tau_{\text{int},\bm{\mathcal{M}}^2} \approx 3.5$ at 
$v=0$. This convergence is faster for larger values of $L$. 
In conclusion, we observe a single peak around the transition point that 
behaves in the expected way for a critical point. Outside the critical 
region, it behaves rather smoothly in $v$ and $L$, and converges to fixed 
values at $v=-1$ and $v=0$. This is empirical evidence that WSK is ergodic 
on the BH graphs $G_{\text{BH},L}$ for $q=5$ at $T=0$. 

%
%

We have used the 64-bit linear congruential pseudo-random-number 
generator proposed by Ossola and Sokal \cite{Ossola_Sokal_04a,Ossola_Sokal_04b}.
This generator is very simple, is relatively fast, and has been 
successfully used to obtain high-precision data for the MC simulations of the 
three-dimensional Ising model ($q=2$) with the Swendsen--Wang algorithm. 

In this paper, we have made several fits of the finite-size data of some 
quantity in order to obtain the relevant physical information. We have 
used {\sc Mathematica}'s function {\tt NonlinearModelFit} to perform the 
weighted least-squares method for both linear and nonlinear \emph{Ans\"atze}. 
In order to detect corrections to scaling not taken into account in the 
\emph{Ansatz}, we have repeated each fit by only allowing data with 
$L\ge L_\text{min}$. We then study the behavior of the estimated
parameters as a function of $L_\text{min}$. For each fit we report  
the observed value of the $\chi^2$, the number of degrees of freedom (DF), 
and the confidence level (CL). This quantity is the probability that the 
$\chi^2$ would exceed the observed value, assuming that the underlying 
statistical model is correct. A ``reasonable'' CL corresponds to 
$\text{CL}\gtrsim 10-20\%$, and a very low $\text{CL}\lesssim 5\%$ suggests 
that this underlying statistical model is incorrect. Commonly, this is due to 
additional corrections to scaling not taken into account. 

Finally, we have performed some simple checks at $v=0$ and $v=-1$ (assuming
that WSK is ergodic at this latter temperature). 

The values of the observables at $v=0$ have been checked by performing MC
simulations on systems of linear sizes $L=3,6,12,24,48$. For the internal
energy $E(0;L)$ \eqref{def_E_mean}, the FSS corrections are so 
small, that a fit to a constant is enough: for $L_\text{min}=3$, we 
already obtain a good fit  
\be
E(0;\infty) \;=\; 0.199999(3)\,, 
\label{def_E_v=0_exp}
\ee
with $\chi^2/\text{DF} = 1.75/4$ and $\text{CL} = 78\%$. 
For the specific heat $C_H(0;L)$ \eqref{def_CH}, we use the same constant
\emph{Ansatz}, and for $L_\text{min}=3$, we obtain 
\be
C_H(0;\infty) \;=\; 0.1601(2)\,, 
\label{def_CH_v=0_exp}
\ee
with $\chi^2/\text{DF} = 1.53/4$ and $\text{CL} = 82\%$. 
Finally, the fit for the Binder-like cumulant $U_4(0;L)$ \eqref{def_U4} needs
an additional FSS correction:
\be
U_4(0;L) \;=\; U_4(0,\infty) + A \, L^{-2} \,.
\label{def_Ansatz_U4}
\ee
For $L_\text{min}=3$, we obtain
\be
U_4(0;\infty)  \;=\; 3.002(4) \,,
\label{def_U4_v=0_exp}
\ee
with $\chi^2/\text{DF} = 1.95/3$ and $\text{CL} = 58\%$. 
In all cases, the estimates 
\eqref{def_E_v=0_exp}, \eqref{def_CH_v=0_exp}, and \eqref{def_U4_v=0_exp} 
agree very well with the exact values 
\eqref{def_E_v=0}, \eqref{def_CH_v=0}, and \eqref{def_U4_v_neq_vc}. 
 
The susceptibility $\chi(0;L)$ \eqref{def_chi}, can be well described by a fit 
to a constant; for $L_\text{min}=3$, we obtain the estimate 
\be
\chi(0;\infty) \;=\; 1.0011(9) \,,
\label{def_chi_v=0_exp}
\ee
with $\chi^2/\text{DF} = 1.20/4$ and $\text{CL} = 88\%$.  
The fit to the Binder cumulant $R(0;L)$ \eqref{def_R} needs the \emph{Ansatz} 
\eqref{def_Ansatz_U4}. For $L_\text{min}=3$, we get 
\be
R(0;\infty) \;=\; 1.5007(9)\,,
\label{def_R_v=0_exp}
\ee
with $\chi^2/\text{DF} = 2.71/3$ and $\text{CL} = 44\%$. Again, the 
agreement among the estimates \eqref{def_chi_v=0_exp} and \eqref{def_R_v=0_exp}
and the exact values \eqref{def_chi_v=0} and \eqref{def_R_v=0} 
is also very good.  

The values of the observables at $v=-1$ have been obtained from MC
simulations on systems of linear sizes $L=3,6,12,24,48,96$. 
We have fitted the susceptibility $\chi(-1;L)$ \eqref{def_chi} to a power-law 
\emph{Ansatz}: for $L_\text{min}=24$, we get
\be
\chi(-1;L) \;=\; 0.8890(3) \, L^{1.99985(7)}\,,
\ee
with $\chi^2/\text{DF} = 0.017/1$ and $\text{CL} = 89\%$. 
The fit for the correlation length $\xi(-1;L)$ \eqref{def_chi} is good
if we consider a power-law plus a constant. For $L_\text{min}=12$, we
obtain
\be
\xi(-1;L) \;=\; 0.340(2) \, L^{2.001(1)} + 2.73(2) \,, 
\ee
with $\chi^2/\text{DF} = 0.28/1$ and $\text{CL} = 59.8\%$. 
Finally, the fit of the Binder cumulant $R(-1;L)$ \eqref{def_R} to the 
\emph{Ansatz} \eqref{def_U4}, gives for $L_\text{min}=24$: 
\be
R(-1;\infty) \;=\; 1.0000002(3) \,, 
\label{def_R_v=-1_exp}
\ee
with $\chi^2/\text{DF} = 0.122/1$ and $\text{CL} = 72\%$. In first two cases,
we have checked that the behavior at $v=-1$ is the expected one,
namely $\propto L^2$, and the value \eqref{def_R_v=-1_exp} agrees well with 
the value \eqref{def_R_v=-1}. It is worth noting that these three results give 
support to the existence of an ordered phase at low temperature.  

%
%
\section{Numerical results} \label{sec.res}

In this section we will discuss the MC simulations we have done, and the 
results for the physical quantities of interest. We have followed a similar
methodology as in Ref.~\cite{KSS}. 

%
%
\subsection{Summary of the MC simulations} \label{sec.summ.MC}

At every performed MC simulation, we have measured some basic observables
$\mathcal{E}$ \eqref{def_E}, $\bm{\mathcal{M}}^2$ \eqref{def.M2.OK}, and  
$\mathcal{F}$ \eqref{def.F.OK}. From these measurements, we obtain the 
basic (static) physical quantities \eqref{def_thermal.obs}, 
\eqref{def_magnetic.obs}, and \eqref{def_xi}. However,
in order to compute their correct error bar, we need to estimate the 
corresponding integrated autocorrelation times $\tau_{\text{int},\mathcal{O}}$
for $\mathcal{O}\in \{\mathcal{E}, \bm{\mathcal{M}}^2, \mathcal{F}\}$. 
We have achieved this computation by using the self-consistent 
algorithm introduced by Madras and Sokal \cite{Madras_Sokal,Sokal_97}
(see also Ref.~\cite{SS_97}). In this paper, we will call 
$\tau_{\text{int}}$ the maximum of the measured autocorrelation times
$\tau_{\text{int},\mathcal{O}}$.

We have performed several preliminary sets of MC simulations to isolate the
regions of interest in $v$, to determine a rough description of the phase 
diagram, and to check that the dynamic behavior of the WSK algorithm is 
correct. In these simulations, we focused on the thermal quantities 
\eqref{def_thermal.obs}, and on the mean magnetization quadratic form 
$\mathsf{M}$ [c.f., \eqref{def_mab}] (we are interested in its dominant 
eigenvalue and its corresponding eigenvector; see Appendix~\ref{appen.stag}
for details). These runs were split in several groups: 

\begin{itemize}
 \item[a)] We made $101$ runs at equidistant values of $v \in [-1,0]$ for 
       $L=3,6,12,24,48$. They showed that the interesting range was 
       $v\in [-1,-0.9]$, as for $v>-0.9$ the system behaved as it were in
       a disordered phase (e.g., $U_4 \approx 3$). In each run, we discarded 
       $\gtrsim 1.3\times 10^3 \tau_{\text{int}}$ MCS, and took  
       $\gtrsim 1.1\times 10^4 \tau_{\text{int}}$ measurements.

 \item[b)] We made $101$ runs at equidistant values of $v \in [-1,-0.9]$ for 
       $L=3,6,12,24,48,96$. We obtained that the phase transition was very
       close to the previous estimate $v_c=-0.95132(2)$ \cite{union-jack}.
       We could also monitor the behavior of 
       $\tau_{\text{int},\bm{\mathcal{M}}^2}$; the result is practically 
       identical to Fig.~\ref{fig.tauM2.fine}. In each run, we discarded   
       $\gtrsim 1.2\times 10^3 \tau_{\text{int}}$ MCS, and took
       $\gtrsim 1.1\times 10^4 \tau_{\text{int}}$ measurements.
\end{itemize}

%
%
\begin{figure}[htb]
  \includegraphics[width=0.8\columnwidth]{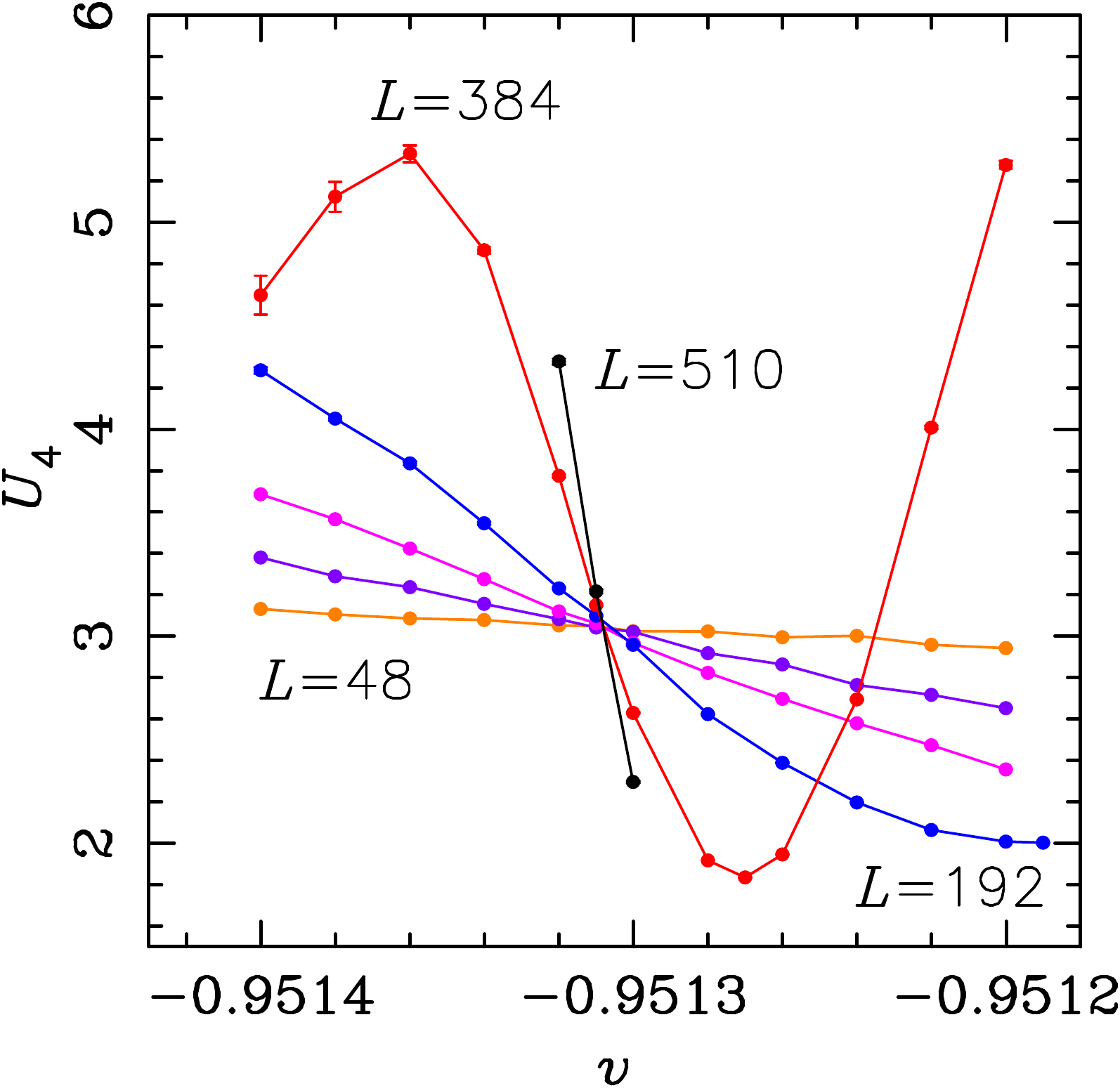} 
  \vspace*{-2mm}
  \caption{%
   Binder cumulant $U_4$ for the five-state BH-lattice AF Potts model 
   in the interval $v\in[-0.95143,-0.95118]$.
   We show data for $L=48$ (brown), $L=96$ (violet), $L = 132$ (pink), 
   $L = 192$ (blue), $L = 384$ (red), and $L = 512$ (black). 
   Points have been joined with lines to guide the eyes.
   }
  \label{fig.Sfine.u4}
\end{figure}

%
%
\begin{figure}[htb]
  \includegraphics[width=0.8\columnwidth]{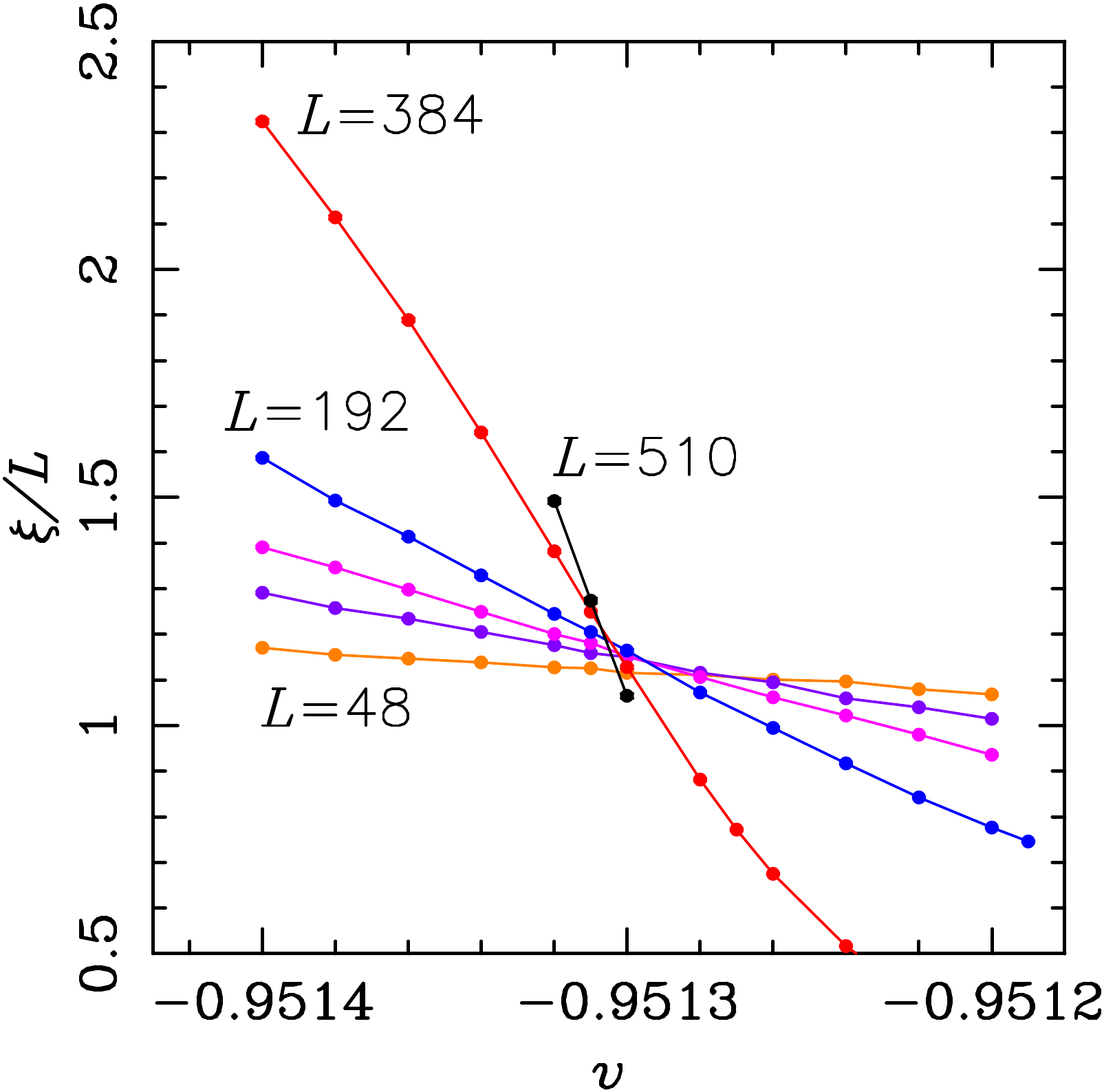} \\
  \vspace*{-2mm}
  \caption{%
   The ratio $\xi/L$ for the five-state BH-lattice AF Potts model 
   in the interval $v\in[-0.95143,-0.95118]$.
   Color code is as in Fig.~\ref{fig.Sfine.u4}. 
   }
  \label{fig.Sfine.xioverL}
\end{figure}

\begin{itemize}
 \item[c)] We made $11$ runs at equidistant values of $v \in [-0.952,-0.951]$ 
       for $L=24,48,96$. We obtained a good choice for the right staggering to
       use [c.f., \eqref{def.M.OK}/\eqref{def_mag_gen_OK}]. Details can 
       be found in Appendix~\ref{appen.stag}. We have improved the statistics:
       we discarded  $\gtrsim 10^4 \tau_{\text{int}}$ MCS, and took
       $\gtrsim 9.5\times 10^4 \tau_{\text{int}}$ measurements.
\end{itemize}

After these preliminary simulations, we then repeated the simulations
in (b); but this time, we measured the whole set of thermal 
\eqref{def_thermal.obs} and magnetic \eqref{def_magnetic.obs} and 
\eqref{def_xi} observables, as well as their dynamic counterparts 
(see Fig.~\ref{fig.tauM2.fine}). The final set of simulations consisted 
in $11$ equidistant runs in the interval $v \in [-0.952,-0.951]$ for 
$L=24,48,96,132,192,384$. For $L\le 132$, we performed MC simulations of 
lengths in the range $(0.5 - 2)\times 10^8$ MCS, for $L=192$,
in the range $(1.1 - 4.0)\times 10^8$ MCS, and for $L=384$, in
the range $(2.3 - 8.7)\times 10^8$ MCS. In all cases, we discarded the first 
$(1.1 - 1.8)\times 10^4 \, \tau_{\text{int}}$ MCS; and we took 
$(1.0 - 1.6)\times 10^5\, \tau_{\text{int}}$ measurements. 
The results shown in Figs.~\ref{fig.Sfine.u4}  and~\ref{fig.Sfine.xioverL}
for two universal amplitudes, 
support that the transition was in the interval $v\in [-0.95132,-0.95130]$, 
so we performed three additional MC simulations at the endpoints and the
center of this interval for $L=512$. In these cases, the statistics was 
smaller: we performed $4\times 10^8$ MCS, discarded 
$\gtrsim 3.6\times 10^3 \, \tau_{\text{int}}$ MCS, and took 
$\gtrsim 3.2\times 10^4 \, \tau_{\text{int}}$ measurements in each of these
runs. Notice that the number of spins for $L=510$ is $1.56 \times 10^6$. 
The large amount of CPU time needed to perform $10^8$ MCS for $L=512$ 
prevented us from performing more simulations at other values of $v$ for 
$L=510$, or simulate larger systems for $v\in [-0.95132,-0.95130]$. 
The numerical MC data can be obtained by request from the corresponding
author.

The total amount of CPU time needed for these simulations (plus those 
reported in Sec.~\ref{sec.histo}) was approximately $9.6$ years 
normalized to an Intel Xeon CPU E5-2687W running at $3.10$ GHz.

%
%
\subsection{Determination of the critical point} \label{sec.critical.point}

The goal of this section is to estimate the critical point $v_c$ for this model.
As mentioned at the end of Sec.~\ref{sec.summ.MC}, the interesting interval 
is $v\in [-0.95132,-0.95130]$. So we have only considered the data points 
inside it, where the physical quantities are approximately linear functions 
of $v$.  

\begin{figure}[htb]
  \includegraphics[width=0.8\columnwidth]{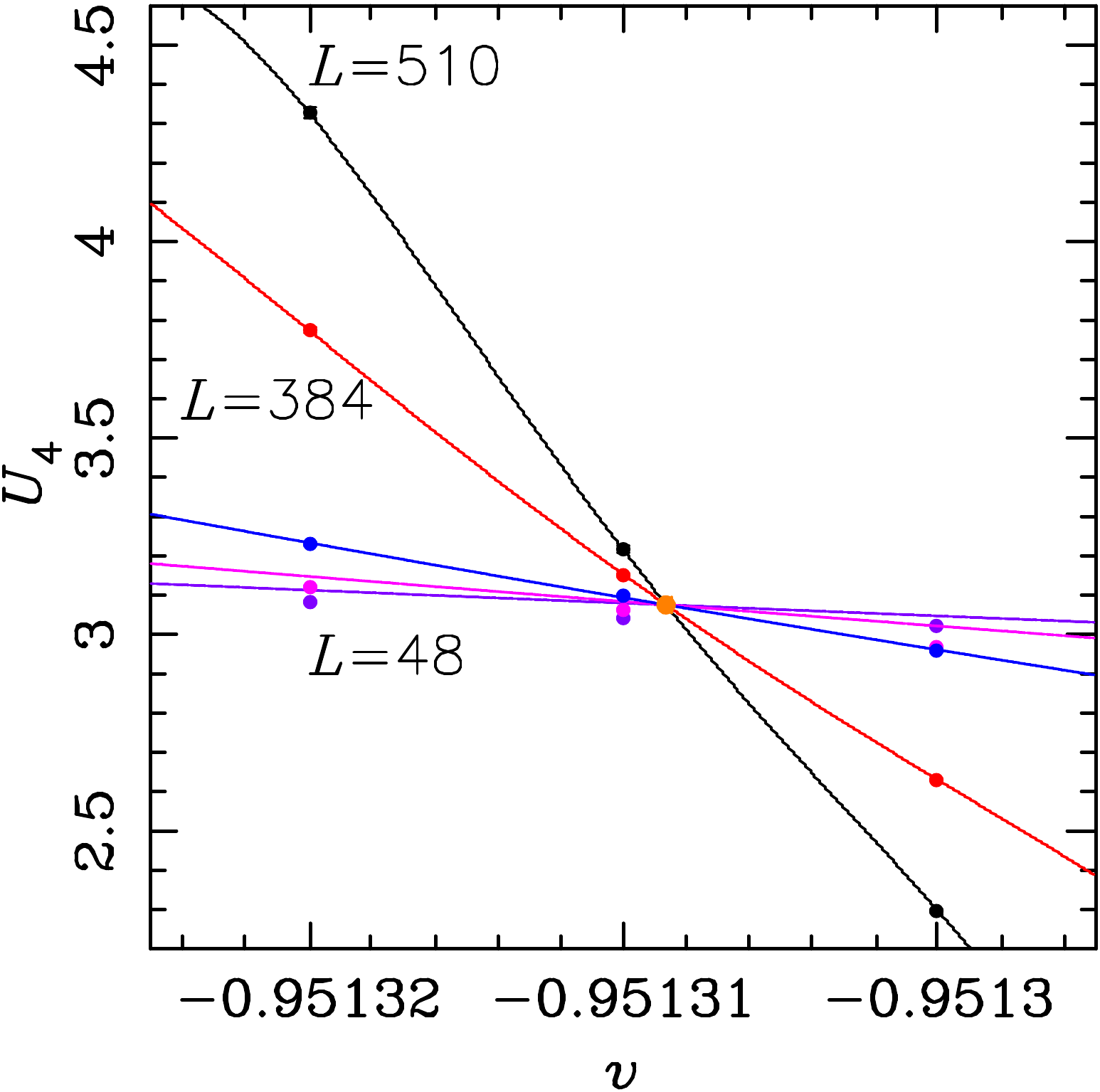} \\
  \vspace*{-2mm}
  \caption{%
   MC results for the Binder ratio $U_4$ \eqref{def_U4} in the interval 
   $v \in [-0.951325, -0.951295]$. We show only data for $L \ge 96$ with
   the color code as in Fig.~\ref{fig.Sfine.u4}. The solid curves
   correspond to the actual nonbiased fit, and the golden dot, 
   to our preferred result \eqref{sols.U4.NB}. Note that the values of 
   $U_4$ behave monotonically with  $L$. 
   }
  \label{fig.SSfine.U4}
\end{figure}

\begin{figure}[htb]
  \includegraphics[width=0.8\columnwidth]{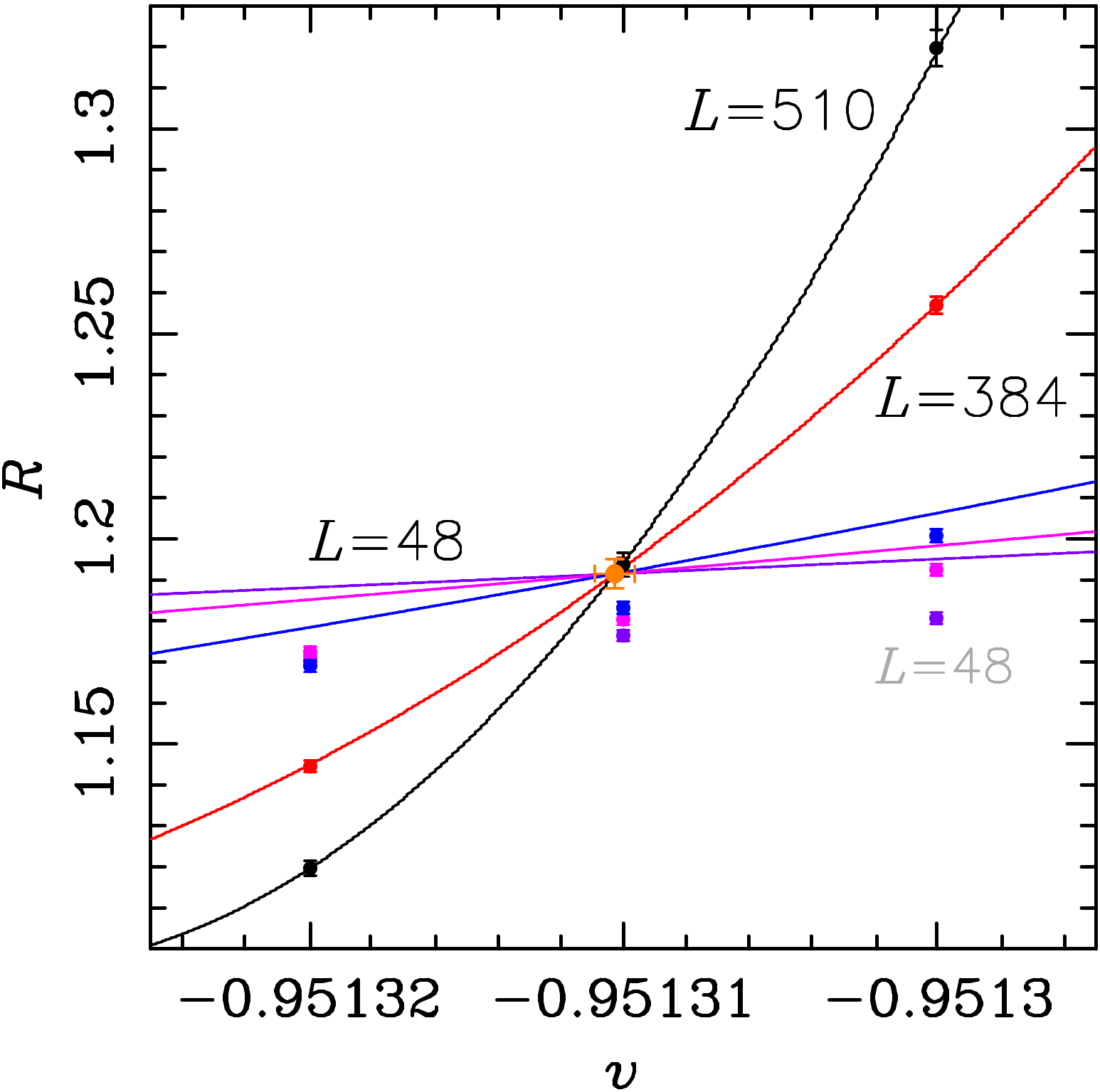} \\
  \vspace*{-2mm}
  \caption{%
   MC results for the Binder ratio $R$ \eqref{def_R} in the interval 
   $v \in [-0.951325, -0.951295]$. We show only data for $L \ge 96$ with
   the color code as in Fig.~\ref{fig.Sfine.u4}. The solid curves 
   correspond to the actual nonbiased fit, and the golden dot, to our 
   preferred result \eqref{sols.R.NB}. The black (resp.\/ gray) labels
   refer to the curves (resp.\/ data points). 
   }
  \label{fig.SSfine.R}
\end{figure}

\begin{figure}[htb]
  \includegraphics[width=0.8\columnwidth]{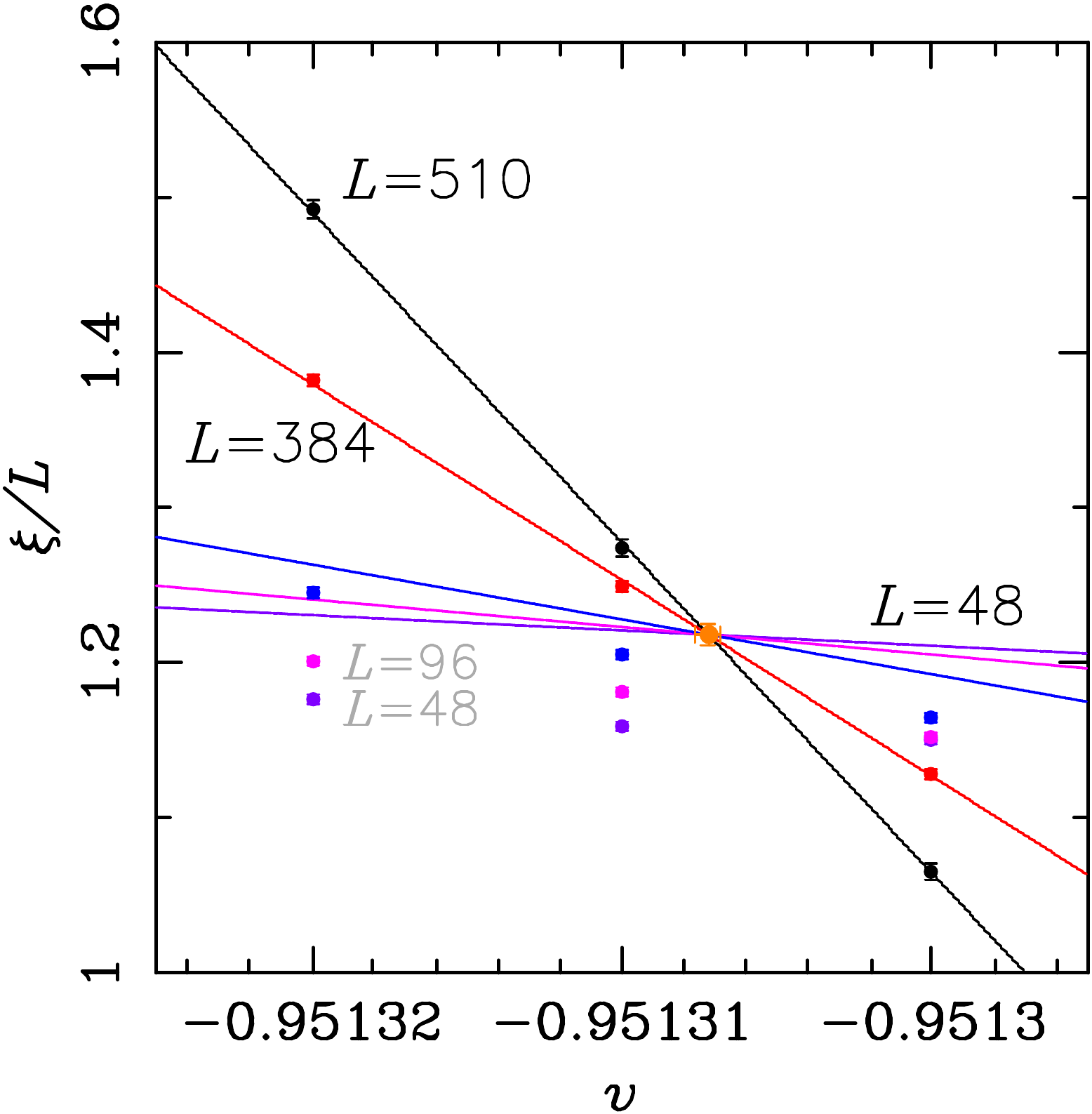} \\
  \vspace*{-2mm}
  \caption{%
   MC results for the ratio $\xi/L$ \eqref{def_xi} in the interval 
   $v \in [-0.951325, -0.951295]$. We show only data for $L \ge 96$ with
   the color code as in Figs.~\ref{fig.Sfine.u4} and~\ref{fig.SSfine.R}. 
   The solid curves correspond to the actual nonbiased fit, and the golden 
   dot, to our preferred result \eqref{sols.xi.NB}. 
   }
  \label{fig.SSfine.xioverL}
\end{figure}

For each quantity $O = U_4, R, \xi/L$ shown in
Figs.~\ref{fig.SSfine.U4}--\ref{fig.SSfine.xioverL}, we have performed a 
\emph{simultaneous} fit of the data with $L\ge L_\text{min}$ to the 
generic \emph{Ansatz}:
\begin{multline}
O(v;L) \;=\; O_c + \sum\limits_{k=1}^{k_\text{max}} a_k \, (v-v_c)^k \, 
L^{k y_t} \\ 
     + b_1 \, L^{-\omega_1} + \cdots \,,
\label{def_FSS_Ansatz}
\end{multline}
where the dots represent higher-order FSS corrections. 
We have varied the number of terms in the \emph{Ansatz} \eqref{def_FSS_Ansatz},
and the number of data points entering the fit $L\ge L_\text{min}$ in order
to detect further FSS corrections. In this section, we will not assume any 
hypothesis on any of the parameters in the \emph{Ansatz} 
\eqref{def_FSS_Ansatz}.  

%
%

From Fig.~\ref{fig.Sfine.u4}, it is clear that $U_4$ behaves close to the 
crossing point in a more complicated way than $R$ or $\xi/L$. 
This explains why we had to use $k_\text{max}=4$ in the
\emph{Ansatz} \eqref{def_FSS_Ansatz}. The best fit is obtained by fixing
$b_1=0$ and taking $L_\text{min}=192$. The results are
\begin{subequations}
\label{sols.U4.NB}
\begin{align}
v_c   &\;=\; -0.951\, 308\, 64(9)\,, \\
y_t   &\;=\; \phantom{-}2.05(5)\,, \\
U_{4,c} &\;=\; \phantom{-}3.075(3)\,,
\end{align}
\end{subequations}
with $\chi^2/\text{DF} = 1.92/2$ and $\text{CL}=37.7\%$. 

%
%

For the Binder ratio $R$, the best fit is obtained for $k_\text{max}=2$,
$b_1 = 0$ and $L_\text{min} = 384$. The results are:
\begin{subequations}
\label{sols.R.NB}
\begin{align}
v_c   &\;=\; -0.951\, 310\, 3(6)\,, \\
y_t   &\;=\; \phantom{-}2.0(1)\,, \\ 
R_c   &\;=\; \phantom{-}1.191(4)\,,
\end{align}
\end{subequations}
with $\chi^2/\text{DF} = 0.30/1$ and $\text{CL}=58.5\%$.  

Finally, for the ratio $x = \xi/L$, the best fit is obtained for  
$k_\text{max}=1$, $b_1=0$, and $L_\text{min}=384$. The results are:
\begin{subequations}
\label{sols.xi.NB}
\begin{align}
v_c  &\;=\; -0.951\, 307\, 2(5)\,, \\
y_t  &\;=\; \phantom{-}1.84(9)\,, \\
x_c  &\;=\; \phantom{-}1.218(7)\,,
\end{align}
\end{subequations}
with $\chi^2/\text{DF} = 7.35/5$, and $\text{CL}=19.6\%$. 

Notice that our preferred fits for each quantity have a reasonable value 
of CL in the range $\approx 20- 59\%$. For the Binder ratio $U_4$,
we needed an \emph{Ansatz} with more terms than usual (e.g., Ref.~\cite{KSS}). 
And in all cases, our data did not allow us to determine the leading 
correction-to-scaling exponent $\omega$ in \eqref{def_FSS_Ansatz}. 

Figures~\ref{fig.Sfine.xioverL} and~\ref{fig.SSfine.xioverL}
show clearly that $v=-0.95130$ is in the disordered phase, as $\xi/L$ 
decreases as $L$ increases (contrary to what happens for the other two 
values of $v$ in Fig.~\ref{fig.SSfine.xioverL}). Let us recall that our
definition of $\chi$ \eqref{def_chi} does not contain the term
$\<\bm{\mathcal{M}}\>^2$, so $\chi$ grows with $L$ in the low-$T$ phase. 
We have $\xi(-0.95130;510)=543(3) \gtrsim L=510$. Therefore, for the linear 
sizes that we are able to simulate, the correlation length for 
$v = v_c + \epsilon$ is slightly larger than the linear size, so we 
should expect large corrections to scaling. 

%
%

Looking at the three distinct estimates for $v_c$, we see that the dispersion
among them is larger that the error bars. Therefore, if we took into account
the statistic and systematic errors, we arrive at the conservative  
subjective estimate 
\be
v_c \;=\; -0.951\, 308(2) \,.
\label{value_vc.NB}
\ee
This value agrees within errors with the result obtained in 
Ref.~\cite{union-jack} 
using MC simulations $v_c=-0.95132(2)$; but our error bar is one order of 
magnitude smaller. This is a significant improvement.

If we look at the estimates of the thermal RG parameter $y_t = 1/\nu$, we
observe that they all agree within two standard deviations and they are 
consistently larger than the previous MC estimate $y_t=2-X_t=1.505(5)$ 
\cite{union-jack}. A conservative estimate taking into account both the 
statistic and systematic errors is 
\be
y_t \;=\; 1/\nu \;=\; 2.0(1) \,.
\label{value_yt}
\ee
This result agrees well with the value expected for a first-order
phase transition $y_t=2$ in a 2D system \cite{Nauenberg_74,Klein_76,Fisher_82}.
If we fix $y_t=2$ in the \emph{Ansatz} \eqref{def_FSS_Ansatz}, we do not get 
any sizable improvement in the determination of both $v_c$ and the 
universal amplitudes. Finally, $y_t=2$ implies the following critical 
exponents in 2D systems: 
\be
\nu \;=\; 1/2 \,, \qquad 
\alpha \;=\; 2 - 2\nu \;=\; 1 \,, \quad 
\alpha/\nu \;=\; 2\,. 
\label{exp.crit}
\ee 

%
%
\subsection{Determination of the static critical exponents} 
\label{sec.critical.exp}

The goal of this section is to estimate the critical- exponent ratios
$\alpha/\nu$ and $\gamma/\nu$ for this model. We are going to use the
same interval $v\in [-0.95132,-0.95130]$. 

\begin{figure}[htb]
  \includegraphics[width=0.8\columnwidth]{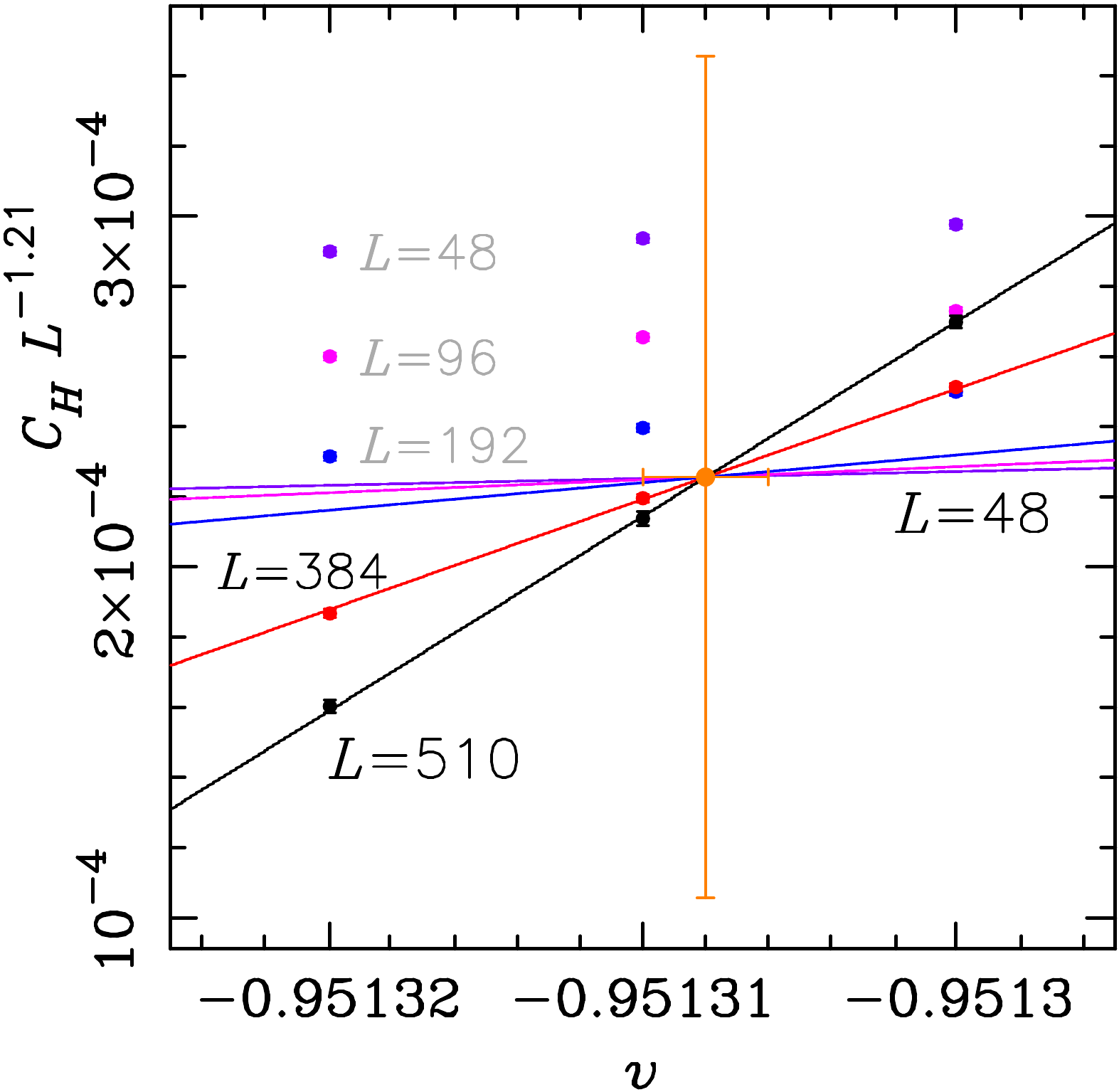} \\
  \vspace*{-2mm}
  \caption{%
   MC results for the ratio $C_H/L^{1.21}$ in the interval 
   $v \in [-0.951325, -0.951295]$. We show only data for $L \ge 96$ with
   the color code as in Figs.~\ref{fig.Sfine.u4} and~\ref{fig.SSfine.R}. 
   The solid curves correspond to the actual biased fit \eqref{sols.CH.B}, 
   and the golden dot, to our preferred result \eqref{sols.CH.B.OK}. The 
   values of $C_H/L^{1.21}$ behave monotonically with $L$ on the lhs 
   of the crossing point.
   }
  \label{fig.SSfine.CH}
\end{figure}

\begin{figure}[htb]
  \includegraphics[width=0.8\columnwidth]{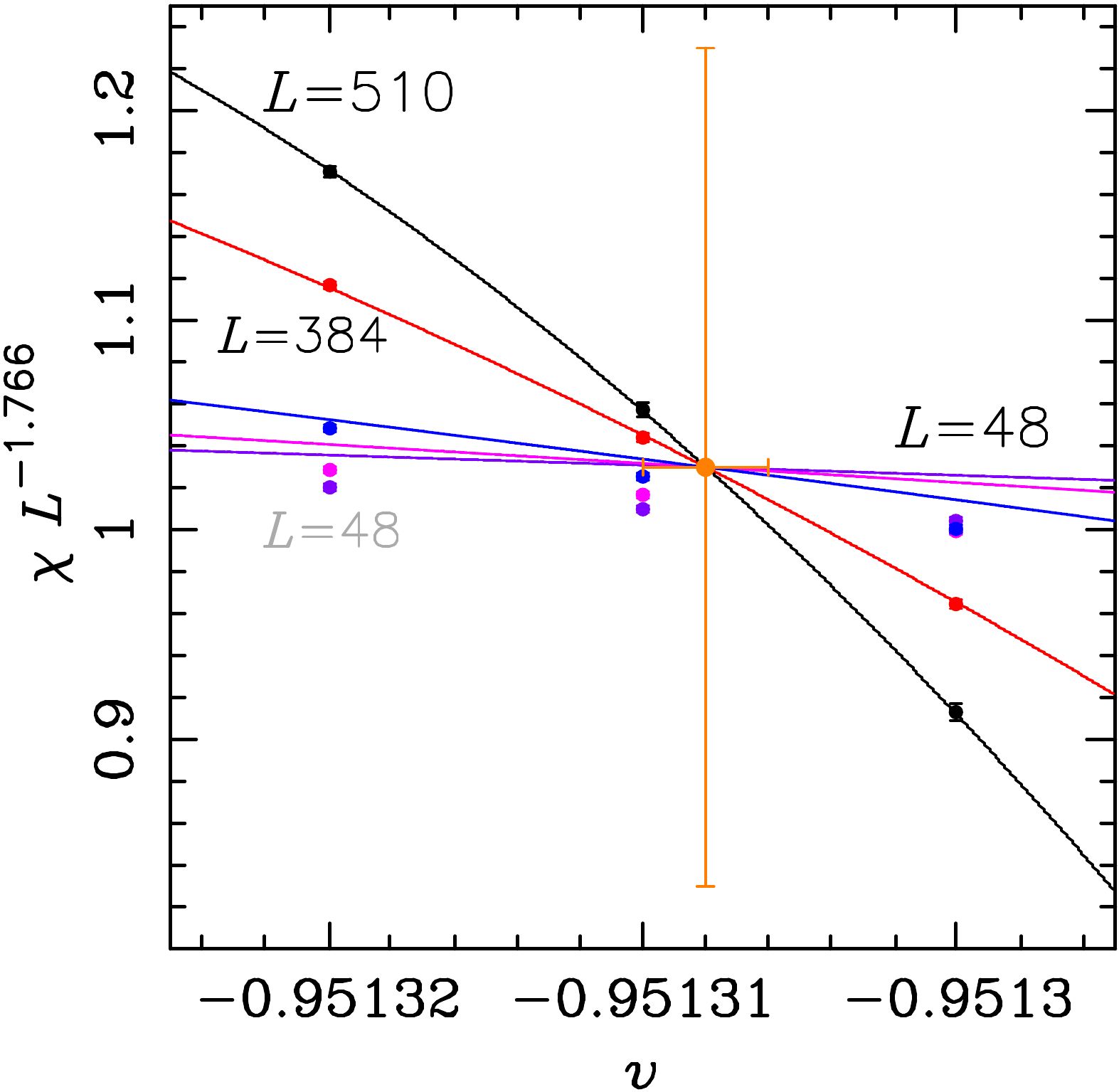} \\
  \vspace*{-2mm}
  \caption{%
   MC results for the ratio $\chi/L^{1.766}$ in the interval 
   $v \in [-0.951325, -0.951295]$. We show only data for $L \ge 96$ with
   the color code as in Figs.~\ref{fig.Sfine.u4} and~\ref{fig.SSfine.R}. 
   The solid curves correspond to the actual biased fit \eqref{sols.chi.B}, 
   and the golden dot, to our preferred result \eqref{sols.chi.B.OK}. 
   }
  \label{fig.SSfine.chi}
\end{figure}

For each quantity $O = C_H, \chi$ shown in
Figs.~\ref{fig.SSfine.CH} and~\ref{fig.SSfine.chi}, we have performed a 
\emph{simultaneous} fit of the data with $L\ge L_\text{min}$ to the 
generic \emph{Ansatz}:
\begin{multline}
O(v;L) \;=\; L^{\rho_O} \Bigl[ O_c + 
   a_1 \, (v-v_c) \, L^{y_t} \\ + a_2 \, (v-v_c)^2 \, L^{2y_t} 
        + b_1 \, L^{-\omega_1} + \cdots \Bigr] \,,
\label{def_FSS_Ansatz.Bis}
\end{multline}
where the dots represent higher-order FSS corrections, and $\rho_O$ is the
corresponding critical exponent. Again, we have varied both the
number of terms in the \emph{Ansatz} \eqref{def_FSS_Ansatz.Bis}, and the
number of data points $L\ge L_\text{min}$ entering the fit as a precaution 
against further FSS corrections. 

Let us start with the specific heat $C_H$. For this quantity, the best 
results are always obtained from the simplest \emph{Ansatz} $a_2=b_1=0$. 
If we perform this fit \emph{without} assuming any value for $v_c$ and $y_t$, 
the best result corresponds to
$L_\text{min}=192$:  
\begin{subequations}
\label{sols.CH.NB}
\begin{align}
v_c        &\;=\; -0.951\, 313(1)\,, \\
y_t        &\;=\; \phantom{-}2.01(8)\,, \\
\alpha/\nu &\;=\; \phantom{-}1.03(2) \,, \\ 
C_{H,c}    &\;=\; \phantom{-}6.0(6)\times 10^{-4} \,, 
\end{align}
\end{subequations}
with $\chi^2/\text{DF} = 1.18/4$, and $\text{CL}=88.2\%$. The value for 
$v_c$ is $2.5$ standard deviations away from our preferred estimate 
\eqref{value_vc.NB}; while the value for $y_t$ agrees within errors with the
estimate \eqref{value_yt}. Notice that the estimate for the ratio $\alpha/\nu$
is not compatible with the value of $y_t$ [cf., \eqref{exp.crit}].  
 
%
%

We can try to improve the estimate for $\alpha/\nu$ by performing \emph{biased}
fits by \emph{assuming} that $v_c=-0.951308$ [cf., \eqref{value_vc.NB}] and 
$y_t=2$ [cf., \eqref{value_yt} and~\eqref{exp.crit}]. In this case, the best 
fit comes from $L_\text{min}=384$:
\be 
\label{sols.CH.B}
\alpha/\nu \;=\; 1.21(2) \,, \quad C_{H,c} \;=\; 2.3(3)\times 10^{-4}\,,  
\ee
with $\chi^2/\text{DF} = 1.63/3$, and $\text{CL}=65.1\%$. As the error bar in
$v_c$ is $2\times 10^{-6}$, we have repeated the fits with the values 
$v_c=-0.951306$ and $v_c=-0.951310$. We observe that the dispersion
in the previous estimates is larger than the corresponding error bars. 
By taking into account these systematic errors, our best (conservative) 
estimates are: 
\be 
\label{sols.CH.B.OK}
\alpha/\nu \;=\; 1.21(7) \,, \quad C_{H,c} \;=\; 2.3(12) \times 10^{-4}\,. 
\ee
The estimate for $\alpha/\nu$ is $\approx 20\%$ larger than
in the nonbiased fit \eqref{sols.CH.NB}; but it is still far away from
the expected value for a first-order phase transition $\alpha/\nu=2$.  
Our final estimate is larger than the previous MC estimate 
$\alpha/\nu=1.01(1)$ \cite{union-jack}. Finally, the estimate for the
amplitude $C_{H,c}$ is not very precise. 
The results are depicted in Fig.~\ref{fig.SSfine.CH}.  

We now consider the susceptibility $\chi$ \eqref{def_chi}. For this quantity, 
the best results are always obtained for $b_1=0$. If we perform this fit 
\emph{without} any further assumption, the best one corresponds to
$L_\text{min}=192$:  
\begin{subequations}
\label{sols.chi.NB}
\begin{align}
v_c      &\;=\; -0.951\, 309(1)\,, \\
y_t      &\;=\; \phantom{-}1.75(6)\,, \\
\gamma/\nu &\;=\; \phantom{-}1.786(9) \,, \\ 
\chi_{c}  &\;=\; \phantom{-}0.92(4)\,,
\end{align}
\end{subequations}
with $\chi^2/\text{DF} = 2.81/3$, and $\text{CL}=42.2\%$. The value for 
$v_c$ agrees within errors with our preferred estimate \eqref{value_vc.NB}.
However,  our estimate for $y_t$ is $\approx 4$ standard deviations from the 
estimate \eqref{value_yt}. 
 
%
%

Again, we may improve the estimate for $\gamma/\nu$ by performing \emph{biased}
fits by \emph{assuming} that $v_c=-0.951308$ [cf., \eqref{value_vc.NB}] and 
$y_t=2$ [cf., \eqref{value_yt} and~\eqref{exp.crit}]. In this case, the best 
fit comes from $L_\text{min}=384$:
\be 
\label{sols.chi.B}
\gamma/\nu \;=\; 1.766(9) \,, \qquad  \chi_c     \;=\; 1.03(6)\,,
\ee
with $\chi^2/\text{DF} = 1.09/2$, and $\text{CL}=58.0\%$. Again, by repeating 
the fits with $v_c=-0.951306$ and $v_c=-0.951310$, we observe that the 
dispersion among the estimates is larger than their statistical errors.
By taking into account this dispersion, our best estimates are:
\be 
\label{sols.chi.B.OK}
\gamma/\nu \;=\; 1.77(4) \,, \qquad  \chi_c     \;=\; 1.0(2) \,. 
\ee
The estimate for $\gamma/\nu$ agrees within errors with that from the  
nonbiased fit \eqref{sols.chi.NB}, but it is still $\lesssim 6$ 
standard deviations from the expected value $\gamma/\nu=2$ for a first-order
phase transition. On the other hand, our final estimate agrees within errors
with the value quoted in the literature $\gamma/\nu=1.774(8)$ \cite{union-jack}.
Finally, our estimate for the amplitude $\chi_c$ is not very precise
(see Fig.~\ref{fig.SSfine.chi}).

%
%
\subsection{Determination of the dynamic critical exponents} 
\label{sec.dyn.critical.exp}

The goal of this section is to study how the integrated autocorrelation time
for the slowest mode of the WSK algorithm (i.e., 
$\tau_{\text{int},\bm{\mathcal{M}}^2}$) behaves close to the critical 
point $v_c$. In particular, we would like to tell whether 
$\tau_{\text{int},\bm{\mathcal{M}}^2} \sim 
L^{z_{\text{int},\bm{\mathcal{M}}^2}}$, or 
$\tau_{\text{int},\bm{\mathcal{M}}^2} \sim L^{2\sigma_{o,d} \, L}$.
Again, we will focus on the interval  $v\in [-0.95132,-0.95130]$. 

In this case, we have performed \emph{biased} fits to two different 
\emph{Ans\"atze}: a power-law one like in \eqref{def_FSS_Ansatz.Bis}, 
and an exponential one in which the term $L^{\rho_O}$ in 
\eqref{def_FSS_Ansatz.Bis} has been replaced by $e^{2\sigma_{o,d} L}$.
 
\begin{figure}[b]
  \includegraphics[width=0.8\columnwidth]{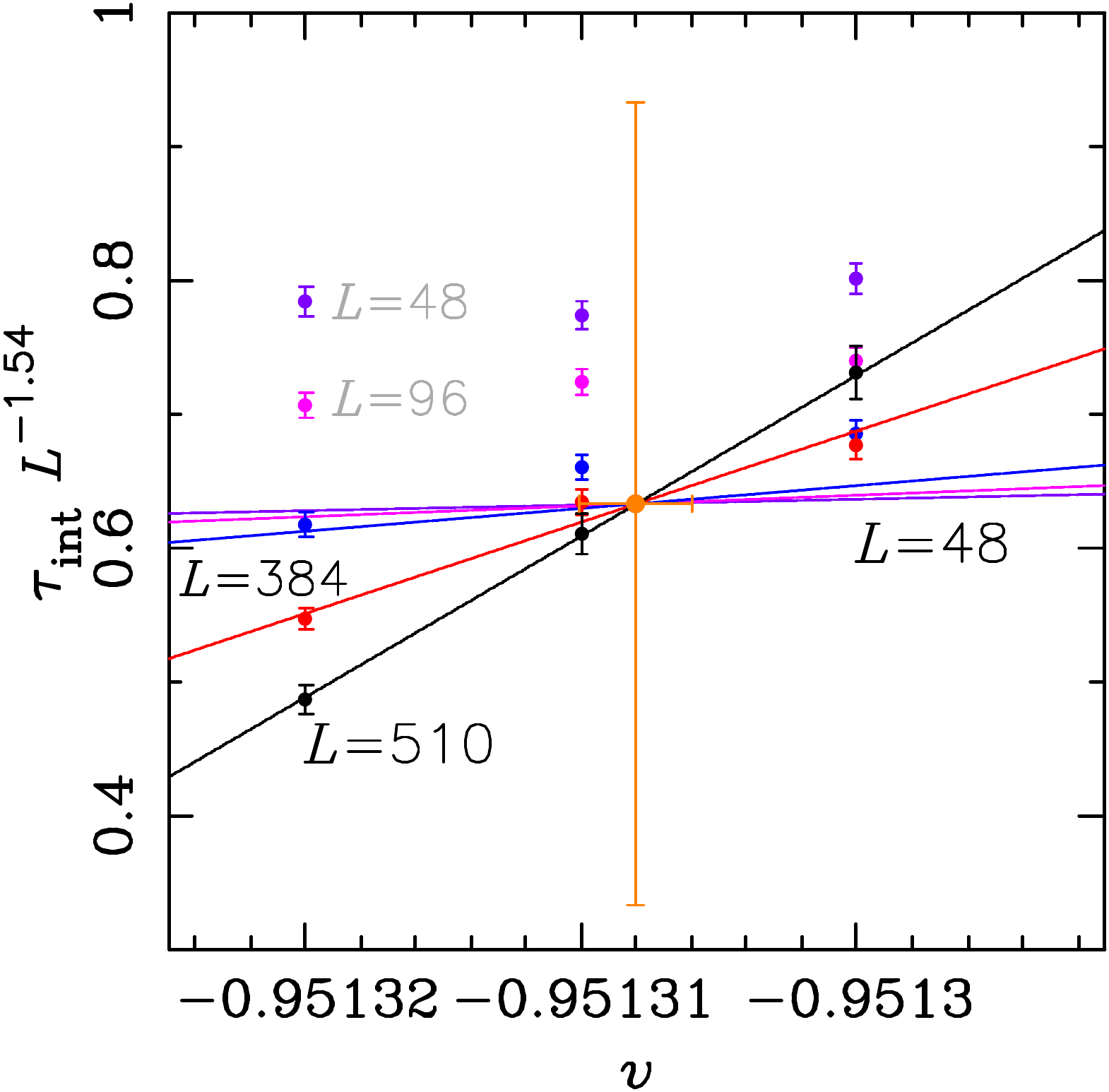} \\
  \vspace*{-2mm}
  \caption{%
   MC results for the integrated autocorrelation time 
   $\tau_{\text{int},\bm{\mathcal{M}}^2}/L^{1.54}$ in the interval 
   $v \in [-0.951325, -0.951295]$. We show only data for $L \ge 96$ with
   the color code as in Figs.~\ref{fig.Sfine.u4} and~\ref{fig.SSfine.R}.
   The solid curves correspond to the actual biased fit \eqref{sols.tau.B}, 
   and the golden dot, to our preferred result \eqref{sols.tau.B.OK}. 
   }
  \label{fig.SSfine.tau}
\end{figure}

In the first case, our preferred fit corresponds to $a_2=b_1=0$ and 
$L_\text{min}=384$: 
\be
\label{sols.tau.B}
z_{\text{int},\bm{\mathcal{M}}^2} \;=\; 1.54(6) \,, \qquad  
\tau_c     \;=\; 0.6(2)\,,
\ee
with $\chi^2/\text{DF} = 3.86/3$, and $\text{CL}=27.7\%$. We have repeated 
the fits with $v_c=-0.951306$ and $v_c=-0.951310$, and in this case, the
dispersion of the estimates is only slightly larger than the error bars:
\be
\label{sols.tau.B.OK}
z_{\text{int},\bm{\mathcal{M}}^2} \;=\; 1.54(6) \,, \qquad  
\tau_c     \;=\; 0.6(3)\,.
\ee
The results are shown in Fig.~\ref{fig.SSfine.tau}.

%
%

\begin{figure}[htb]
  \includegraphics[width=0.8\columnwidth]{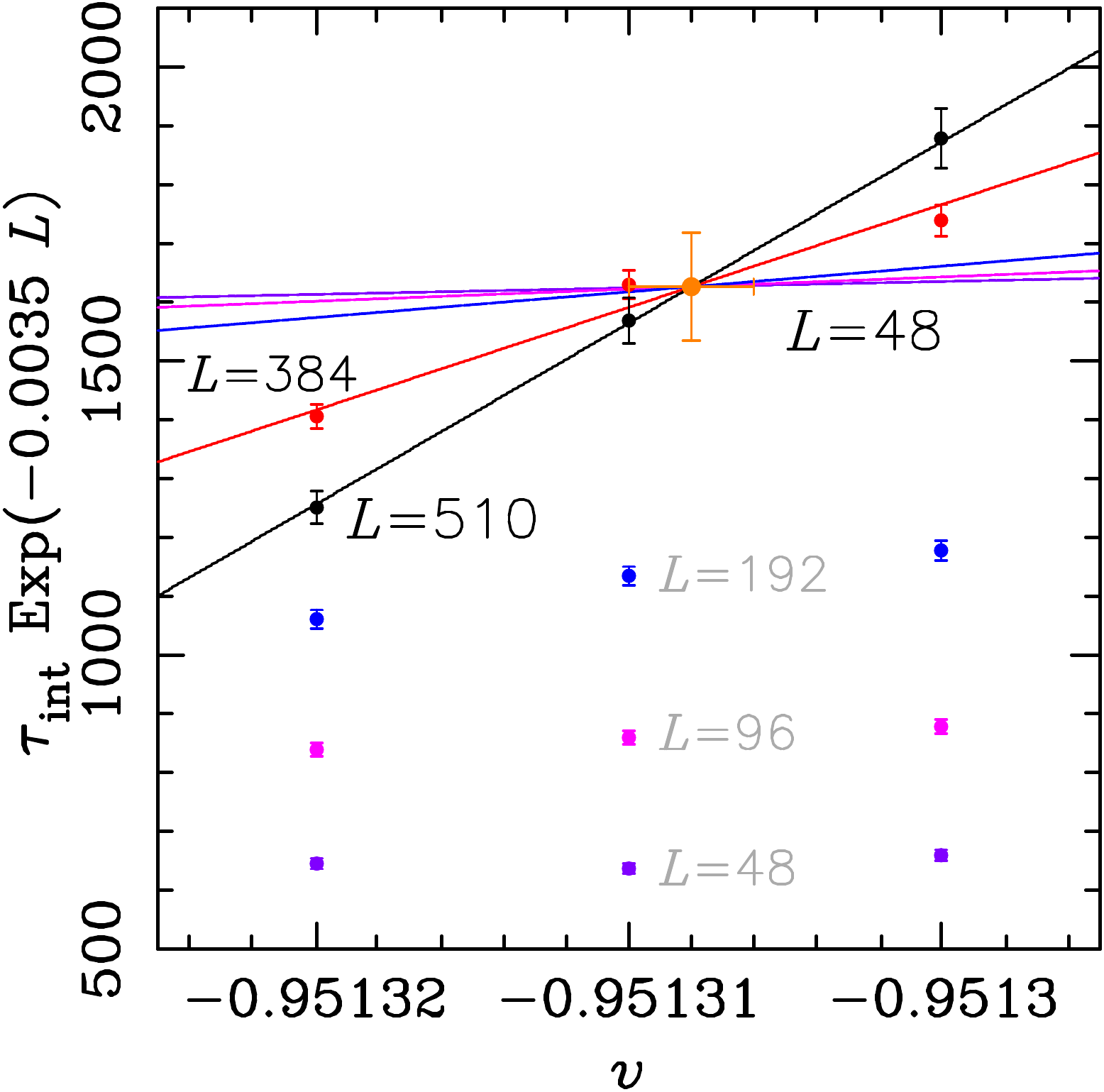} \\
  \vspace*{-2mm}
  \caption{%
   MC results for the integrated autocorrelation time 
   $\tau_{\text{int},\bm{\mathcal{M}}^2}e^{-0.0035L}$ in the interval 
   $v \in [-0.951325, -0.951295]$. We show only data for $L \ge 96$ with
   the color code as in Figs.~\ref{fig.Sfine.u4} and~\ref{fig.SSfine.R}. 
   The solid curves correspond to the actual biased fit \eqref{sols.tau2.B}, 
   and the golden dot, to our preferred result \eqref{sols.tau2.B}. 
   }
  \label{fig.SSfine.tauB}
\end{figure}

If we consider the exponential \emph{Ansatz}, we also obtain the best estimates
for $a_2=b_1=0$ and $L_\text{min}=384$:
\be
\label{sols.tau2.B}
2\sigma_{o,d}  \;=\; 0.0035(1) \,, \qquad
\tau_c   \;=\; 1.63(9)\times 10^{4} \,,
\ee
with $\chi^2/\text{DF} = 3.86/3$, and $\text{CL}=27.7\%$. Again, we have 
repeated the fits with $v_c=-0.951306$ and $v_c=-0.951310$, and in this case, 
the dispersion of the estimates is smaller than the error bars quoted
in \eqref{sols.tau2.B}. The results are displayed in 
Fig.~\ref{fig.SSfine.tauB}. 

%
%

The quality of the fits \eqref{sols.tau.B}/\eqref{sols.tau2.B} is the same,
so we do not have arguments in favor of any of the two \emph{Ans\"atze}. The
power-law \emph{Ansatz} implies that the dynamic critical exponent 
\eqref{sols.tau.B} is somewhat smaller than the exponent associated to 
single-site algorithms $\gtrsim 2$. On the other hand, the exponential
\emph{Ansatz} predicts a very small interface tension \eqref{sols.tau2.B}. 

%
%
\section{Histogram analysis} \label{sec.histo}

In the previous section we have seen several indications that the transition 
undergone by the five-state BH-lattice AS Potts model was first order, but 
there were other estimates pointing to the second-order nature of that 
transition. The use of histograms for analyzing simulations has a long story
(see e.g., the list of references in Ref.~\cite{Lee_91}). 
Actually, for certain class of models 
(that include the $q$-state FM Potts model at large $q$) there is a 
rigorous theory \cite{Borgs_90,Borgs_91} (see also Ref.~\cite{Lee_91}). In
particular, it was proven that for the 2D $q$-state FM Potts model at large 
$q$, the partition function can be written as 
\begin{multline}
Z(\beta;L) \;=\; e^{-L^2 \beta f_d(\beta)} + q e^{-L^2 \beta f_o(\beta)} \\
    + O\Bigl( e^{-bL} \Bigr) e^{-L^d \beta f(\beta)} \,,
\label{Z.BKMS}
\end{multline}
for some positive constant $b>0$. In Eq.~\eqref{Z.BKMS}, periodic boundary 
conditions are assumed, $f_d(\beta)$ and $f_o(\beta)$ are smooth functions 
independent of $L$, and $f(\beta) = \min(f_d(\beta),f_o(\beta))$.  
The free energy of the model is given by $f_o(\beta)$ (resp.\/ $f_d(\beta)$)
when $\beta \ge \beta_c$ (resp.\/ $\beta \le \beta_c$).
By using an inverse Laplace transform \cite{Billoire_94}, one can recover the 
phenomenological two-Gaussian \emph{Ansatz} introduced by Binder, Challa, 
and Landau \cite{Binder_84,Challa_86}, with the correct weight for 
the Gaussians. 

%
%
\begin{figure}[htb]
  \includegraphics[width=0.8\columnwidth]{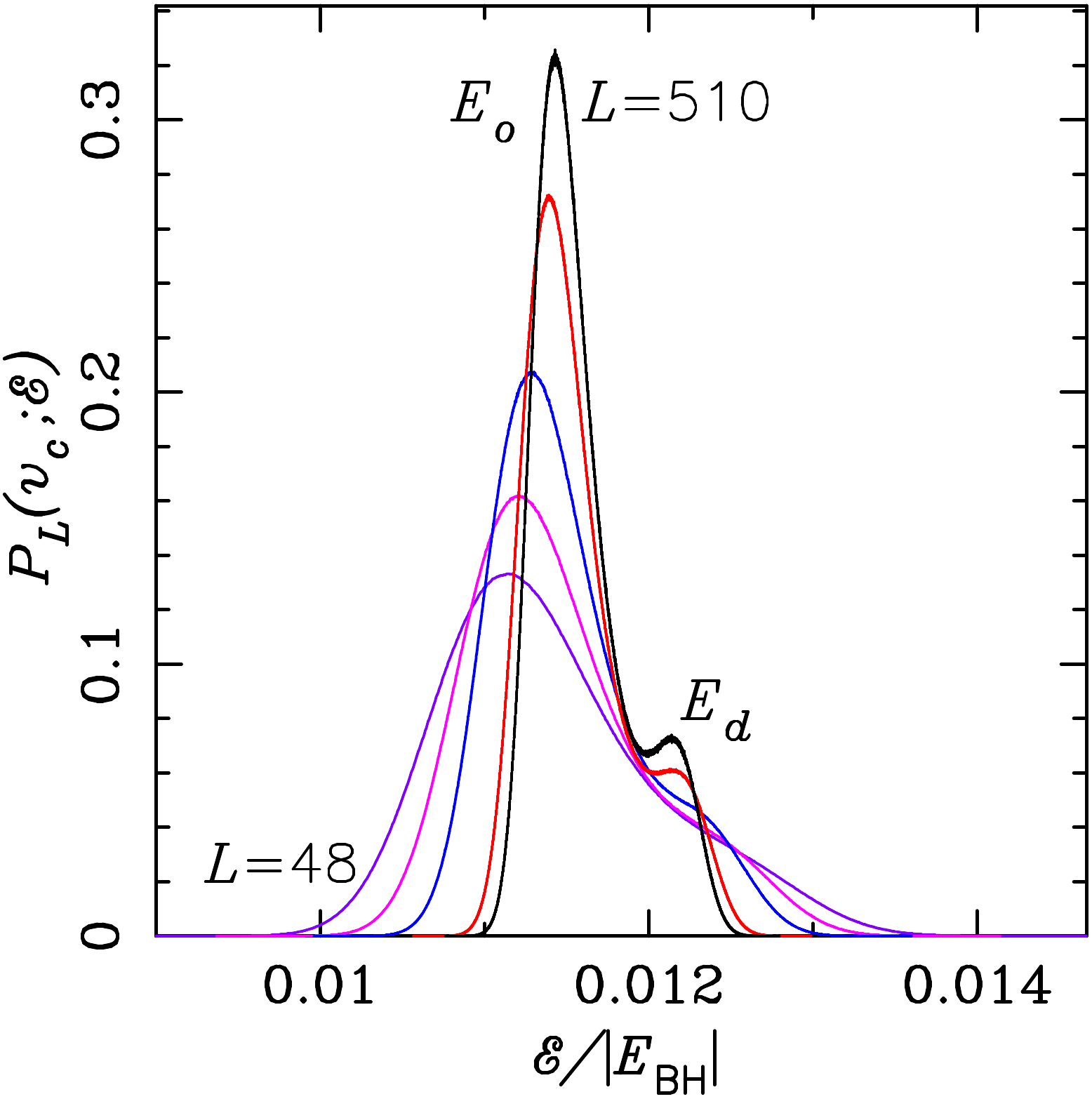} \\
  \vspace*{-2mm}
  \caption{%
  Energy histograms for the five-state BH-lattice AF Potts model at 
  the critical point $v_c=-0.951\, 308$ [cf., \eqref{value_vc.NB}]  
  for $L=96$ (violet), $L=132$ (pink), $L=192$ (blue), 
  $L=384$ (red), and $L=510$ (black).  
  The label $E_o$ (resp.\/ $E_d$) shows the energy peak for the ordered 
  (resp.\/ disordered) phase. The peak at $E_o$ becomes higher and narrower as
  $L$ increases. 
}
\label{fig_histograms_e0}
\end{figure}

One may think that \emph{a priori} the five-state BH-lattice AF Potts model 
does not allow the use of the histogram method: it is an AF model with a 
nonzero entropy density at $T=0$. However, this might not be the case: 
the results discussed in Appendix~\ref{appen.ground} indicate that this model 
has an ordered low-$T$ phase in which five distinct phases coexist. This 
situation is similar to the one found for the three-state diced-lattice AF 
Potts model \cite{KSS}. For that model, a contour analysis was used to prove 
the existence of a finite-$T$ phase-transition point (see also 
Ref.~\cite{Kotecky-Sokal-Swart} for more details and generalizations). 
In conclusion, we consider it makes sense to use the histogram method to study 
the nature of the phase transition of our model. Furthermore, we were able
to find two peaks in the energy probability distribution
$P_L(v;\{\mathcal{E}\})$ for some
of the MC simulations described in Sec.~\ref{sec.res}. However, as shown in 
Ref.~\cite{Lee_90}, a two-peak signal does not always imply a first-order 
phase transition: one has to analyze carefully how these peaks behave as 
$L$ is increased.

First, we have extrapolated the MC histograms at $v=-0.95131$  for $L\ge 96$
to the bulk critical point $v_c=-0.951\, 308$ [cf., \eqref{value_vc.NB}]  
using the Ferrenberg--Swendsen \cite{FS_89} method. 
The probability distributions shown in Fig.~\ref{fig_histograms_e0} 
have been normalized in the following way. First, the energy $\mathcal{E}$ 
has been normalized as in Eq.~\eqref{def_E_mean}, so it varies in the 
interval $[0,1]$. On the other hand, 
the number of values the energy can take in each simulation increases with $L$
(see below). Therefore, the ``raw'' probability distributions cannot be 
compared directly. In order to be able to do so, we have divided the 
whole interval $\mathcal{E}/|E_\text{BH}| \in [0,0.2]$ (where the internal 
energy is constrained close to the critical temperature) into $100$ bins,
and recompute the probability distributions with this new \emph{common} 
bin size. In this way we obtain the right normalization factor for each 
histogram, such that they can be compared when plotted all together as in 
Fig.~\ref{fig_histograms_e0}. It is clear that there is a single peak 
for $L\lesssim 192$; but for $L=384$ we see a small \emph{plateau} that 
develops into a small ``bump'' for $L=510$. So, from the study of the 
histograms at $v=v_c$, we see that $L=510$ is actually too small for our
model to show a clear double-peak structure (if any). 

%
%
\subsection{Analysis when the peaks have the same height} 
\label{sec.histo.equal}

As noted in Ref.~\cite{Borgs_92}, at the point $v_{C_{H,\text{max}}}$ where 
the specific heat $C_H(v_{C_{H,\text{max}}};L)$ attains its maximum value, 
the energy probability distribution $P_L(v_{C_{H,\text{max}}};\mathcal{E})$ 
typically shows two peaks of similar heights. This phenomenon is also 
observed in our system, but only when $L\ge 96$. For $L=48$, we find only 
a broad peak; but not two peaks. This means that if the transition is 
first order, then we 
need to study systems with linear sizes $L\ge 96$. In practice, we made a 
long MC simulation close to that point, and then used the Ferrenberg--Swendsen 
method to find the point $v_\circ(L)$ for which the 
heights of these two peaks were equal. This method of locating the temperature
for which the energy histogram displays two equal peak has been successfully 
used in the literature to study first-order phase transitions 
\cite{Lee_90,Lee_91,Janke_93,Janke_94}. One good property of this method is 
that we do not need the \emph{exact} position of the bulk critical temperature 
$v_c$.  
 
%
%
\begin{figure}[htb]
  \includegraphics[width=0.8\columnwidth]{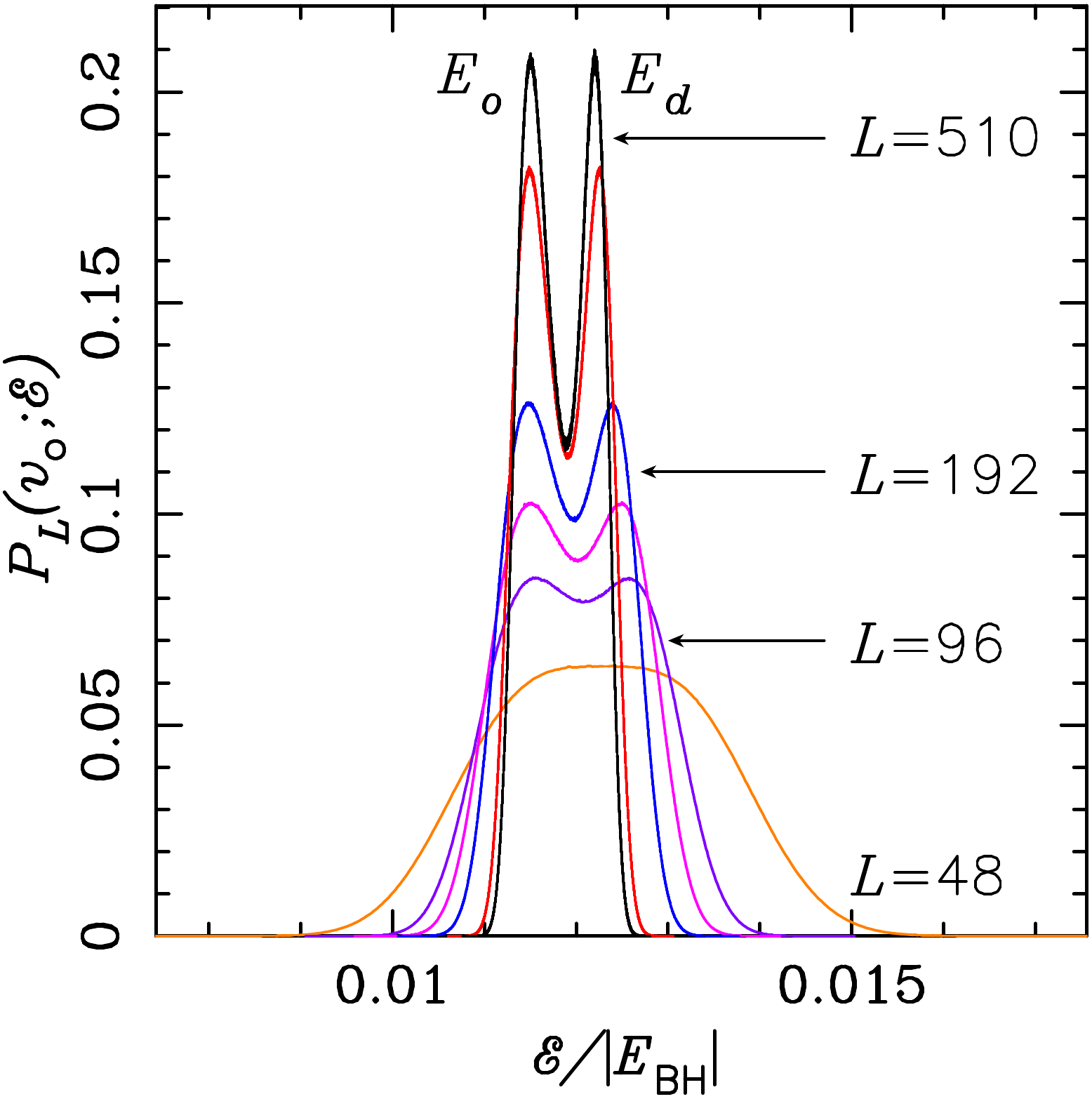} \\
  \vspace*{-2mm}
  \caption{%
  Energy histograms for the five-state BH-lattice AF Potts model at 
  $v=v_\circ(L)$ 
  for $L\ge 48$. The color code is as in Fig.~\ref{fig_histograms_e0}. 
  See Table~\ref{table_histograms_e} for the values of $v_\circ(L)$.  
  The label $E_o$ (resp.\/ $E_d$) shows the energy peak for the ordered 
  (resp.\/ disordered) phase. 
}
\label{fig_histograms_e}
\end{figure}

For $L=48$ we made a long MC simulation at the point $v_\circ=-0.95029$ 
where the histogram displayed a broad peak (see Fig.~\ref{fig_histograms_e}). 
For the other values of $L=96, 132, 192, 384, 510$, we have performed a long
MC simulation at a temperature $v_1$, for which the histogram showed
a two-peak structure, but with unequal heights; we then used the 
Ferrenberg--Swendsen method
to locate the nearby temperature $v_\circ(L)$ for which the histogram had 
two peaks of identical height. 
The probability distributions shown in Fig.~\ref{fig_histograms_e} 
have been normalized as in Fig.~\ref{fig_histograms_e0}.  

The values of $v_1$ and $v_\circ$ are displayed on the second and third  
columns of Table~\ref{table_histograms_e}. The length of each MC simulation 
(``MCS''), the number of discarded MCS (``Disc.''), and the number of 
measurements (``Meas.'') are also displayed in Table~\ref{table_histograms_e}.
For $48 \le L \le 384$, we have discarded 
$\gtrsim 1.1\times 10^4\, \tau_{\text{int}, \bm{\mathcal{M}}^2}$, 
and taken $\gtrsim 10^5\, \tau_{\text{int}, \bm{\mathcal{M}}^2}$ measures.
The statistics for $L=510$ is smaller: we have discarded  
$\gtrsim 3.6\times 10^3\, \tau_{\text{int}, \bm{\mathcal{M}}^2}$,
and taken $\gtrsim 3.2\times 10^4\, \tau_{\text{int}, \bm{\mathcal{M}}^2}$, 
measures.

\def\kk{\phantom{1}}
%
%
\begin{table}[htb]
\centering
\begin{tabular}{rlllll}
\hline\hline
\multicolumn{1}{c}{$L$}   &
\multicolumn{1}{c}{$v_1$} &
\multicolumn{1}{c}{$v_\circ$} &
\multicolumn{1}{l}{$\,$MCS}   &
\multicolumn{1}{c}{Disc.} &
\multicolumn{1}{c}{$\!\!\!$Meas.} \\
\hline \\[-4mm]
48 & $-0.95029$             &
   & $2 \times 10^8$ & $\kk2 \times 10^7$
   & $1.8 \times 10^8$      \\
96 & $-0.95095$             & $-0.950955(1)$
   & $2 \times 10^8$ & $\kk2 \times 10^7$
   & $1.8\times 10^8$       \\
132& $-0.95110$             & $-0.951098(1)$
   & $2 \times 10^8$ & $\kk2 \times 10^7$
   & $1.8\times 10^8$       \\
192& $-0.95119$             & $-0.951193(1)$
   & $4 \times 10^8$ & $\kk4 \times 10^7$
   & $3.6\times 10^8$       \\
384& $-0.95127$             & $-0.9512722(2)$
   & $9\times 10^8$  & $10 \times 10^7$
   & $8.0\times 10^8$       \\
510& $-0.95130$             & $-0.9512864(2)$
   & $4\times 10^8$  &$\kk4 \times 10^7$
   & $3.6\times 10^8$       \\
\hline\hline  
\end{tabular} 
\caption{For each value of $L=48, 96, 132, 192, 384, 510$,  
we show the temperature $v_1$ where the MC simulation was performed,  
the temperature $v_\circ(L)$ for which the extrapolated probability 
distribution $P_L(v_\circ;\{\mathcal{E}\})$ shows a two-peak structure with 
peaks of the same height, the total number of MCS, the number of discarded 
initial steps (``Disc.''), and the number of performed measures (``Meas.'').  
} 
\label{table_histograms_e} 
\end{table} 

We have followed a procedure to estimate the quantities displayed in
Table~\ref{table_histograms_e} which is based on the Ferrenberg--Swendsen 
algorithm and on the jackknife method to estimate the error bars (see e.g. 
Ref.~\cite{Weigel,Young} and references therein).

First, for each MC simulation performed at $v_1=v_1(L)$, we store the 
measured energy $\mathcal{E}$ into $N=20$ histograms 
$H_L^{(i)} = H_L^{(i)}(v_1(L);\{\mathcal{E}\})$. The $j$-th bin 
$H_L^{(i)}(\mathcal{E}_j)$ of the $i$-th histogram $H^{(i)}_L$ contains 
the number of times the energy value 
$\mathcal{E}_j \in \Delta{\mathcal E}(L)=
[\mathcal{E}_\text{min}(L),\mathcal{E}_\text{max}(L)]$ 
has appeared in the $i$-th part of that MC simulation. Indeed, the probability
distribution of the energy is given by
\be
P^{(i)}_L(v_1;\mathcal{E}_j) \;=\; 
\frac{H_L^{(i)}(v_1;\mathcal{E}_j)}
{\sum\limits_{\mathcal{E}_k \in \Delta{\mathcal E}(L)} 
      H_L^{(i)}(v_1;\mathcal{E}_k) } \,. 
\label{def_PL}
\ee 
Each histogram $H_L^{(i)}$ contains 
$\gtrsim 5\times 10^3 \, \tau_{\text{int}, \bm{\mathcal{M}}^2}$ measures for
$L\le 384$, and $\gtrsim 1.6\times 10^3\,\tau_{\text{int}, \bm{\mathcal{M}}^2}$ 
measures for $=510$.  
Due to the large values for the autocorrelation times for the WSK algorithm, 
this choice is a compromise between a number of blocks of order $10^2$, and 
a size of each block of order $10^4 \tau_{\text{int}}$ needed to apply 
the jackknife method according to Ref.~\cite{Weigel}.  
The number of energy values 
$\delta E(L)=\mathcal{E}_\text{max}(L)-\mathcal{E}_\text{min}(L)$ 
increases with $L$: $\delta E(96)=991$, $\delta E(132)=1451$, 
$\delta E(192)=2459$, $\delta E(384)=6500$, and $\delta E(510)=9521$.
 
Let us fix $L$, and suppose that, if we extrapolate the histograms
$H^{(i)}_L(v_1)$ to a nearby temperature $v_\circ$, we get a two-peak 
structure with peaks of equal height. Then we performed the following 
procedure to extract the relevant physical information:
\begin{enumerate}
\item[(a)] First, we obtained the probability distributions 
      $P_L^{(i)}(v_1;\{\mathcal{E}\})$ using \eqref{def_PL}. The  
      total probability distribution is given by  
\be
P_L(v_1;\{\mathcal{E}\}) \;=\; \frac{1}{N} \, \sum\limits_{i=1}^N 
               P^{(i)}_L(v_1;\{\mathcal{E}\})
\ee 
      because the normalization of every histogram is the same and equal to
      the total number of measurements. The computations in this procedure 
      have been carried out using {\sc Mathematica} with 50-digit precision.  
\end{enumerate} 

%
%
\begin{figure}[htb]
  \includegraphics[width=0.8\columnwidth]{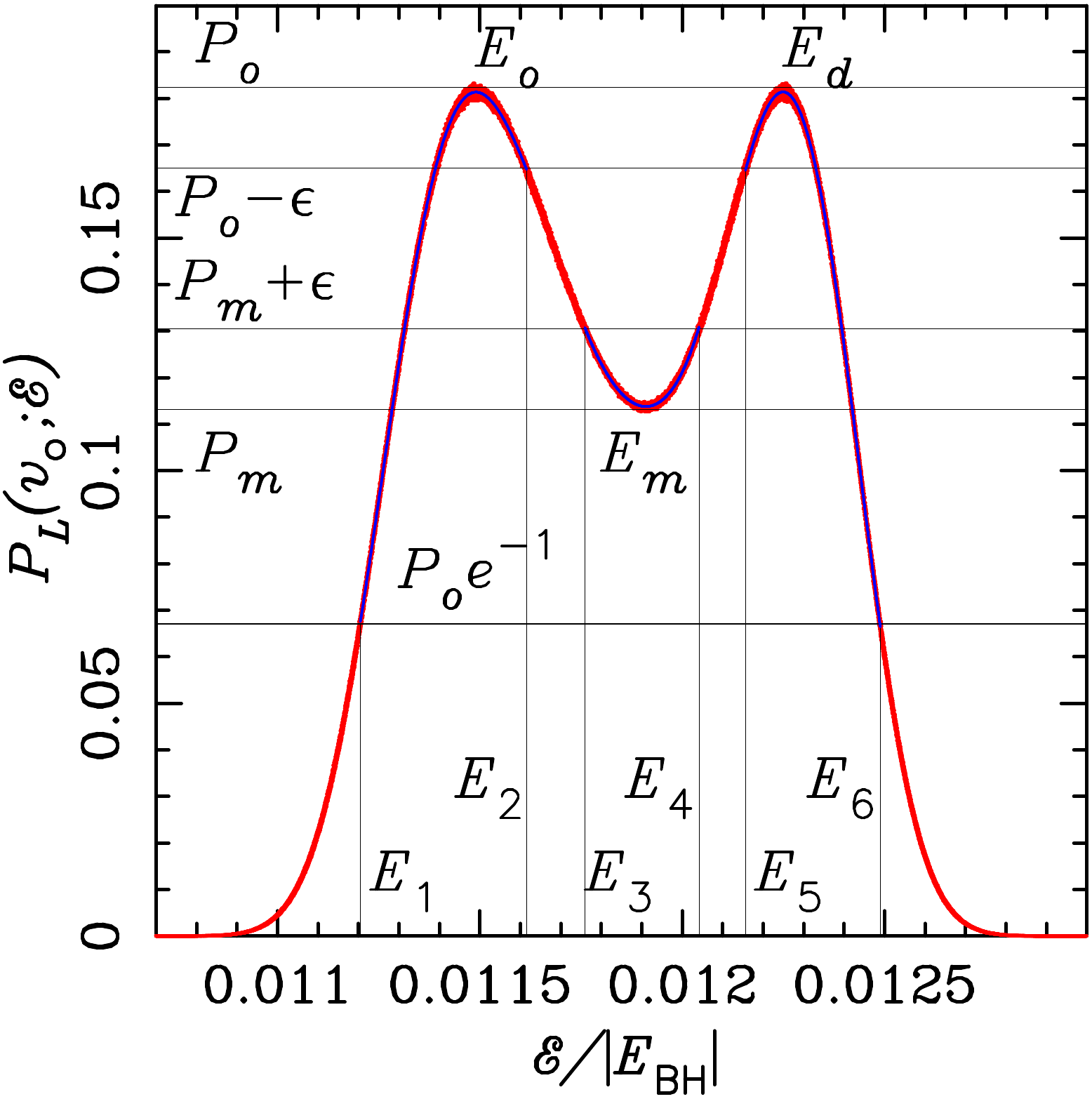} \\
  \vspace*{-2mm}
  \caption{%
  Energy histogram for the five-state BH-lattice AF Potts model at 
  $v=-0.9512722$ for $L=385$ (red). The blue curves are the actual fits 
  around the two peaks and the valley. We show the limits where these
  fits were performed. See point~(d) for an explanation of the labels.  
}
\label{fig_histograms_L=384}
\end{figure}
 
\begin{enumerate}
\item[(b)] We then obtained the extrapolated probability distributions 
      $P_L^{(i)}(v_\circ;\{\mathcal{E}\})$ at $v=v_\circ$ using the 
      Ferrenberg--Swendsen method. Likewise, we obtain the 
      total probability distribution $P_L(v_\circ;\{\mathcal{E}\})$. 

\item[(c)] We also needed an estimate of the error bars in the values of 
      $P_L(v_\circ;\mathcal{E}_j)$ for each 
      $\mathcal{E}_j\in \Delta\mathcal{E}(L)$. 
      We used the jackknife method to compute such error bars. Due to the 
      large amount of data involved, we \emph{assumed} that 
      each bin was statistically independent of the other bins, i.e.,  
      $\cov(P_L(v_\circ;\mathcal{E}_j),P_L(v_\circ;\mathcal{E}_k))=0$ for
      any $j\neq k$. Indeed, the jackknife estimate for the mean value of
      the $j$-th bin coincides with $P_L(v_\circ;\mathcal{E}_j)$, so we 
      recorded the corresponding error bars  
      $\sigma(P_L(v_\circ;\mathcal{E}_j))$.
      The assumption of neglecting the off-diagonal terms of the covariance
      matrix is \emph{not true} in general, so our error bars might be either 
      underestimated or overestimated. This is a warning in order to 
      interpret correctly the following fits.  

\item[(d)] The next step was to locate the position of both peaks and the 
      valley of the probability distribution $P_L(v_\circ;\{\mathcal{E}\})$, 
      and certain energy intervals containing them 
      (see Fig.~\ref{fig_histograms_L=384}). If the peaks of the 
      probability distribution have heights $P_o = P_d$, and the valley 
      in between them, $P_m$, then we define the quantities $P_{o,w} = P_o/e$
      and $\epsilon = (P_o-P_m)/4$. As seen in 
      Fig.~\ref{fig_histograms_L=384} for the case $L=384$, the fit to 
      obtain the peak at $E_o$ should be performed in the energy interval 
      $[E_1,E_2]$, the fit for the valley, in the interval $[E_3,E_4]$,
      and the fit for the peak at $E_d$, in the interval $[E_5.E_6]$.  

\item[(e)] For each energy interval, we fitted the logarithm of the 
     probability distribution to a \emph{cubic} \emph{Ansatz} 
     \cite{Billoire_94}: 
\be
     -\log\bigl( P_L(v_\circ;\{\mathcal{E})\}\bigr) \;=\; 
      A + B \, \mathcal{E} + C \mathcal{E}^2 + D \mathcal{E}^3 \,.
\ee
     As discussed in Ref.~\cite{Billoire_94}, Fig.~1, close to a peak, a 
     quadratic \emph{Ansatz} is not enough: a cubic term is needed. 
     These three fits allowed us to obtain the location of the three 
     extremal points $E_o(L)$, $E_d(L)$, and $E_m(L)$, the values 
     of the probability distribution at them [namely, $P_o(L)$, $P_d(L)$, 
     and $P_m(L)$, respectively], and three derived 
     quantities: the latent heat $\Delta E(L) = E_d(L)-E_o(L)$, the ratio  
\be
R_2(L) \;=\; \sqrt{ \frac{P_o(L)\, P_d(L)}{P_m(L)^2} } \,,  
\label{def_R2}
\ee
     (which is related to the interface tension \cite{Billoire_95}), 
     and the ratio $R_1(L) = P_o(L)/P_d(L)$, which is expected to be $1$ 
     if both peaks have the same height. Notice that both ratios $R_i(L)$  
     are \emph{independent} of the histogram normalization. For all these 
     fits, we obtained $0.13 \lesssim \chi^2/\text{DF} \lesssim 0.57$. The
     fact that this ratio is \emph{always} smaller than $1$ indicates that
     our error bars are a bit overestimated. 

\item[(f)] In order to compute the errors of the above estimates, we first
    tried a jackknife approach, but the error bars were unrealistically small: 
    e.g., $E_i(L) \sim 10^{-2}$ and $\sigma(E_i(L))\sim 10^{-8}$ for 
    $i\in\{o,d,m\}$. These error bars make no sense, as the relative error 
    in the probability distribution around the peaks or the valley is of order 
    $\gtrsim 10^{-3}$. 

    Therefore, we estimated the above physical quantities for each probability 
    distribution $P^{(i)}_L$ ($1\le i \le N$), and then we 
    estimated the corresponding error bars assuming that the results obtained 
    from each $P^{(i)}_L$ were statistically independent of the rest. Indeed,
    the errors for the $P^{(i)}_L(v_\circ,\{\mathcal{E}\})$ should be 
    multiplied by a factor $\sqrt{N}$ to take into account that the number of 
    measurements in each histogram $P^{(i)}_L$ is $N$ times smaller than the
    number of measurements in $P_L$. For all the performed fits, we obtained 
    $0.10 \lesssim \chi^2/\text{DF} \lesssim 0.57$; these bounds are similar 
    to those obtained in (e), and again are \emph{always} smaller than $1$.  

\item[(g)] Finally, in order to find the error bar of the extrapolated 
   temperature $v_\circ$ shown in Table~\ref{table_histograms_e}, we varied 
   $v_\circ$ so that the ratio $R_1(L)$ differ from our best estimate by 
   one standard deviation. Actually, we checked that the dispersion of our 
   estimates was very similar to the error bars found for our preferred value
   for $v_\circ$.  
\end{enumerate}

The final results for our procedure are summarized in 
Table~\ref{table_histograms_e2}. 

\def\kk{\phantom{1}}
%
%
\begin{table}[htb]
\centering
\begin{tabular}{rlll}
\hline\hline
\multicolumn{1}{c}{$L$}   &
\multicolumn{1}{c}{$E_o$} &
\multicolumn{1}{c}{$E_d$} &
\multicolumn{1}{c}{$E_m$} \\  
\hline \\[-4mm]
96 & $0.011\, 565(5)$ & $0.012\, 568(3)$     & $0.012\, 083(9)$ \\ 
132& $0.011\, 504(3)$ & $0.012\, 484(2)$     & $0.012\, 017(5)$ \\
132& $0.011\, 481(2)$ & $0.012\, 394(2)$     & $0.011\, 974(3)$ \\ 
384& $0.011\, 491(1)$ & $0.012\, 249\, 3(7)$ & $0.011\, 910(2)$ \\
510& $0.011\, 501(1)$ & $0.012\, 203(1)$     & $0.011\, 891(3)$ \\
\hline\hline 
\multicolumn{1}{c}{$L$}   &
\multicolumn{1}{c}{$R_1$} &
\multicolumn{1}{c}{$R_2$} &
\multicolumn{1}{c}{$\Delta E$} \\
\hline \\[-4mm]
96 & $1.001(7)$  & $1.072(1)$  & $0.001\, 403(6)$ \\
132& $1.000(7)$  & $1.152(4)$  & $0.000\, 980(3)$ \\ 
192& $1.00(1)$   & $1.277(3)$  & $0.000\, 913(2)$ \\ 
384& $1.00(1)$   & $1.595(5)$  & $0.000\, 758(1)$ \\
510& $1.00(1)$   & $1.78(1)$   & $0.000\, 702(2)$ \\
\hline\hline  
\end{tabular} 
\caption{For each value of $L=96, 132, 192, 384, 510$,  
we show on the top part, the estimates for the position of the two peak 
$E_o(L)$ and $E_d(L)$, and the position of the valley $E_m(L)$ for 
$v=v_\circ(L)$ given in Table~\ref{table_histograms_e}. In the lower part,
we show the estimates for the ratios $R_1$, $R_2$ \eqref{def_R2}, and the 
latent heat $\Delta E$. 
} 
\label{table_histograms_e2} 
\end{table} 

\begin{figure}[htb]
  \includegraphics[width=0.8\columnwidth]{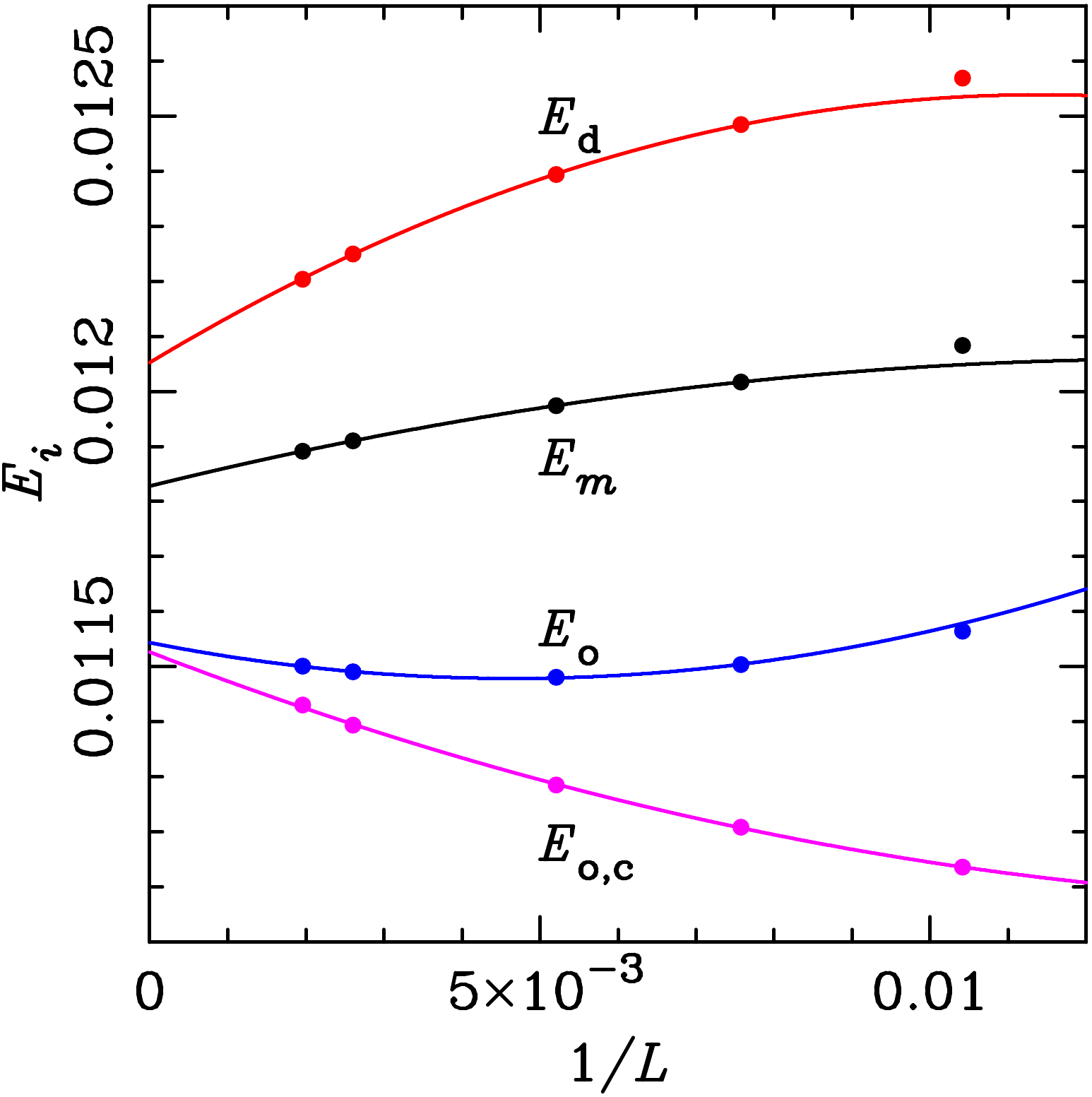} \\
  \vspace*{-2mm}
  \caption{%
  Estimates for the energies $E_d(L)$ (red), $E_m(L)$ (black) and $E_o(L)$ 
  (blue) obtained from the histograms at $v_\circ(L)$ for $L\ge 96$. The data 
  are displayed in Table~\ref{table_histograms_e2}, and the error bars 
  are smaller than the size of the points. The curves correspond to the 
  actual fits to the \emph{Ansatz} \eqref{fit.peaks.ansatz}. The pink points 
  correspond to the estimates for $E_{o,c}(L)$ obtained from the histograms 
  at $v=v_c$ (see Table~\ref{table_histograms_ec}), and the pink curve 
  is the actual fit to those points \eqref{fit.peaks.Eoc}. 
}
\label{fig_histo_peaks}
\end{figure}

Let us analyze the data shown in Table~\ref{table_histograms_e2}. Even though 
the phenomenological two-Gaussian \emph{Ansatz} \cite{Binder_84,Challa_86}
predicted that the position of the peaks $E_o(L)$ and $E_d(L)$ had no 
$L$ dependence, it was shown in Refs.~\cite{Lee_91,Janke_93} that these 
positions 
\emph{at the bulk critical point} varied linearly in $1/L$ for 2D FM Potts
models with $q=8,10$. Moreover, in Ref.~\cite{Billoire_94}, it was argued 
that this dependence was actually linear in $1/L^2$. In our case, we are 
\emph{not} at the bulk critical temperature, but it is easy to show using the  
phenomenological two-Gaussian framework, that imposing 
$P_L(v_\circ,E_o)=P_L(v_\circ,E_d)$ leads to $E_i(L) = E_i + O(1/L^2)$ for
$i=o,d$, and $v_\circ(L) = v_c + O(1/L^2)$ for a 2D system. Therefore,
looking at Fig.~\ref{fig_histo_peaks}, we have used the \emph{Ansatz}  
\be
E_i(L) \;=\; E_i + A/L + B/L^2 \,, \qquad i\in \{o,d,m\} \,.
\label{fit.peaks.ansatz}
\ee
%
%

The best results correspond to $L_\text{min}=132$ (DF = 1):  
\begin{subequations}
\label{fit.peaks}
\begin{align}
\label{fit.peaks.Eo}
E_o &\;=\; 0.011\, 544(5)\,, \\ 
E_d &\;=\; 0.012\, 052(5)\,, \\
E_m &\;=\; 0.011\, 828(9)\,. 
\end{align}
\end{subequations}
The values of $\chi^2$ are $0.70$, $2.46$, and $4.87\times 10^{-3}$, 
respectively. The corresponding confidence levels are 40.4\%, 11.7\%, 
and 94.4\%. 

%
%
\begin{figure}[htb]
  \includegraphics[width=0.8\columnwidth]{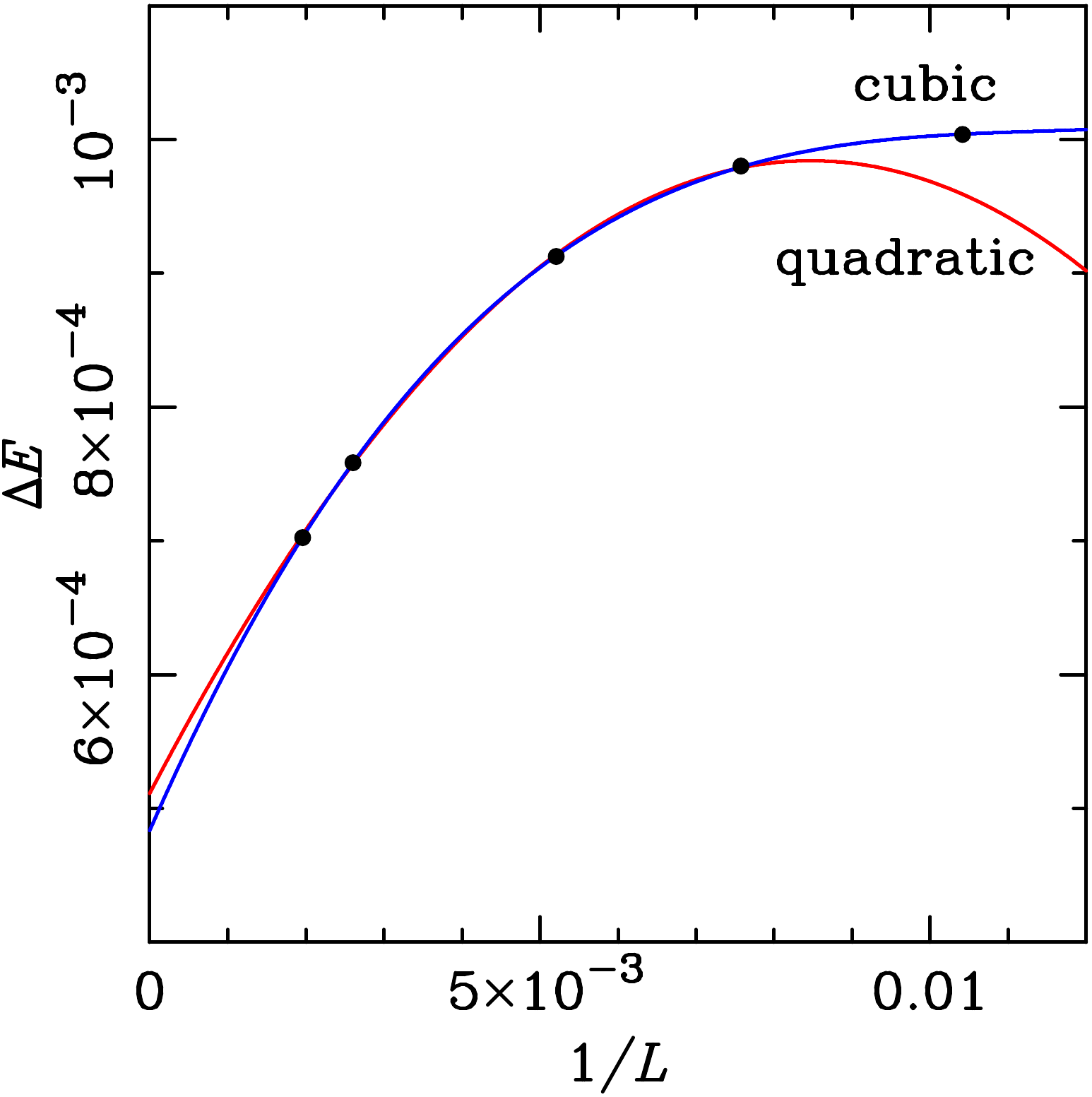} \\
  \vspace*{-2mm}
  \caption{%
  Estimates for the latent heat $\Delta E(L)$ obtained by
  using the histogram method for $L\ge 96$. The data are displayed in 
  Table~\ref{table_histograms_e2}, and the error bars are smaller than the 
  size of the points. The curve labeled ``quadratic'' corresponds to the 
  \emph{Ansatz} \eqref{fit.peaks.ansatz} and the results \eqref{fit.Lh0}.   
  The other curve labeled ``cubic'' corresponds to a cubic \emph{Ansatz}
  and the results \eqref{fit.LhOK}.
}
\label{fig_histo_lh}
\end{figure}

For the latent heat $\Delta E(L)$, we have first used the same \emph{Ansatz}
\eqref{fit.peaks.ansatz}, and the best results correspond to 
$L_\text{min}=132$:
\be
\Delta E \;=\; 0.000\, 511(6) \,, 
\label{fit.Lh0}
\ee
with $\chi^2/\text{DF}= 3.57/1$ and $\text{CL} = 5.89\%$. 
This last value is rather poor, so we tried to add a cubic term $C/L^3$ to the 
\emph{Ansatz} \eqref{fit.peaks.ansatz}. In this case, we get a better 
fit for $L_\text{min}=96$:
\be
\Delta E \;=\; 0.000\, 48(1) \,,
\label{fit.LhOK}
\ee
with $\chi^2/\text{DF}= 0.22/1$ and $\text{CL} = 66.6\%$. Both fits 
\eqref{fit.Lh0} and \eqref{fit.LhOK} are shown in Fig.~\ref{fig_histo_lh}.

It is clear from either the study of the position of the peaks $E_o$ and $E_d$
\eqref{fit.peaks}, or equivalently, from the study of the latent heat 
\eqref{fit.Lh0} and \eqref{fit.LhOK} that the transition is of first order.
In the latter case, the estimates for the latent heat are many standard 
deviations away from $\Delta E=0$. Notice that the value of the latent heat
(per edge) is two orders of magnitude smaller than the latent heat for
the five-state square-lattice FM Potts model 
$\Delta E = 0.026\, 459\, 378\ldots$ \cite{Baxter_73}, which is one canonical
example of a very weak first-order phase transition.  
 
%
%
\begin{figure}[h]
  \includegraphics[width=0.8\columnwidth]{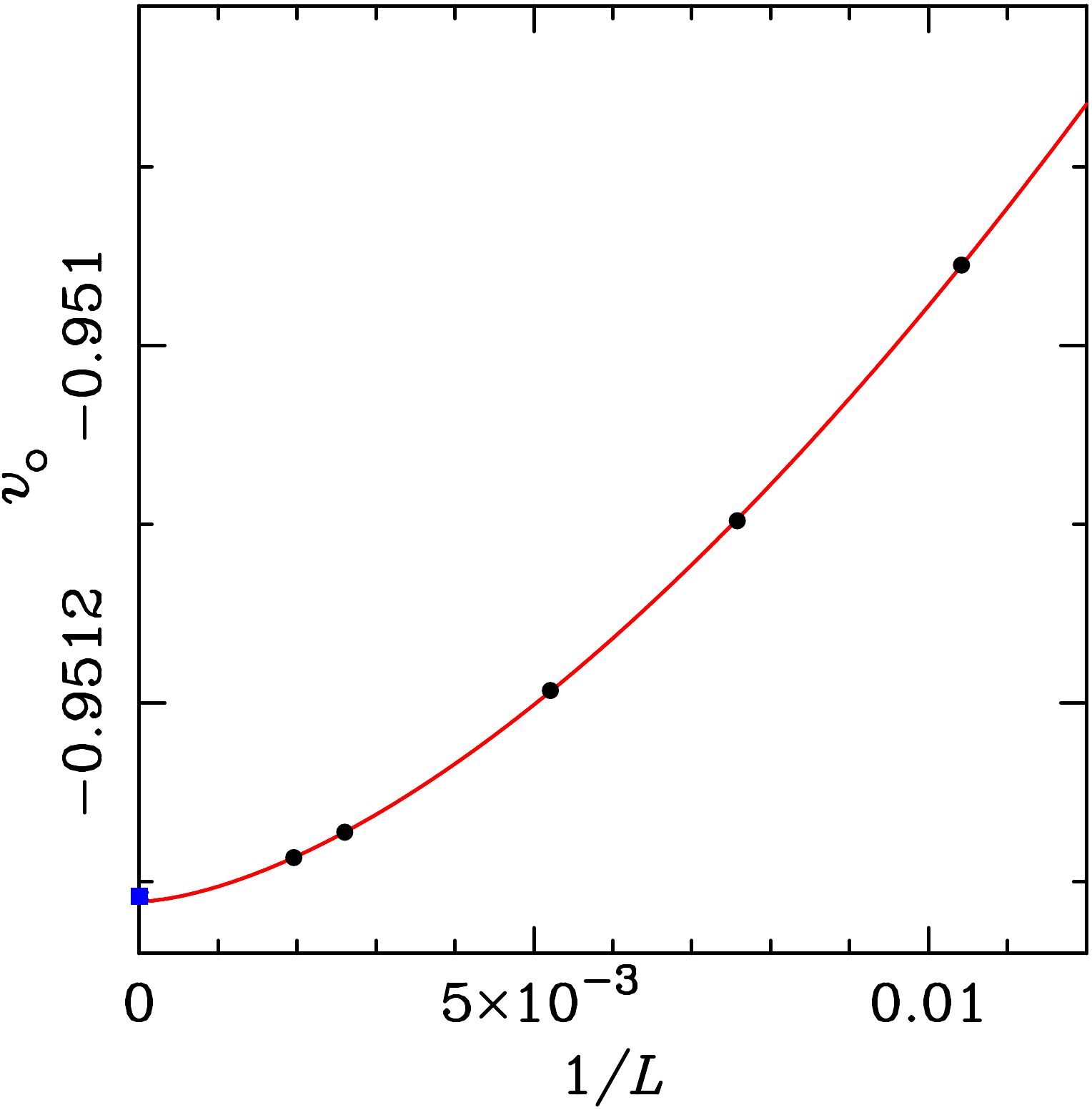} \\
  \vspace*{-2mm}
  \caption{%
  Estimate of the pseudocritical temperature $v_\circ(L)$ for $L\ge 96$ 
  displayed in Table~\ref{table_histograms_e}.
  The corresponding curve corresponds to the \emph{Ansatz} 
  \eqref{fit.vo.ansatz}. 
  The blue point at $1/L=0$ is the MC estimate \eqref{value_vc.NB}.  
}
\label{fig_histo_v1}
\end{figure}

We have also fitted the quantity $v_\circ(L)$ displayed in 
Table~\ref{table_histograms_e} to a power-law \emph{Ansatz}:
\be
v_\circ(L) \;=\; v_c + L^{-\omega} \,. 
\label{fit.vo.ansatz}
\ee
The best result corresponds to $L_\text{min}=96$:
\begin{subequations}
\label{fit.vo}
\begin{align}
v_c    &\;=\; -0.951\, 311\, 0(7) \,, \\
\omega &\;=\; \phantom{-}1.60(1) \,,   
\end{align}
\end{subequations}
with $\chi^2/\text{DF}= 1.06/2$, and $\text{CL} = 58.9\%$  
(see Fig.~\ref{fig_histo_v1}).

This result for $v_c$ is slightly smaller than the MC estimate
$v_c=-0.951\, 308(2)$ \eqref{value_vc.NB}; but the difference is 
only $1.5$ standard deviations of the latter estimate. Notice that the 
exponent in \eqref{fit.vo} is expected to be $\omega = 2$. This difference 
can be explained by the fact that our lattices are too small to attain the
expected asymptotic behavior.  

%
%
\begin{figure}[htb]
  \includegraphics[width=0.8\columnwidth]{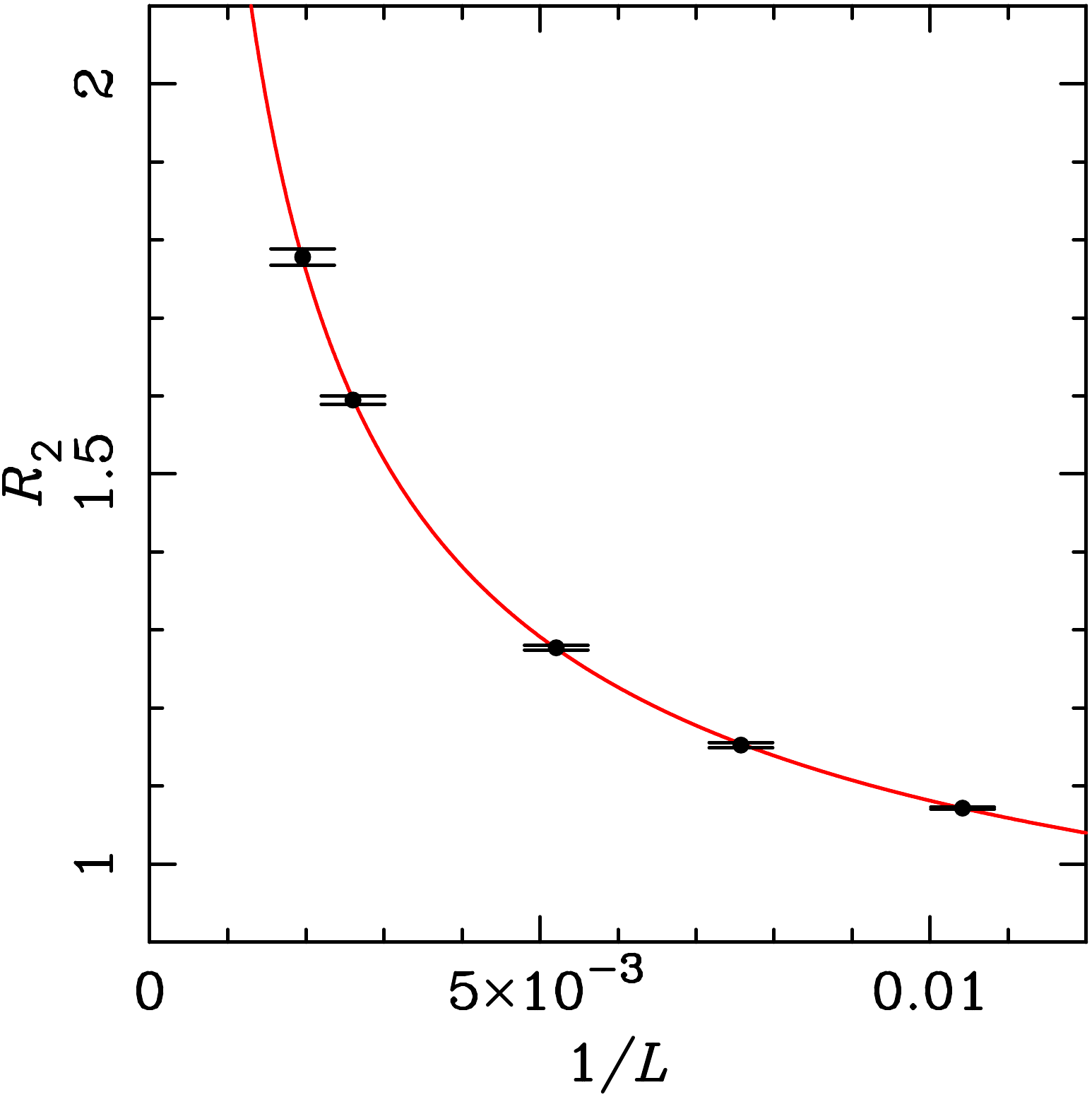} \\
  \vspace*{-2mm}
  \caption{%
  Estimates for the ratio $R_2(L)$ for $L\ge 96$ 
  and displayed in Table~\ref{table_histograms_e}. The curve corresponds to the 
  \emph{Ansatz} \eqref{fit.R2.ansatz}.  
}
\label{fig_histo_ratio2}
\end{figure}

Let us now consider the ratio $R_2(L)$ \eqref{def_R2}. If this ratio grows as
$L$ increases, it means that the value of $P_m(L)$ becomes exponentially small
with respect to the values $P_o(L)=P_d(L)$. The data are displayed in 
Table~\ref{table_histograms_e}. We have tried a power-law fit with a 
constant:
\be
R_2(L) \;=\; A \, L^{\omega} + B \,.
\label{fit.R2.ansatz}
\ee 
We obtain a good fit for $L_\text{min}=96$:
\be
\omega \;=\; 0.65(3) \,,
\label{fit.omega.R2}
\ee
with $\chi^2/\text{DF}=0.65/2$ and $\text{CL} = 72.2\%$  
(see Fig.~\ref{fig_histo_ratio2}). 

\begin{figure}[h]
  \includegraphics[width=0.8\columnwidth]{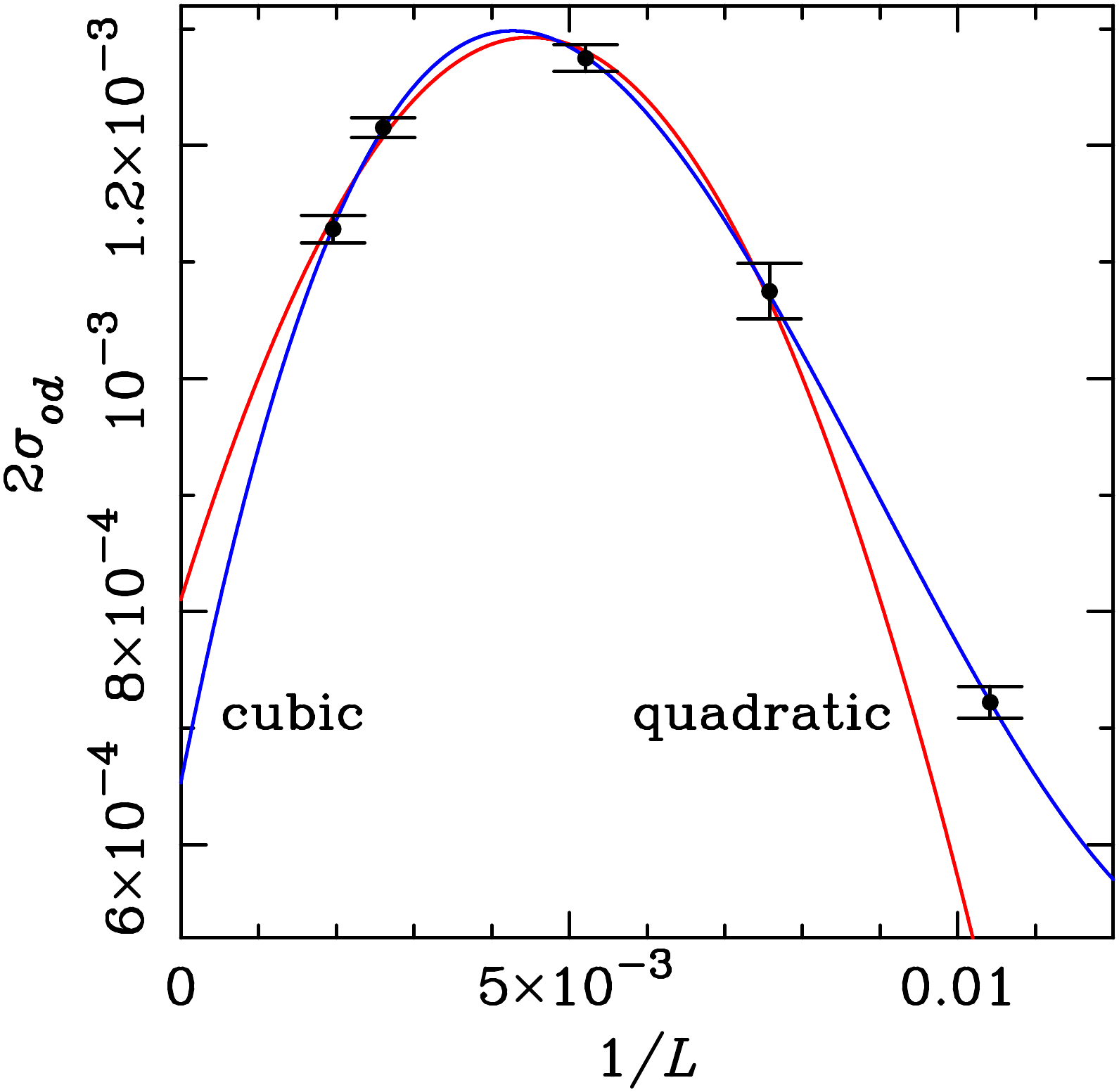} 
  \vspace*{-2mm}
  \caption{%
  Estimates for the interface tension $2\sigma_{od}(L)$ \eqref{def_sigmaod}
  for $L\ge 96$ obtained from the ration $R_2(L)$ displayed in 
  Table~\ref{table_histograms_e}. The blue curve corresponds to the 
  quadratic \emph{Ansatz} \eqref{fit.tension.ansatz}. The red curve 
  corresponds to a cubic \emph{Ansatz} in $1/L$. 
}
\label{fig_histo_sigmaod}
\end{figure}

Finally, we are going to consider the interface tension $\sigma_{od}(L)$. 
Notice that this study is usually performed at the 
\emph{bulk critical temperature} by defining the estimate for 
a 2D system of linear size $L$ as 
\cite{Billoire_94,Billoire_95}:
\be
2 \sigma_{od}(L) \;=\; \frac{1}{L} \, \log (R_2(L)) \,.
\label{def_sigmaod}
\ee
However this method has also been applied at the pseudocritical temperature
$v_\circ(L)$, where the two peaks attain the same height \cite{Berg_93}.
(See also the discussion in Ref.~\cite[end of Sec.~3]{Billoire_94}.)
The data are depicted in Fig.~\ref{fig_histo_sigmaod}. The behavior of
$2 \sigma_{od}(L)$ is nonmonotonic with $L$, so we do not expect a very good
fit due to the small number of data points we have. We have tried a quadratic 
fit in $1/L$:
\be
2 \sigma_{od}(L) \;=\; 2 \sigma_{od} + A/L + B/L^2 \,.
\label{fit.tension.ansatz} 
\ee

%
%

The best result corresponds to $L_\text{min}=132$:
\be
2 \sigma_{od} \;=\; 0.000\, 81(4) \,,
\label{fit.sigmaod}
\ee
with $\chi^2/\text{DF}=2.08/1$ and $\text{CL}=14.9\%$. 
If we add a cubic term $C/L^3$ to the \emph{Ansatz} \eqref{fit.tension.ansatz},
we find a good result for $L_\text{min}=96$:
\be
2 \sigma_{od} \;=\; 0.000\, 65(8) \,,
\label{fit.sigmaod.cubic}
\ee
with $\chi^2/\text{DF}=0.059/1$ and $\text{CL}=80.7\%$.

For a first-order phase transition, we expect that $2 \sigma_{od}(L)$ has
a positive limit when $L\to\infty$. If the transition is second-order, this
limit is expected to be zero. For the five-state FM Potts model on the square 
lattice, the interface surface is given by 
$\sigma_{od} = 0.000\, 398\, 050\ldots$ \cite{Borgs_92b,Janke_94}.    
This number is one order of magnitude smaller than the estimate we obtained 
using the growth of the autocorrelation time [c.f., \eqref{sols.tau2.B}], 
but is of the same order of magnitude than the previous estimates
\eqref{fit.sigmaod} and \eqref{fit.sigmaod.cubic}. Therefore, even though our
results are small, they are not consistent with zero, and there are well-known
first-order phase transitions with interface tension this small.  

In any case, our results 
\eqref{fit.sigmaod} and~\eqref{fit.sigmaod.cubic} are consistent
with a first-order phase transition; but the difference between both 
estimates indicate that we have not attained yet the asymptotic regime. 
Moreover, the exponent $\omega$ found for the growth of the ratio $R_2(L)$ 
\eqref{fit.omega.R2} is not consistent with the expected result $\omega=1$
for a 2D system \cite{Billoire_94}. Again, these results should be 
interpreted with a grain of salt, as the linear sizes of the simulated 
systems are not large enough.  

%
%
\subsection{Analysis at the critical temperature} 
\label{sec.histo.vc}

We have followed the same procedure to estimate the value of $E_o(L)$ 
from the histograms at $v=v_c$ (see Fig.~\ref{fig_histograms_e0}).
This is a consistency check for the procedure followed in the previous 
section. In this case, we have fitted the data close to the peak 
at $E_o(L)$, which in this case we have chosen to be in the interval
$[P_o(L),P_o(L)/2]$. [Choosing $P_o(L)/e$ will lead to points too close to the
incipient small bump at $E_d(L)$.] The mean value comes from the estimates
obtained at $v_c=-0.951\, 308(2)$ \eqref{value_vc.NB}; and the error bars 
take into account the statistical errors, as well as the dispersion 
obtain by repeating the procedure at $v=-0.951\, 306$ and 
$v_c=-0.951\, 310$. The data are displayed in 
Table~\ref{table_histograms_ec} and depicted in the lowermost curve 
of Fig.~\ref{fig_histo_peaks}. 
 
%
%
\begin{table}[htb]
\centering
\begin{tabular}{rl}
\hline\hline
\multicolumn{1}{c}{$L$}   &
\multicolumn{1}{c}{$E_o$} \\  
\hline \\[-4mm]
96 & $0.011\, 136(2)$  \\ 
132& $0.011\, 209(2)$  \\
192& $0.011\, 286(3)$  \\ 
384& $0.011\, 394(5)$  \\
510& $0.011\, 430(6)$  \\
\hline\hline 
\end{tabular} 
\caption{For each value of $L=96, 132, 192, 384, 510$,  
we show the estimates for the position of the ordered peak 
$E_o(L)$ obtained from the histogram at $v=v_c$ (see 
the lower curve labeled ``$E_{o,c}$'' in Fig.~\ref{fig_histograms_e0}).  
} 
\label{table_histograms_ec} 
\end{table} 

We have used the quadratic \emph{Ansatz} in $1/L$ \eqref{fit.peaks.ansatz},
and we obtain a good result for $L_\text{min}=96$: 
\be
E_o \;=\; 0.011\, 527(8)\,, 
\label{fit.peaks.Eoc}
\ee
with $\chi^2/\text{DF}=1.34/2$ and $\text{CL}=51.1\%$. 
This value differs from the previous estimate \eqref{fit.peaks.Eo}
by $\sim 1.7\times 10^{-5}$, i.e., roughly two standard deviations given in 
\eqref{fit.peaks.Eoc}.    

Finally, we could find the position of the small ``bump'' displayed by
the probability distribution for $L=510$ (see Fig.~\ref{fig_histograms_e0}).
Using this value, we could estimate the latent heat
\be
\Delta E(510) \;=\; 0.000\, 708(6) \,,
\ee
which agrees well with the estimate obtained in the previous section 
(see Table~\ref{table_histograms_e2}). 

%
%
\subsection{Other physical quantities} 
\label{sec.histo.other}

It is well known 
\cite{Binder_84,Challa_86,Borgs_90,Borgs_91,Lee_91,Billoire_Lacaze_Morel_92}
that the specific heat for a 2D system undergoing a first-order phase
transition has a maximum value $C_{H,\text{max}}(L)$ at a point 
$v_{C_{H,\text{max}}}(L)$ such that 
\begin{subequations}
\label{fit.cvmax}
\begin{align}
\label{fit.cvmax1}
v_{C_{H,\text{max}}}(L)  &\;=\; v_c + A \, L^{-2} + \cdots \,, \\ 
C_{H,\text{max}}(L)      &\;=\; A\, L^2 + B + C\, L^{-2} + \cdots \,, 
\label{fit.cvmax2}
\end{align}
\end{subequations}
where the dots represent higher-order powers in $L^{-2}$. Similar results 
are obtained for the minimum of the Binder cumulant $U_4$:
\begin{subequations}
\label{fit.u4min}
\begin{align}
\label{fit.u4min1}
v_{U_{4,\text{min}}}(L)  &\;=\; v_c + A \, L^{-2} + \cdots \,, \\
U_{4,\text{min}}(L)      &\;=\; A\, L^2 + B + C\, L^{-2} + \cdots \,,
\label{fit.u4min2}
\end{align}
\end{subequations}

%
%
\begin{table}[htb]
\centering
\begin{tabular}{rllll}
\hline\hline
\multicolumn{1}{c}{$L$}   &
\multicolumn{1}{c}{$v_{C_{H,\text{max}}}$} & 
\multicolumn{1}{c}{$C_{H,\text{max}}$} & 
\multicolumn{1}{c}{$v_{U_{4,\text{min}}}$} & 
\multicolumn{1}{c}{$U_{4,\text{min}}$} \\ 
\hline \\[-4mm]
96 &$-0.950\, 962(2)$    & $0.094\, 7(1)$&$-0.950\, 961(3)$    &$2.203(3)$\\ 
132&$-0.951\, 101(1)$    & $0.127\, 8(2)$&$-0.951\, 096(2)$    &$2.104(3)$\\
192&$-0.951\, 193\, 4(7)$& $0.186\, 7(2)$&$-0.951\, 189\, 7(9)$&$2.009(2)$\\ 
384&$-0.951\, 271\, 4(2)$& $0.404\, 8(6)$&$-0.951\, 270\, 8(3)$&$1.831(2)$\\
510&$-0.951\, 285\, 7(2)$& $0.571(1)$    &$-0.951\, 284\, 8(4)$&$1.775(3)$\\
\hline\hline 
\end{tabular} 
\caption{For each value of $L=96, 132, 192, 384, 510$,  
we show the estimates for the position of the maximum of the specific heat
$v_{C_{H,\text{max}}}$, the value of such maximum $C_{H,\text{max}}$, 
the position of the minimum of the Binder cumulant $v_{U_{4,\text{min}}}$,
and the value of such minimum $U_{4,\text{min}}$. 
} 
\label{table_cvmax} 
\end{table} 

%
%
\begin{figure}[htb]
  \includegraphics[width=0.8\columnwidth]{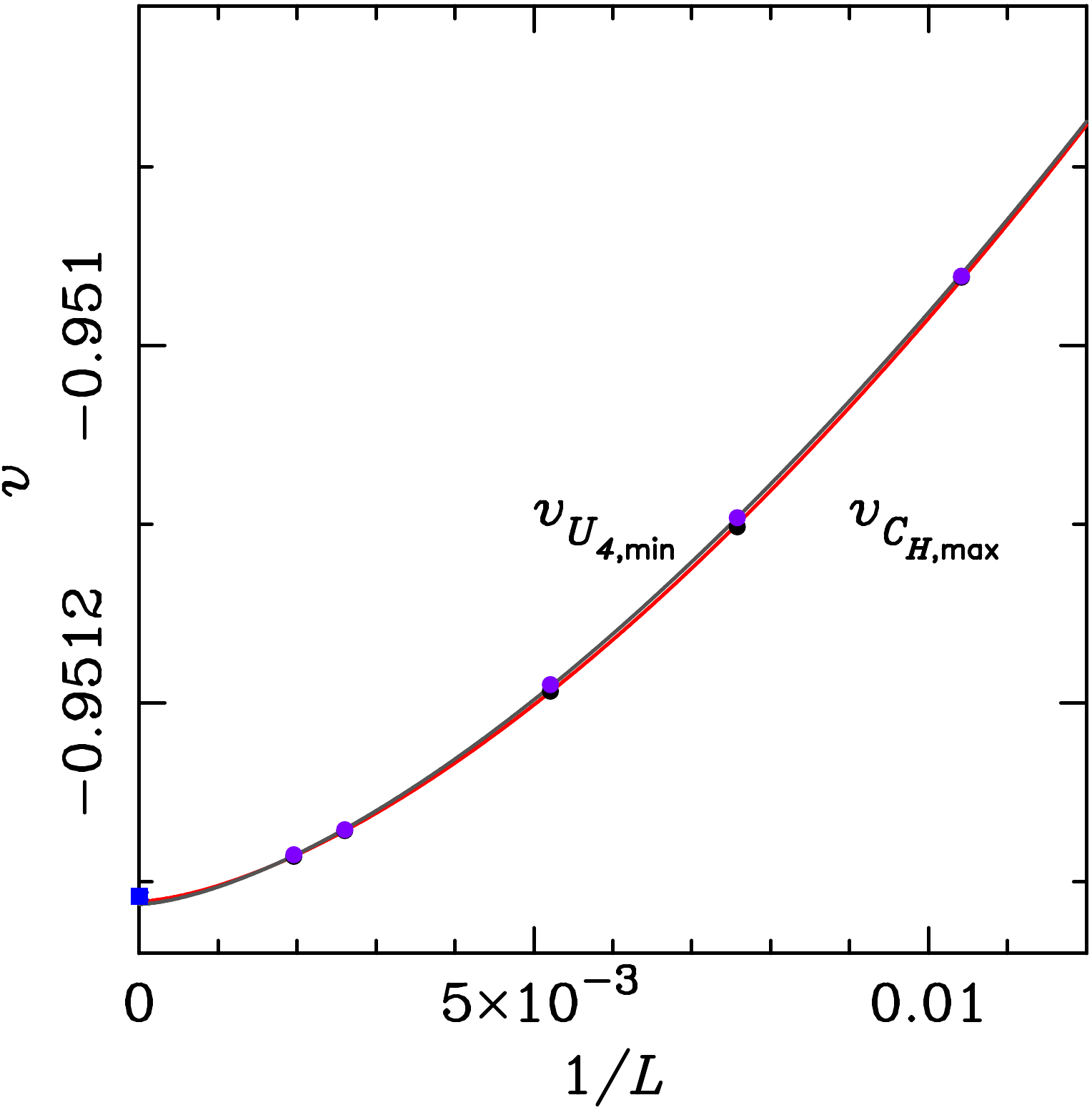} \\
  \vspace*{-2mm}
  \caption{%
  Estimates of the pseudocritical temperatures 
  $v_{C_{H,\text{max}}}$ (black points) and 
  $v_{U_{4,\text{min}}}$ (violet points) for $L\ge 96$
  displayed in Table~\ref{table_cvmax}.
  The corresponding curves (that are almost indistinguishable)  correspond
  to the \emph{Ansatz} \eqref{fit.vmax.ansatz} for the latter (red curve),
  and \eqref{fit.vmin.ansatz} for the former (dark gray curve). 
  The blue point at $1/L=0$ is the MC estimate \eqref{value_vc.NB}.
}
\label{fig_histo_vcvmax}
\end{figure}

By using the Ferrenberg--Swendsen extrapolation method, we have been able to 
locate these maxima
of the specific heat for $L\ge 96$. We have used the same method as in 
Sec.~\ref{sec.histo.equal}: (1) Obtain the mean values for 
$v_{C_{H,\text{max}}}(L),C_{H,\text{max}}(L)$ by using the whole set of
measurements; and (2) in order to estimate the error bars, split this set
of measurements into $N=20$ groups with identical size, compute estimates 
for these quantities within each group, and obtain its mean value and 
standard deviation \emph{assuming} that the results coming from each group are 
statistically independent from the rest. These results are displayed in
Table~\ref{table_cvmax}. We also performed a jackknife analysis of these
data, but again the error bars were too small: e.g., for $L=192$, the
jackknife estimate for the error bar is $\approx 10^{-5}$ (i.e., $20$ times
smaller that the error quoted in Table~\ref{table_cvmax}). However, the 
value of the specific heat for the original MC simulation is
$C_H(-0.95119;192) = 0.187(2)$. This error bar is $10$ times the error
quoted in Table~\ref{table_cvmax} but $200$ times larger than the jackknife 
one. Therefore, we conclude that the latter errors are too small, and we
prefer to consider the more conservative ones displayed in 
Table~\ref{table_cvmax}. In a similar way we have obtained the corresponding
results for the mimima of $U_4(L)$.     

Let us consider the values of $v_{C_{H,\text{max}}}(L)$. If we use the
\emph{Ansatz} [inspired by the exact result \eqref{fit.cvmax1}]: 
\be
v_{C_{H,\text{max}}}(L) \;=\; v_c + A\, L^{-\omega} \,,
\label{fit.vmax.ansatz}
\ee
we obtain a good fit for $L_\text{min}=132$: 
\begin{subequations}
\label{fit.vmax}
\begin{align}
v_c    &\;=\; -0.951\, 312(1) \,, \\
\omega &\;=\; \phantom{-}1.54(3) \,,
\end{align}
\end{subequations}
with $\chi^2/\text{DF}= 0.32/1$, and $\text{CL} = 57.0\%$.
See the red curve in Fig.~\ref{fig_histo_vcvmax}. These results are 
very similar to the ones we obtained for $v_\circ(L)$ [cf., \eqref{fit.vo}]. 

\begin{figure}[htb]
  \includegraphics[width=0.8\columnwidth]{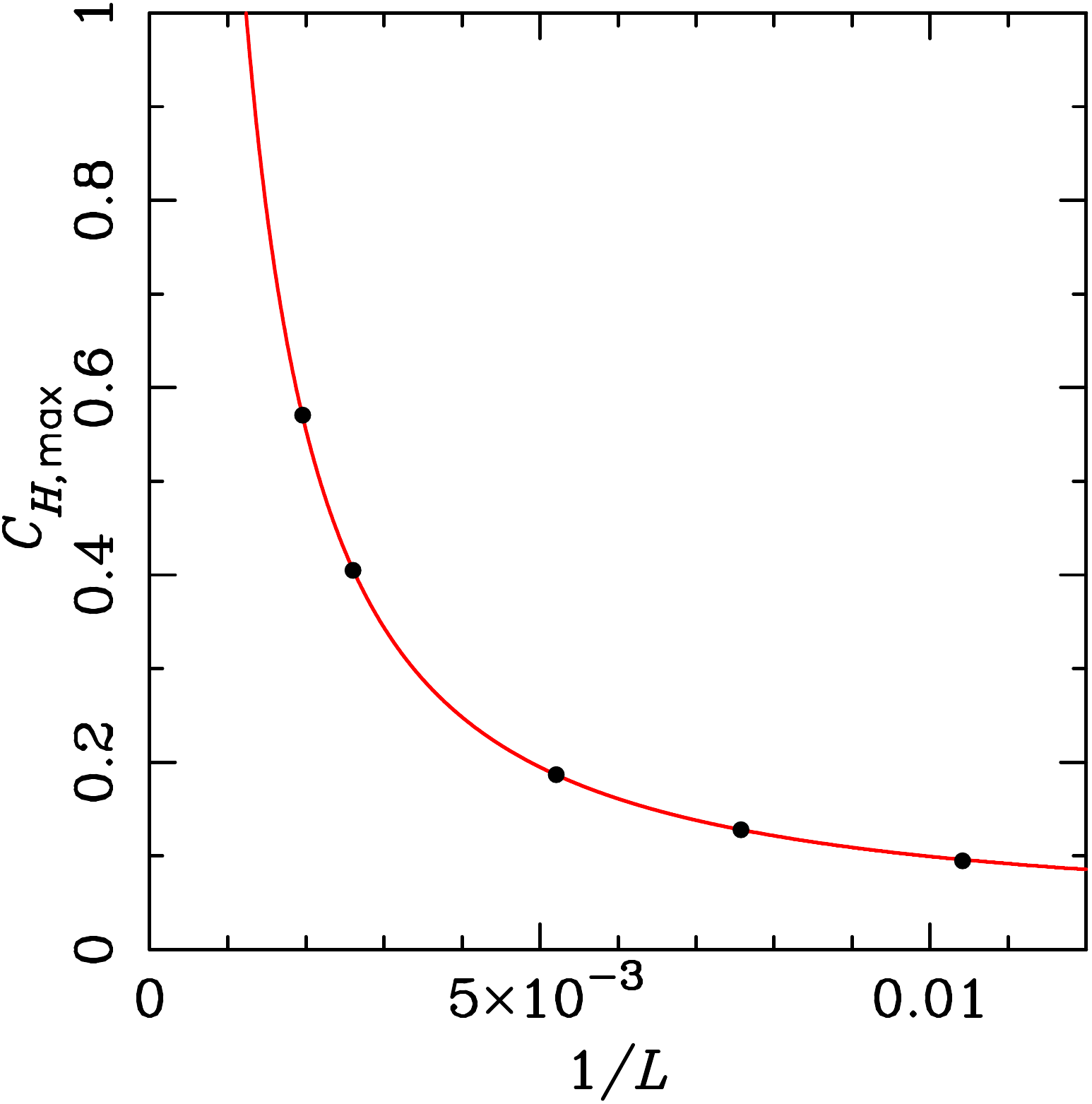} \\
  \vspace*{-2mm}
  \caption{%
  Estimates for the maximum value of the specific heat $C_{H,\text{max}}(L)$ 
  as a function of $L\ge 96$. The data is displayed in Table~\ref{table_cvmax}.
  The curve corresponds to the \emph{Ansatz} \eqref{fit.cvmax3}. Error bars
  are smaller than the symbols.  
}
\label{fig_histo_cvmax}
\end{figure}

Similarly, we used the \emph{Ansatz} 
\be
C_{H,\text{max}}(L) \;=\; A\, L^\omega + B\,,  
\label{fit.cvmax3}
\ee
which is inspired by the exact result \eqref{fit.cvmax2}. 
The ``best'' result is obtained for $L_\text{min}=132$:
\be
\omega \;=\; 1.275(5)\,, 
\ee 
with $\chi^2/\text{DF}= 5.97/1$, and $\text{CL} = 1.45\%$. The confidence level
is rather poor. The best explanation we have is that the sizes of our
systems are still too small compared to the infinite-volume (\emph{finite})
correlation length $\xi_c$ at $v=v_c$. Notice that $\omega = 1.275(5)$ is  
compatible within errors with the estimate for $\alpha/\nu = 1.21(7)$ 
\eqref{sols.CH.B} obtained in Sec.~\ref{sec.critical.exp} from the 
standard FSS analysis of the MC data (see Fig.~\ref{fig_histo_cvmax}). 

%
%

Let us now repeat the same analysis for the Binder cumulant $U_4$. If we
consider the values of $v_{U_{4,\text{min}}}(L)$ and use the  
\emph{Ansatz} 
\be
v_{U_{4,\text{min}}}(L) \;=\; v_c + A\, L^{-\omega} \,,
\label{fit.vmin.ansatz}
\ee
the ``best'' fit is obtained for for $L_\text{min}=96$: 
\begin{subequations}
\label{fit.vmin}
\begin{align}
v_c    &\;=\; -0.951\, 313(1) \,, \\
\omega &\;=\; \phantom{-}1.53(2) \,,
\end{align}
\end{subequations}
with $\chi^2/\text{DF}= 4.67/2$, and $\text{CL} = 9.68\%$.
These results are very similar to the ones we obtained for \eqref{fit.vmax},
and to those for $v_\circ(L)$ [cf., \eqref{fit.vo}].

%
%
\begin{figure}[htb]
  \includegraphics[width=0.8\columnwidth]{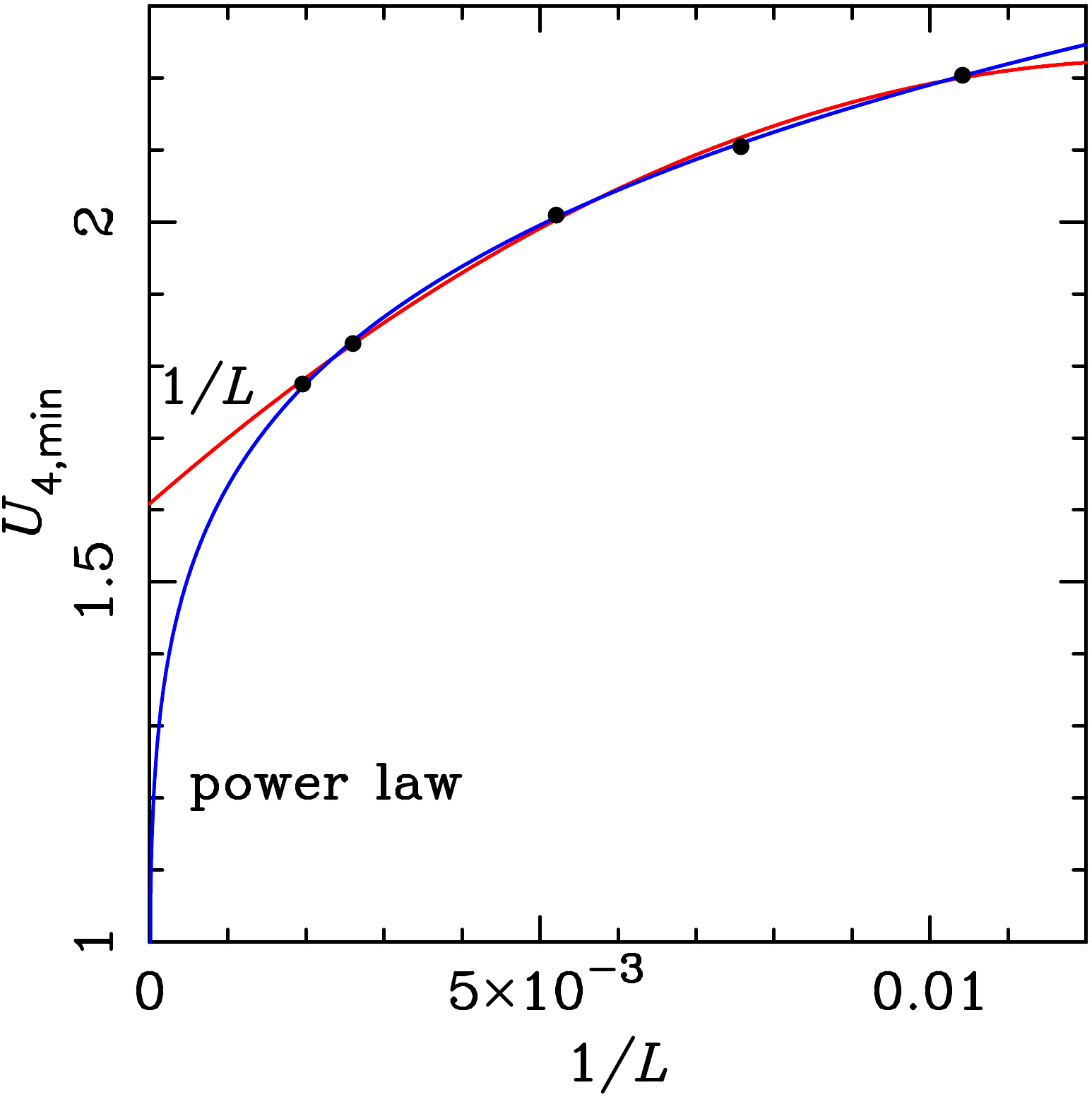} \\
  \vspace*{-2mm}
  \caption{%
  Estimates for the minimum value of the Binder cumulant $U_{4,\text{min}}(L)$ 
  as a function of $L\ge 96$. The data are displayed in Table~\ref{table_cvmax}.
  The lower blue curve corresponds to the \emph{Ansatz} 
  \eqref{fit.u4min.ansatz},
  and the upper red curve to the \emph{Ansatz} \eqref{fit.u4min.ansatz2}. 
  Error bars are smaller than the symbols.  
}
\label{fig_histo_u4min}
\end{figure}

For the mimimum value of $U_4$, we first used the \emph{Ansatz} 
\be
U_{4,\text{min}}(L) \;=\; U_{4,c} + B\, L^{-\omega} \,.  
\label{fit.u4min.ansatz}
\ee
The ``best'' result is obtained for $L_\text{min}=96$:
\begin{subequations}
\label{fit.umin}
\begin{align}
U_{4,c} &\;=\; 0.7(2) \,, \\ 
\omega  &\;=\; 0.20(3) \,, 
\end{align}
\end{subequations}
with $\chi^2/\text{DF}= 8.77/2$, and $\text{CL} = 1.25\%$. The confidence level
is rather poor. This is the blue curve on Fig.~\ref{fig_histo_u4min}. One
flaw of this result is that the value for $U_{4,c} < 1$, which is impossible
\cite{Billoire_Lacaze_Morel_92}: $U_{4,c} \ge 1$, and it attains the value
$U_{4,c}=1$ when it is the sum of two delta functions. Therefore, we tried
the following \emph{Ansatz}:
\be
U_{4,\text{min}}(L) \;=\; U_{4,c} + A \, L^{-1} + B\, L^{-2} \,,
\label{fit.u4min.ansatz2}
\ee 
as the behavior of $U_{4,\text{min}}(L)$ seems rather smooth as a function
of $1/L$ on Fig.~\ref{fig_histo_u4min}. The result in this case is
\be
U_{4,c} \;=\; 1.575(8)\,,
\ee 
for $L_\text{min}=132$, $\chi^2/\text{DF}=9.02\times 10^{-8}/1$, and
$\text{CL} =100\%$. This result corresponds to the red curve on
Fig.~\ref{fig_histo_u4min}. Probably the right value is somewhere in 
between these two values, as we have seen in other fits that the 
expected asymptotic regime has not been attained due to the fact that 
$L \ll \xi_c$.

%
%
\section{Conclusions} \label{sec.conclusions}

In this paper we have performed a high-precision MC study of the five-state 
AF Potts model on the BH lattice. We have found that there is a low-$T$
phase where five distinct phases coexist (see Appendix~\ref{appen.ground}),
and a standard paramagnetic disordered phase at high temperature. 
In between there is
a phase transition at $v_c=-0.951\, 308(2)$, which is one order of magnitude 
more precise than previous determinations \cite{union-jack}. 

The ``standard'' analysis of the MC data gives a value for the exponent 
$y_t = 1/\nu = 2.0(1)$, which is compatible within errors with the 
expected value for a first-order phase transition in a 2D system  
\cite{Nauenberg_74,Klein_76,Fisher_82}. However the values of the 
critical-exponent ratios are far from the expected ones: namely, 
$\alpha/\nu=1.21(2)$ and $\gamma/\nu=1.766(9)$. (The expected values of 
these ration is $2$ for a 2D system undergoing a first-order phase 
transition). Notice that there is a clear inconsistency between the 
values of $1/\nu=2.0(1)$ and $\alpha/\nu=1.21(2)$: below the upper 
critical dimension $D<4$, the ``hyperscaling'' relation for a $D$-dimensional
system is expected to hold (see Ref.~\cite{Pelissetto_02}, Sec.~1.3):
\be
(2-\alpha)/\nu \;=\; D \,. 
\ee 
The lhs is $\approx 2.79$, while the rhs is equal to $D=2$.  

Concerning the dynamics of the WSK algorithm, its slowest mode turns out to be
the square magnetization $\bm{\mathcal{M}}^2$ \eqref{def.M2.OK}. It is clear 
from the MC data (see Fig.~\ref{fig.tauM2.fine}) that the system suffers from
critical slowing down: the autocorrelation times diverges close to the bulk 
critical point.
If we assume that this divergence is like a power (as in second-order 
critical points \cite{Sokal_97}), we obtain the dynamic critical exponent
$z_{\text{int},\bm{\mathcal{M}}^2}=1.54(6)$. This is smaller than the 
expected value ($\gtrsim 2$) for single-site MC algorithms. On the other 
hand, if we assume the exponential growth  
$\tau_{\text{int},\bm{\mathcal{M}}^2} \sim \exp\bigl(\sigma_{od}\, L\bigr)$   
which characterizes the behavior for a first-order phase transition on a
2D system, we find a very small interface tension 
$\sigma_{od} = 0.0035(1)$.  

It is worth noticing that the fits for the universal amplitudes $R$, $U_4$,
and $\xi/L$ (see Figs.~\ref{fig.SSfine.U4}--\ref{fig.SSfine.xioverL}),
even though they needed $L_\text{min}=192,384$, the curves for smaller values
of $L$ were not too far from the corresponding data points. However, the 
fits for $C_H$, $\chi$, and $\tau_{\text{int},\bm{\mathcal{M}}^2}$ did 
need $L_\text{min}=384$, and the curves for smaller values of $L$ look 
quite far away from the corresponding points (especially for the 
dynamic data; see Figs.~\ref{fig.SSfine.CH}--\ref{fig.SSfine.tauB}). 

Due to the fact that some results were in good agreement with a first-order
phase transition, while others pointed in the other direction, we analyzed 
the data using the histogram method. Even though one can get only some
partial results from the energy histograms extrapolated to the bulk
critical point $v_c$, the whole picture can be studied by considering  
these histograms extrapolated to a pseudocritical temperature $v_\circ(L)$
defined as the one such that the histogram has two peaks of equal height.
Notice that this two-peak structure appears only for $L\gtrsim 96$, while
it is absent for $L\lesssim 48$. 
The first result is that there is a tiny latent heat (per edge) for this system 
$\Delta E = 0.000\, 48(1)$ (and the position of the peak corresponding to the
low-$T$ ordered phase $E_o$ was in good agreement with the estimate obtained
from the histograms at $v_c$). We call it ``tiny'' because it is two orders
of magnitude smaller than the latent heat for other weak first-order 
phase transitions \cite{Baxter_73,Adler_95}. From this result, we 
conclude that the transition is an extremely weak first-order one. 
Indeed, the extrapolation of the pseudocritical points  $v_\circ(L)$ 
agrees well with the value obtained in the ``standard'' analysis of the MC
data.

We could also estimate the interface tension $\sigma_{od}$ from these
histograms extrapolated at $v_\circ(L)$. Its behavior with $1/L$ is not
monotonic, and we do not have many data points. So the results are not
very conclusive: $\sigma_{od} = 0.000\, 7(1)$. This value is five times
smaller from the estimate obtained from the growth of the autocorrelation 
time $\tau_{\text{int},\bm{\mathcal{M}}^2}$ (see Fig.~\ref{fig_histo_sigmaod}). 

Finally, we also considered the position of the maximum of the specific
heat and the minimum of $U_4$, leading to new pseudocritical 
temperatures $v_{C_{H,\text{max}}}$ and $v_{U_{4,\text{min}}}$, respectively 
(see Fig.~\ref{fig_histo_vcvmax}). Their behavior is very similar to that
of $v_\circ(L)$, and the extrapolated values are close to the bulk 
critical point $v_c$, although in general the estimates coming from
the histogram method are slightly smaller than that coming from the 
standard MC analysis [e.g., $-0.951\, 311(2)$ versus $-0.951\, 308(2)$].   
We also tried to analyze how the maximum of the specific heat 
$C_{H,\text{max}}$ and the minimum of the Binder cumulant $U_{4,\text{min}}$
behave with $L$. 
Notice that the confidence levels for the power-law fits to these 
quantities are rather poor 
(see Figs.~\ref{fig_histo_cvmax} and~\ref{fig_histo_u4min}).  

The result that the transition is an extremely weak first order implies that
the correlation length at the bulk critical point is very large, but 
finite. For instance, for the five-state FM Potts model on the square 
lattice, we have that \cite{Borgs_92b,Janke_94} 
\be
\sigma_{od} \;\approx\; 3.981 \times 10^{-4} \,, \quad 
\xi_d \;=\; \frac{1}{\sigma_{od}} \;\approx\; 2.512 \times 10^3 \,,   
\ee
where $\xi_d$ is the critical correlation length on the disordered phase. 
Even though these figures do not correspond to our system, they can give 
us an order of magnitude. For instance, the interface tension is
of the same order as the one obtained from the histogram method. The 
fact that $\xi_d \sim 10^3$ implies that the linear size of our systems 
are fairly small to ``see''  the correct scaling for a first-order 
transition. This is probably the reason why we do not get the expected 
values for $\gamma/\nu$ and  $\alpha/\nu$, and the estimates for the 
interface tension do not agree among them.  

For future work, the main problem left is to find a lattice $\mathcal{G}$ such
that the five-state $\mathcal{G}$-lattice AF Potts model is critical. 
With respect to the five-state BH-lattice AF Potts model, there are 
two interesting technical problems to be solved:
(1) Find the height representation for this model at $T=0$. As it is not
disordered at all temperatures, it should have one such representation, 
as claimed by Henley \cite{Henley_unpub}. 
(2) Even though the numerical data support the idea that WSK is ergodic 
for this lattice at $T=0$, we need a rigorous proof of such a claim. This 
would be a step towards proving the ergodicity of the WSK algorithm for
the five-state AF Potts model on the triangular lattice. 

\appendix
%
%
\section{The right staggering} \label{appen.stag}

In this Appendix we will discuss the choice we have made for the 
staggering. If $q=3$, then the `obvious' staggering would be to
choose $\phi_A=0$, $\phi_B= 2\pi/3$, and $\phi_C= -2\pi/3$. 
This choice is motivated by the fact that
at $T=0$, the BH graph has a unique 3-coloring modulo global permutations 
of colors. Moreover, for $q=4$ and $T=0$, previous studies \cite{union-jack} 
indicate that sublattice~$A$ (the one with the largest-degree vertices) 
is ordered, while the other two sublattices are disordered. For $q=5$,
one would expect naively similar behavior. 

One way to find the right staggering is the following. We first define the 
magnetization \emph{density} $\bm{\mathcal{M}}_{k}$ 
[cf., \eqref{def_bsigma} and \eqref{eq_e}] for each sublattice 
$k\in \{A,B,C\}$ as 
\be
\bm{\mathcal{M}}_{k} \;=\; \frac{1}{|V_k|}\,
                      \sum\limits_{\bm{x}\in V_k} \bm{\sigma}_{\bm{x}} \,.
\label{def_sublattice_mag}
\ee
We now define the mean value of the product of two of such magnetizations 
densities  
\begin{subequations}
\label{def_mab}
\begin{align}
 M_{ab} &\;=\; 
    \left\< \bm{\mathcal{M}}_a \cdot \bm{\mathcal{M}}_b \right\>  \\
    &\;=\;  \frac{q}{q-1} \frac{1}{|V_a| \, |V_b|}  
    \sum\limits_{\alpha=1}^q 
    \left\< \sum\limits_{x\in V_a} \delta_{\sigma_x,\alpha}   
            \sum\limits_{y\in V_b} \delta_{\sigma_y,\alpha} \right\> \qquad 
    \nonumber \\
    & \qquad \qquad  - \frac{1}{q-1} \,,   
\end{align}
\end{subequations}
so that each diagonal entry is bounded in the interval $[0,1]$, while the
nondiagonal ones are bounded in the interval $\bigl[-\frac{1}{q-1},1\bigr]$. 
In terms of these element, we can compute the mean square magnetization 
density quadratic form $\mathsf{M} = (M_{ab})_{a,b \in \{A,B,C\}}$
\be
\mathsf{M} \;=\; \left( \begin{array}{ccc}
    M_{AA} & M_{AB} & M_{AC} \\
    M_{AB} & M_{BB} & M_{BC} \\
    M_{AC} & M_{BC} & M_{CC} \\
\end{array} \right) \,,
\label{def_M_quadratic}
\ee
where we have used the obvious symmetry $M_{ab}=M_{ba}$.
 
We can now compute the eigenvalues and eigenvectors of \eqref{def_M_quadratic}.
The eigenvector associated to the largest eigenvalue will correspond to the
right staggering. As $\mathsf{M}$ is a real and symmetric matrix, its 
eigenvalues are real and the corresponding eigenvectors have real entries and 
can be chosen to form an orthogonal basis of $\R^3$. 

Given the quadratic form \eqref{def_M_quadratic}, we can compute the
susceptibility associated to any linear combination of the sublattice
magnetizations \eqref{def_sublattice_mag}. In particular, if 
the leading eigenvalue of \eqref{def_M_quadratic} $\lambda_\circ$, and its
associated eigenvector is 
$\bm{\alpha}_\circ =(\alpha_A,\alpha_B,\alpha_C)\in\R^3$, then the linear  
combination
\begin{equation}
\bm{\mathcal{M}}_\circ \;=\; \sum_{a \in\{A,B,C\}} 
\alpha_a\, \bm{\mathcal{M}}_a
\label{def_mag_gen}
\end{equation}
gives the susceptibility 
\begin{equation}
\chi_\circ \;=\; |V_\text{BH}| \, 
       \< \bm{\mathcal{M}}_\circ \cdot \bm{\mathcal{M}}_\circ \> \;=\; 
   |V_\text{BH}| \, \bm{\alpha}^t_\circ \cdot 
                    \mathsf{M} \cdot \bm{\alpha}_\circ \,.   
\label{def_chi_gen}
\end{equation}
Note that the normalization is distinct from that of Eq.~\eqref{def_chi}.

As a check, we can compute the staggering for the zero-temperature three-state 
AF Potts model on a triangular lattice of size $L\times L$. In this case, 
$M_{aa} = 1$, and $M_{ab}=-1/2$ whenever $a\neq b$. The dominant
eigenvalue of the quadratic form \eqref{def_M_quadratic} is 
$\lambda_\circ= 3/2$, 
and its associated eigenvector is $\bm{\alpha}_\circ = (1,-1,0)^t$. Then  
$\< \bm{\alpha}^t_\circ \cdot \mathsf{M} \cdot \bm{\alpha}_\circ \> = 3/2$. 
Notice that the standard staggering 
$\bm{\alpha}_{\text{st}} = \bigl(1,e^{2\pi i/3},e^{-2\pi i/3}\bigr)^t$ 
gives exactly the same result:
$\left\< \bm{\mathcal{M}}_{\text{st}}^* \cdot 
        \bm{\mathcal{M}}_{\text{st}} \right\>  = 
\left\<(\bm{\alpha}^*_{\text{st}})^t \cdot \mathsf{M} \cdot 
        \bm{\alpha}_{\text{st}} \right\> = 3/2$. 

Going back to the five-state BH-lattice AF Potts model, we focused in the 
interval 
$v\in[-0.952,-0.951]$. For each value of $L=24,48,96$, we performed $11$ 
MC simulations at equidistant values of $v$ in that interval. For each one 
of these runs, we discarded at least 
$10^4\, \tau_{\text{int}}$ MCS, and we took at least 
$9.5\times 10^4\, \tau_{\text{int}}$ measures.

%
%
\begin{table}[htb]
\centering
\begin{tabular}{rlll}
\hline\hline
\multicolumn{1}{c}{$L$} &
\multicolumn{1}{c}{$\alpha_A$} &  
\multicolumn{1}{c}{$\alpha_B$} &  
\multicolumn{1}{c}{$\alpha_C$} \\  
\hline
24 & $0.92704$ & $-0.28027$ & $-0.24911$ \\  
48 & $0.92674$ & $-0.28117$ & $-0.24919$ \\  
96 & $0.92654$ & $-0.28181$ & $-0.24923$ \\  
\hline\hline
\multicolumn{1}{c}{Best} 
   & $0.927$   & $-0.282$   & $-0.249$   \\
\hline\hline
\end{tabular}
\caption{For each value of $L=24,48,96$ and for $v=-0.9513$, we display the 
three components $\alpha_i$ of the (unitary) leading eigenvector 
$\bm{\alpha}_\circ$ of the quadractic form \eqref{def_M_quadratic}. 
The lower row (labeled ``Best'') shows our preferred estimates for $\alpha_i$. 
}
\label{table.eigen.M}
\end{table}

We computed the leading eigenvector $\bm{\alpha}_\circ$ of the quadratic 
form \eqref{def_M_quadratic} for each simulation, and we saw that they behaved  
smoothly as a function of $v$. In order to determine the optimal value of 
$\bm{\alpha}_\circ$, we choose $v=-0.9513$, which is the one closest 
to the previous estimate of $v_c=-0.95132(2)$ \cite{union-jack}. 
The numerical results are displayed in Table~\ref{table.eigen.M}.  

The results look rather stable as a function of $L$. As 
$\alpha_A/4 \approx |\alpha_B|,|\alpha_C|$, we could define the 
staggered magnetization as this simple linear combination:  
\begin{equation}
\bm{\mathcal{M}}_\circ \;=\; 4 \bm{\mathcal{M}}_A - \bm{\mathcal{M}}_B - 
                               \bm{\mathcal{M}}_C \,.
\label{def_mag_gen_prop}
\end{equation}
It is clear that the most relevant contribution to $\bm{\mathcal{M}}_\circ$
corresponds to $\bm{\mathcal{M}}_A$ (i.e., the sublattice with vertices of
largest degree, in agreement with previous works \cite{union-jack}).  
Actually, what we need is an observable with a large overlap with 
$\bm{\mathcal{M}}_\circ$ \eqref{def_mag_gen_prop}. The simplest choice is 
\begin{equation}
\bm{\mathcal{M}}_\circ \;=\; \bm{\mathcal{M}}_A \,. 
\label{def_mag_gen_OK}
\end{equation}
This definition, although is not the optimal one, is expected to pick up the 
relevant physics of the system without the technical programming difficulties
of using \eqref{def_mag_gen_prop}, and moreover, it is expected to reduce 
significantly the CPU time of the MC simulations. In the main text, 
we drop for notational simplicity the subindex $\circ$. 

%
%
\section{The ground-state structure} \label{appen.ground}

Let us consider the five-state BH-lattice Potts model at $T=0$, and 
assume that WSK is ergodic for the graphs $G_{\text{BH},L}$ (see the 
discussion in Sec.~\ref{sec.WSK}, in particular, 
Fig.~\ref{fig.tauM2.fine}). In this Appendix, we are going to consider 
first the six sublattices shown in Fig.~\ref{fig.bh.geom} labeled 1 to 6,
so that $A=\{1\}$, $B=\{2,3\}$, and $C=\{4,5,6\}$. 

We have computed the corresponding enlarged mean square magnetization 
density quadratic form 
$\mathsf{M} = (M_{ab})_{a,b \in \{1,2,\ldots,6\}}$ [cf., 
\eqref{def_mab} and~\eqref{def_M_quadratic}]. For the lattice with $L=96$, we
performed $8\times 10^6$ MCS, discarded 
$\gtrsim 2.5\times 10^4 \, \tau_{\text{int}}$,
and took $\gtrsim 2.2\times 10^5 \, \tau_{\text{int}}$ measurement. 
In this case, the slowest mode corresponded to $\mathcal{M}_1^2$.  
First, we checked that the obvious symmetries hold:
\begin{subequations}
\begin{align}
\mathsf{M}_{bb} &\;=\; \phantom{-}0.062\, 344\, 0(3)\,, \quad b\in B\,,\\
\mathsf{M}_{cc} &\;=\; \phantom{-}0.060\, 789\, 1(9)\,, \quad c\in C\,,\\
\mathsf{M}_{1b} &\;=\; -0.235\, 129(3)\,, \quad b\in B\,,\\
\mathsf{M}_{1c} &\;=\; -0.232\, 186(4)\,, \quad c\in C\,,\\
\mathsf{M}_{bc} &\;=\; \phantom{-}0.061\, 404\, 5(6)\,, \quad b\in B\,, 
                       \quad c\in C\,,\\
\mathsf{M}_{cd} &\;=\; \phantom{-}0.060\, 709\, 7(8)\,, \quad c,b\in C\, 
                       \quad  c\neq d \,. 
\end{align}
\end{subequations}
The numerical estimates come from fits to a constant \emph{Ansatz}. 
In each fit, the data points are not statistically independent, so we have 
quoted for each quantity, instead of the error bar obtained from the fit, 
the more conservative error bar of the individual values (which were 
approximately constant). 
In all fits, there is at least one DF, and we obtain an excellent value 
for $\chi^2 \lesssim 0.3$. 

%
%
\begin{table}[ht]
\centering
\def\kk{\phantom{1}}
\begin{tabular}{rlll}
\hline\hline
\multicolumn{1}{r}{$L$} &
\multicolumn{1}{c}{$\mathsf{M}_{AA}$} &
\multicolumn{1}{c}{$\mathsf{M}_{BB}$} &
\multicolumn{1}{c}{$\mathsf{M}_{CC}$} \\
\hline
 3 & $\kk0.909(3)$   & $\kk0.1458(3)$     & $\kk0.1178(2)$     \\
 6 & $\kk0.8944(9)$  & $\kk0.08327(6)$    & $\kk0.07510(4)$    \\
12 & $\kk0.8900(3)$  & $\kk0.06746(1)$    & $\kk0.06427(1)$    \\
24 & $\kk0.8886(1)$  & $\kk0.063536(2)$   & $\kk0.061577(5)$   \\
48 & $\kk0.88855(5)$ & $\kk0.0625549(7)$  & $\kk0.060906(2)$   \\
96 & $\kk0.88843(2)$ & $\kk0.0623084(3)$  & $\kk0.0607362(8)$  \\
\hline
\multicolumn{1}{r}{$\infty$} &
     $\kk0.88841(2)$ & $\kk0.0622266(3)$  & $\kk0.0606800(8)$  \\
\hline
\multicolumn{1}{c}{$L_\text{min}$}&$\kk6$     & $12$        & $\kk6$  \\
\multicolumn{1}{c}{$\chi^2/$DF}   &$\kk2.84/3$& $\kk1.97/2$ & $\kk1.37/3$\\
\multicolumn{1}{c}{CL}            &$41.7\%$   & $37.4\%$    & $71.2\%$ \\
\hline\hline
\end{tabular}
\caption{Values of the diagonal terms of the quadratic form 
\eqref{def_M_quadratic} as a function of $L$. 
The row labeled ``$\infty$'' shows the results obtained by
fitting the data to a \emph{biased} power-law \emph{Ansatz}. 
We also display the values of $L_\text{min}$, $\chi^2/\text{DF}$, and CL 
of our preferred fit (see text). 
\label{table_MSMD_diagonal}
}
\end{table}

Next, we computed the quadratic form \eqref{def_M_quadratic} at $v=-1$, taking
into account all these symmetries to improve the statistics. 
We simulate systems with $L=3,6,12,24,48,96$, and for each of them, we 
simulate $10^6 - 8\times 10^7$ MCS, discarded at least 
$3.1\times 10^3 \tau_{\text{int}}$ MCS, and took 
$2.8 \times 10^4 \tau_{\text{int}}$ measures. For $L\ge 24$, we increased 
the statistics to at least $1.1 \times 10^5 \tau_{\text{int}}$ measures. 
The slowest mode corresponded to $\bm{\mathcal{M}}_A^2$. 
The diagonal terms of the quadratic form \eqref{def_M_quadratic} 
for different values of $L$ 
are given in Table~\ref{table_MSMD_diagonal}. For each diagonal element 
$\mathsf{M}_{aa}$ (with $a\in \{A,B,C\}$), we have performed a power-law fit 
$\mathsf{M}_{aa}(L) = \mathsf{M}_{aa}^* + B_a \, L^{-\omega_a}$. 
In all cases, we found values of $\omega_a$ that agree with $\omega_a=2$ within 
errors: namely, $\omega_A= 2.0(2)$, $\omega_B=1.997(3)$, 
and $\omega_C=2.003(4)$. Therefore, we have performed \emph{biased}
power-law fits with $\omega_a=2$. The values of $\mathsf{M}_{aa}^*$ coming
from these \emph{biased} fits are displayed in Table~\ref{table_MSMD_diagonal}
on the row labeled ``$\infty$''. We also show the values of $L_\text{min}$,
$\chi^2/\text{DF}$, and CL of our preferred fits. 

The diagonal term $\mathsf{M}_{aa}$ of the quadratic form 
\eqref{def_M_quadratic} measures
the FM order of the spins within the sublattice 
$a \in \{A,B,C\}$. The study of these values 
provides some insight on the ground-state structure of our model, as 
shown previously in the literature \cite{selfdual1,selfdual2}. 
In particular, if the spins in sublattice $a$ are fully FM ordered, then 
$\mathsf{M}_{aa}=1$. Moreover, if the spins in this sublattice take $r=2,3,4,5$
values at \emph{random}, then the corresponding values of $\mathsf{M}_{aa}$
are $3/8=0.375$, $1/6\approx 0.16667$, $1/16 = 0.0625$, and $0$, respectively.
The last case corresponds to the spins in sublattice $a$ being completely 
uncorrelated.   

Looking at Table~\ref{table_MSMD_diagonal}, we see that 
$\mathsf{M}_{AA}= 0.88841(3)$. This value is close to $1$, and 
more than twice $3/8$. We conclude that the degree-12 sublattice $A$
is almost (but not completely) FM ordered. 
On the other hand, the other two diagonal entries satisfy 
$\mathsf{M}_{CC} = 0.0606800(8) \lesssim \mathsf{M}_{BB} = 
                   0.0622266(3) \lesssim 1/16 = 0.0625$. 
This means that the spins in the degree-6 sublattice $B$ 
take very approximately four distinct values at random, while those in  
the degree-4 sublattice $C$ are slightly more disordered. 

These results coincide qualitatively with those presented in 
Refs.~\cite{selfdual1,selfdual2} for several quadrangulations and $q=3$: 
the sublattice with vertices of largest degree is the most FM ordered,
and the ordering of the other two sublattices is roughly that of having $q-1$ 
spin values at random. For these two lattices, the one with smaller degree
is less ordered. 

The results for the three-state AF Potts models presented in  
Refs.~\cite{selfdual1,selfdual2} could be understood from a theoretical point 
of view: the zero-temperature 3-state AF Potts model on any plane 
quadrangulation has a ``height'' representation 
\cite{Henley_unpub,Kondev_96,Burton_Henley_97,SS_98,Jacobsen_09}. 
Once the height mapping is known, the next step is to find the so-called 
\emph{ideal states}: i.e., families of ground-state configurations whose 
corresponding height configurations are macroscopically flat, and maximize
the entropy density. The long-distance behavior of this microscopic 
height model is expected to be controlled by an effective Gaussian model, 
which can be in two distinct phases. If the Gaussian model is in its 
\emph{rough} phase (or at the \emph{roughening} transition), then the spin 
model is critical at $T=0$. On the other hand, if the Gaussian model is in 
its \emph{smooth} phase, then the corresponding spin model can be described 
by small fluctuations around the ideal states. In this case, the model 
exhibits long-range order at $T=0$. Roughly speaking, the ground state is 
ordered with with as many ordered and coexisting phases as the number of 
ideal states previously found. 

In the five-state BH-lattice AF Potts model, we have found that sublattice 
$A$ is almost FM ordered, while the spins in the other two sublattices
take the other four values randomly. As the spin configurations at $T=0$
should have zero energy (i.e., it should be a proper 5-coloring of the BH
graph), there are many constraints that make the values of $\mathsf{M}_{aa}$ 
slightly different from the expected values $1$ and $1/16$, respectively. 
Therefore, there are five possible families of ground-state configurations, 
one for each value taken by the spins on sublattice A.  
This behavior is qualitatively the same as the one found in 
Refs.~\cite{selfdual1,selfdual2} for three-state models (see also 
Ref.~\cite{planar_AF_largeq} for four-state models). 

We can therefore describe the ground-state structure of our model
as \emph{five} coexisting ordered phases. Although we are not aware of any 
height representation for this particular model, we conjecture its existence
based on the fact that $5 < q_c(\text{BH})$, and on a conjecture due to 
Henley \cite{Henley_unpub} claiming that any AF Potts model that does not 
admit a height representation at $T=0$ should be disordered at all temperatures 
$T\ge 0$. 

To summarize, the low-temperature phase of the five-state BH-lattice AF 
Potts model can be
well described by five coexisting ordered phases. As there is
obviously a high-temperature disordered phase, there should be a finite-$T$ 
phase-transition point $v_c$. However, we cannot predict from the above 
discussion its nature. 
Finally, if this picture is correct, then the ground-state structure of 
the five-state BH-lattice AF Potts model is identical to that of the 
corresponding FM model, so they should belong to the same universality class. 
By using universality \cite[and references therein]{Fisher_98,Pelissetto_02}
arguments in the FM model, we conclude that the 
transition in both models should be of \emph{first order} for any $q>4$ 
\cite{Baxter_73,Baxter_book}. 

\vspace*{2mm}

%
%
\begin{acknowledgments}
We wish to warmly thank Juan Jes\'us Ruiz--Lorenzo for his participation in an  
early stage of this work, and for useful discussions and correspondence. We
also thank Youjin Deng, Jesper Jacobsen, Jian--Ping Lv, and Alan Sokal for
their collaboration in earlier related works. We thank Alan Sokal for allowing 
us to perform some of the computations reported in this paper on the 
large-memory computers funded by the NSF grant PHY-0424082, and by a computer 
donation from the Dell Corporation. We also thank Gesualdo Delfino for 
drawing our attention to Ref.~\cite{Delfino_17} (which was the main 
motivation of the present paper), and for correspondence. Finally, we 
thank Gesualdo Delfino, Youjin Deng and Jesper Jacobsen for a careful 
reading of the manuscript, and their many valuable suggestions and comments. 
This work has been supported in part by the Spanish Ministerio
de Econom\'{\i}a, Industria y Competitividad (MINECO), Agencia Estatal
de Investigaci\'on (AEI), and Fondo Europeo de Desarrollo Regional
(FEDER) through grant No FIS2017-84440-C2-2-P.
\end{acknowledgments}

%
%


\begin{thebibliography}{99}

\bibitem{Potts_52}  R.B. Potts, Proc. Cambridge Philos. Soc. {\bf 48}, 
    106 (1952).

\bibitem{Wu_82}  F.Y. Wu, Rev. Mod. Phys. {\bf 54}, 235 (1982);
   {\bf 55}, 315 (E) (1983). 

\bibitem{Baxter_book} R.J. Baxter, {\em Exactly Solved Models in Statistical
   Mechanics}\/ (Academic Press, New York, 1982).

\bibitem{Wu_84}  F.Y. Wu, J. Appl. Phys. {\bf 55}, 2421 (1984).

\bibitem{Baxter_73} R.J. Baxter, J. Phys. C {\bf 6}, L445 (1973).

\bibitem{Baxter_82} R.J. Baxter, Proc. R. Soc. Lond. A {\bf 383}, 43 (1982). 

\bibitem{Fisher_98} M.E. Fisher, Rev. Mod. Phys. {\bf 70}, 653 (1998). 

\bibitem{Nienhaus_84} B. Nienhuis, J. Stat. Phys. {\bf 34}, 731 (1984).

\bibitem{DiFrancesco_97}  P. Di Francesco, P. Mathieu and D. S\'en\'echal,
   {\em Conformal Field Theory}\/ (Springer-Verlag, New York, 1997).

\bibitem{Chalker} J.T. Chalker, P.C.W. Holdsworth and E.F. Shender, 
   Phys. Rev. Lett. {\bf 68}, 855 (1992).

\bibitem{Welsh_00} D. Welsh and C. Merino, J. Math. Phys. {\bf 41}, 1127 
   (2000).  

\bibitem{Sokal_00} A.D. Sokal, Physica A {\bf 279}, 324 (2000), 
   \arxiv{cond-mat/9910503}. 

\bibitem{handbook_Tutte} J.A. Ellis-Monaghan and I. Moffatt, editors, 
   \emph{Handbook of the Tutte Polynomial}, Chapman \& Hall/CRC Monographs 
   and Research Notes in Mathematics (Apple Academic Press, Oakville, 
   Canada, 2020).
 
\bibitem{JS_flow} J.L. Jacobsen and J. Salas, J. Combin. Theory B 
   {\bf 103}, 532 (2013), \arxiv{1009.4062}.  

\bibitem{JSS_forests} J.L. Jacobsen, J. Salas, and A.D. Sokal,
   J. Stat. Phys. {\bf 119}, 1153 (2005), \arxiv{cond-mat/0401026}.

\bibitem{SS_97a} J. Salas and A.D. Sokal, J. Stat. Phys. {\bf 86},
     551 (1997), \arxiv{cond-mat/9603068}.

\bibitem{Suto1} A. Sut\"o,  Helv. Phys. Acta {\bf 54}, 191 (1981).

\bibitem{Suto2} A. Sut\"o,  Z. Phys. B {\bf 44}, 121 (1981).

\bibitem{Stephenson} J. Stephenson, J. Math. Phys. {\bf 5}, 1009 (1964).

\bibitem{Adler_95} J. Adler, A. Brandt, W. Janke, and S. Shmulyian, 
   J. Phys. A {\bf 28}, 5117 (1995).

\bibitem{KSS}  R. Koteck\'y, J. Salas, and A.D. Sokal,
   Phys. Rev. Lett. {\bf 101}, 030601 (2008), \arxiv{0802.2270}.

\bibitem{FK1} P.W. Kasteleyn and C.M. Fortuin, J. Phys. Soc. Japan, {\bf 26}
   (Suppl.), 11 (1969). 

\bibitem{FK2} C.M. Fortuin and P.W. Kasteleyn, Physica {\bf 57}, 536 (1972).

\bibitem{JS_unpub}  J.L. Jacobsen and J. Salas, unpublished (2008). 

\bibitem{union-jack}  Y. Deng, Y. Huang, J.L.Jacobsen, J. Salas, and A.D. Sokal,
   Phys. Rev. Lett. {\bf 107}, 150601 (2011), \arxiv{1108.1743}.

\bibitem{planar_AF_largeq}  Y. Huang, K. Chen, Y. Deng, J.L.Jacobsen,
   R. Koteck\'y, J. Salas, A.D. Sokal, and J.M. Swart,
   Phys. Rev. E {\bf 87}, 012136 (2013), \arxiv{1210.6248}.

\bibitem{Saleur_90}  H. Saleur, Commun. Math. Phys. {\bf 132}, 657 (1990). 

\bibitem{Saleur_91}  H. Saleur, Nucl. Phys. B {\bf 360}, 219 (1991). 

\bibitem{transfer1}  J. Salas and A.D. Sokal, J. Stat. Phys {\bf 104},
   609 (2001), \arxiv{cond-mat/0004330}. 

\bibitem{transfer2}  J.L. Jacobsen and J. Salas, J. Stat. Phys {\bf 104},
   701 (2001), \arxiv{cond-mat/0011456}. 

\bibitem{transfer3} J.L. Jacobsen, J. Salas, and A.D. Sokal, 
   J. Stat. Phys. {\bf 112}, 921 (2003), \arxiv{cond-mat/0204587}.  

\bibitem{transfer4} J.L. Jacobsen and J. Salas, J. Stat. Phys. {\bf 122},
   705 (2006), \arxiv{cond-mat/0407444}.   

\bibitem{transfer_torus} J.L. Jacobsen and J. Salas, Nucl. Phys. B {\bf 783},
   238 (2007), \arxiv{cond-mat/0703228}.    

\bibitem{transfer5}  J. Salas and A.D. Sokal, J. Stat. Phys {\bf 135},
   279 (2009), \arxiv{0711.1738}. 

\bibitem{transfer6}  J. Salas and A.D. Sokal, J. Stat. Phys {\bf 144},
   1028 (2011), \arxiv{1002.3761}. 

\bibitem{JSS_tri} J.L. Jacobsen, J. Salas and C.R. Scullard, 
    J. Phys. A {\bf 50}, 345002 (2017), \arxiv{1702.02006}. 

\bibitem{Baxter_86} R.J. Baxter, J. Phys. A {\bf 19}, 2821 (1986).  

\bibitem{Baxter_87} R.J. Baxter, J. Phys. A {\bf 20}, 5241 (1987).

\bibitem{Jacobsen_12} J.L. Jacobsen and C.R. Scullard, J. Phys. A {\bf 45}, 
   494003 (2012), \arxiv{1204.0622} 

\bibitem{Scullard_12}  C.R. Scullard and J.L. Jacobsen, J. Phys. A {\bf 45}, 
   494004 (2012), \arxiv{1209.1451};

\bibitem{Jacobsen_13} J.L. Jacobsen and C.R. Scullard, J. Phys. A {\bf 46}, 
   075001 (2013), \arxiv{1211.4335};

\bibitem{Jacobsen_14} J.L. Jacobsen, J. Phys. A {\bf 47}, 135001 (2014) 
   \arxiv{1401.7847}.

\bibitem{Scullard_16}  C.R. Scullard and J.L. Jacobsen, J. Phys. A {\bf 49}, 
   125003 (2016), \arxiv{1511.04374}.

\bibitem{Vernier} E. Vernier, J.L. Jacobsen and J. Salas, J. Phys. A {\bf 49},
    174004 (2016), \arxiv{1509.02804}. 

\bibitem{selfdual1} J.-P. Lv, Y. Deng, J.L. Jacobsen, J. Salas, and A.D. Sokal,
   Phys. Rev. E {\bf 97}, 040104(R) (2018), \arxiv{1712.07047}. 

\bibitem{selfdual2} J.-P. Lv, Y. Deng, J.L. Jacobsen, and J. Salas,
   J. Phys. A {\bf 51}, 365001 (2018), \arxiv{1804.08911}. 

\bibitem{Delfino_17} G. Delfino and E. Tartaglia, Phys. Rev. E {\bf 96},
   042137 (2017), \arxiv{1707.00998}.

\bibitem{Binder_Landau} D.P. Landau and K. Binder, {\em A Guide to Monte-Carlo 
   Simulations in Statistical Physics}, 3rd ed.\/ (Cambridge University Press,
   Cambridge, 2009). 

\bibitem{Madras} N. Madras, {\em Lectures on Monte Carlo Methods}, Fields 
   Institute Monographs (AMS, Providence, RI, 2002).

\bibitem{Nauenberg_74} M. Nauenberg and B. Nienhuis, Phys. Rev. Lett. {\bf 33},
   944 (1974).

\bibitem{Klein_76} W. Klein, D.J. Wallace, and P.K.P. Zia,  Phys. Rev. Lett. 
   {\bf 37}, 639 (1976).

\bibitem{Fisher_82} M.E. Fisher and A.N. Berker, Phys. Rev. B {\bf 26},
   2507 (1982).

\bibitem{WSK1} J.-S. Wang, R.H. Swendsen, and R. Koteck\'y,
   Phys. Rev. Lett. {\bf 63}, 109 (1989) 

\bibitem{WSK2} J.-S. Wang, R.H. Swendsen, and R. Koteck\'y,
   Phys. Rev. B {\bf 42}, 2465 (1990).

\bibitem{Aizenman}  M. Aizenman and E.H. Lieb, , J. Stat. Phys. {\bf 24},
   279 (1981).

\bibitem{Chow} Y. Chow and F.Y. Wu, Phys. Rev. B {\bf 36}, 285 (1987).

\bibitem{Henley_unpub}  C.L. Henley, Discrete spin models with ``height''
   representations and critical ground states, unpublished manuscript 
   (September, 1993).

\bibitem{Kondev_96} J. Kondev and C.L. Henley, Nucl. Phys. B {\bf 464}, 540
   (1996), \arxiv{cond-mat/9511102}. 

\bibitem{Burton_Henley_97}  J.K. Burton Jr. and C.L. Henley,
   J. Phys. A {\bf 30}, 8385 (1997), \arxiv{cond-mat/9708171}.

\bibitem{SS_98}  J. Salas and A.D. Sokal, J. Stat. Phys {\bf 92},
   729 (1998), \arxiv{cond-mat/9801079}. 

\bibitem{Jacobsen_09}  J.L. Jacobsen, in {\em Polygons, Polyominoes and 
   Polycubes}\/, edited by A.J. Guttmann, Lecture Notes in Physics, vol.~775
   (Springer, Dordrecht, 2009), pp.~347--424.

\bibitem{Privmann} V. Privman, editor, {\em Finite Size Scaling and Numerical 
    Simulation of Statistical Systems} (Word Scientific, Singapore, 1990). 

\bibitem{Madras_Sokal} N. Madras and A.D. Sokal, J. Stat. Phys. {\bf 50}, 
   109 (1988).

\bibitem{Sokal_97} A.D. Sokal, in \emph{Functional Integration: Basics and 
   Applications} (1996 Carg\`ese summer school), edited by C. DeWitt-Morette, 
   P. Cartier, and A. Folacci. (Plenum, New York, 1997), pp.~131--192.  

\bibitem{Lee_91} J. Lee and J.M. Kosterlitz, Phys. Rev. B {\bf 43}, 3265 (1991).

\bibitem{FS_89} A.M. Ferrenberg and R.H. Swendsen, Phys. Rev. Lett. {\bf 61},
   2635 (1988).

\bibitem{Weigel} M. Weigel and W. Janke, Phys. Rev. E {\bf 81}, 066701 (2010),
   \arxiv{1002.4517}.

\bibitem{Young} P. Young, Everything you wanted to know about data analysis 
   and fitting but were afraid to ask, \arxiv{1210.3781} (2012).

\bibitem{Borgs_90} C. Borgs and R. Koteck\'y, J. Stat. Phys. 
   {\bf 61}, 79 (1990). 

\bibitem{Borgs_91} C. Borgs, R. Koteck\'y, and S. Miracle-Sol\'e, 
    J. Stat. Phys. {\bf 62}, 529 (1991). 

\bibitem{Binder_84} K. Binder and D.P. Landau, Phys. Rev. B {\bf 30} 
    1477 (1984).

\bibitem{Challa_86} M.S.S. Challa, D.P. Landau, and K. Binder, Phys. Rev. B 
   {\bf 34} 1841 (1986). 

\bibitem{tilings} B. Gr\"unbaum and G.C. Shephard,
   {\em Tilings and Patterns}\/ (W.H. Freeman and Co., New York, 1987).

\bibitem{Billoire_Lacaze_Morel_92} A. Billoire, R. Lacaze, and A. Morel, Nucl.
   Phys. B {\bf 370}, 773 (1992).

\bibitem{Tsai_98} S.-H. Tsai and S.R. Salinas, Braz. J. Phys. {\bf 28}, 
    58 (1998), \arxiv{cond-mat/9810412}.

\bibitem{SS_00}  J. Salas and A.D. Sokal, J. Stat. Phys {\bf 98},
   551 (2000), \arxiv{cond-mat/9904038}. 

\bibitem{Mohar} B. Mohar,
  Kempe equivalence of colorings, in {\em Graph Theory in Paris}\/,
  edited by J.A.~Bondy, J. Fonlupt, J.L. Fouquet, J.-C. Fournier and 
  J. Ram\'{\i}rez Alfons\'{\i}n (Birkh\"auser, Basel, 2007), pp.~287--297.

\bibitem{Ferreira_Sokal} S.J. Ferreira and A.D. Sokal,
   J. Stat. Phys. {\bf 96}, 461 (1999), \arxiv{cond-mat/9811345}.

\bibitem{SW} R.H. Swendsen and J.-S. Wang, Phys. Rev. Lett. {\bf 58}, 
    86 (1987).

\bibitem{SS_97} J. Salas and A.D. Sokal, J. Stat. Phys. {\bf 87},
     1 (1997), \arxiv{hep-lat/9605018}.

\bibitem{Garoni_11} T.M. Garoni, G. Ossola, M. Polin, and A.D. Sokal,
     J.~Stat. Phys. {\bf 144}, 459 (2011), \arxiv{1105.0373}.

\bibitem{JS_98} J. Salas, J. Phys. A {\bf 31}, 5969 (1998), 
   \arxiv{cond-mat/9802145}. 

\bibitem{MS1} B. Mohar and J. Salas, J. Phys. A {\bf 42}, 225204 (2009),
  \arxiv{0901.1010}.  

\bibitem{MS2} B. Mohar and J. Salas, J. Stat. Mech. P05016 (2010),  
  \arxiv{1002.4279}. 

\bibitem{Lubin_Sokal}  M. Lubin and A.D. Sokal, Phys. Rev. Lett. 
   {\bf 71}, 1778 (1993).

\bibitem{Jerrum} M. Jerrum, private communication. 

\bibitem{McDonald} J. McDonald, B. Mohar, and D. Scheide, 
   J. Graph Theory {\bf 70}, 226 (2012), \arxiv{1005.2248}. 
 
\bibitem{Bonamy_19} M. Bonamy, N. Bousquet, C. Feghali, and M. Johnson,
   J. Combin. Theory B {\bf 135}, 179 (2019), \arxiv{1510.06964}.
 
\bibitem{SS_tri} J. Salas and A.D. Sokal, unpublished (1998).

\bibitem{Ossola_Sokal_04a} G. Ossola and A.D. Sokal, Phys. Rev. E {\bf 70},
    027701 (2004), \arxiv{hep-lat/0403010}.  

\bibitem{Ossola_Sokal_04b} G. Ossola and A.D. Sokal, Nucl. Phys. B {\bf 691},
    259 (2004), \arxiv{hep-lat/0402019}. 

\bibitem{Billoire_94} A. Billoire, T. Neuhaus, and B. Berg, Nucl. Phys. B 
   {\bf 413} 795 (1994), \arxiv{hep-lat/9307017}.   
 
\bibitem{Kotecky-Sokal-Swart}  R. Koteck\'y, A.D. Sokal, and J.M. Swart,
   Commun. Math. Phys. {\bf 330}, 1339 (2014), \arxiv{1205.4472}.

\bibitem{Lee_90} J. Lee and J.M. Kosterlitz, Phys. Rev. Lett. {\bf 65}, 137 
   (1990).

\bibitem{Borgs_92} C. Borgs and W. Janke, Phys. Rev. Lett. {\bf 68}, 1738  
   (1992).

\bibitem{Janke_93} W. Janke,  Phys. Rev. B {\bf 47}, 14757 (1993).   

\bibitem{Janke_94} W. Janke, Recent Developments in Monte-Carlo Simulations 
of First-Order Phase Transitions, in \emph{Computer Simulation Studies in 
Condensed-Matter Physics VII}, edited by D.P. Landau, K.K. Mon, and 
H.-B. Sch\"uttler (Springer-Verlag, Berlin, 1994), pp. 29--43,
\arxiv{hep-lat/9410021}.  

\bibitem{Billoire_95} A. Billoire, Nucl. Phys. B (Proc. Suppl.)  
   {\bf 42} 21 (1995), \arxiv{hep-lat/9501003}.   
 
\bibitem{Berg_93} B.A. Berg and T. Neuhaus, 
   Phys. Rev. Lett. {\bf 68}, 9 (1992), \arxiv{hep-lat/9202004}. 

\bibitem{Borgs_92b} C. Borgs and W. Janke,  J. Phys. I France {\bf 2},
    2011 (1992).

\bibitem{Pelissetto_02} A. Pelissetto and E. Vicari, Phys. Rep. {\bf 368},
    549 (2002), \arxiv{cond-mat/0012164}.

\end{thebibliography}
\end{document}